\documentclass[acmsmall,screen]{acmart}

\AtBeginDocument{%
  \providecommand\BibTeX{{%
    \normalfont B\kern-0.5em{\scshape i\kern-0.25em b}\kern-0.8em\TeX}}}

\setcopyright{acmcopyright}
\copyrightyear{2025}
\acmYear{2025}
\acmDOI{10.1145/XXXXX}

\acmJournal{TOSEM}
\acmVolume{1}
\acmNumber{1}
\acmMonth{6}
\newcommand{\todo}[1]{}
\renewcommand{\todo}[1]{{\color{red} TODO: {#1}}}

\usepackage{caption}
\usepackage{subcaption}
\usepackage{multirow}
\usepackage{amsmath}
\usepackage{graphicx}

\usepackage{amssymb}
\usepackage{tabularx}
\usepackage{colortbl}
\usepackage{enumitem}
\usepackage[most]{tcolorbox}
\usepackage{csquotes}
\usepackage{fontawesome}
\usepackage{tikz}
\usepackage[]{mdframed}
\usepackage{soul}
\usepackage{tasks}
\sethlcolor{yellow}
\renewcommand{\arraystretch}{1.15}

\newcommand{\vcenteredincludep}[1]{\begingroup
\setbox0=\hbox{\includegraphics[height=1.1em]{#1}}%
\parbox{\wd0}{\box0}\endgroup}

\definecolor{signif}{RGB}{228, 245, 246}
\definecolor{greyb}{RGB}{238, 238, 238}
\definecolor{greya}{RGB}{244, 244, 244}
\definecolor{dgrey}{RGB}{224, 224, 224}
\definecolor{deepc}{RGB}{204, 204, 204}
\definecolor{signifl}{RGB}{229, 229, 200}
\definecolor{experience}{RGB}{236, 233, 237}
\definecolor{health}{RGB}{222, 232, 252}

\definecolor{darkgreen}{rgb}{0.0, 0.3, 0.0} 
\definecolor{darkblue}{rgb}{0.0, 0.0, 0.45}

\newcommand{\hcode}[1]{\sethlcolor{greya}\hl{#1}}
\newcommand{\hconcep}[1]{\sethlcolor{greyb}\hl{#1}}
\newcommand{\hcate}[1]{\sethlcolor{dgrey}\hl{#1}}

\def\mybarhhigh#1#2{%%
   {\color{black}\rule{#1mm}{5pt}}  #2}
\definecolor{steBoxLine}{rgb}{0.0, 0.0, 0.0}

\usepackage{enumitem}

\newcommand{\boxquote}[1]{
\vspace{0em}
\hspace{0em}
        \tcbox[on line, 
        boxsep=2pt, left=6pt,right=6pt,top=1pt,bottom=1pt,
        colframe=white,colback=white]{%
            \parbox{0.9\linewidth}{%
            \vspace{-0.4em}
            \hspace{0.2em}\developericon
\textit{“#1”}
                \vspace{0.2em}
            }
        }\vspace{0em}
}%

\newcommand{\developericon}{{\includegraphics[width=14pt]{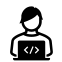}}}

\newcommand{\boxquoter}[1]{
\vspace{0em}
\hspace{0em}
        \tcbox[on line, 
        boxsep=2pt, left=6pt,right=6pt,top=1pt,bottom=1pt,
        colframe=white,colback=white]{%
            \parbox{0.9\linewidth}{%
            \vspace{0.3em}
            \hspace{0.2em}\faCommentO
\textit{ “#1”}
                \vspace{0.2em}
            }
        }\vspace{0em}
}%

\mdfdefinestyle{stebox1}{%
linecolor=steBoxLine,linewidth=2pt,%
leftmargin=0cm,rightmargin=0cm,%
roundcorner=5pt,
topline=false,bottomline=false,rightline=false,leftline=true,%
backgroundcolor=gray!10
}

\newcommand{\linkphd}[2]
{   
    \medskip
    \vspace{0.1cm}
    \begin{mdframed}[style=stebox1,align=center]
    {#2}
    \end{mdframed}
    \vspace{0.3cm}
}

\newtcbtheorem{Summary}{\bfseries Summary}{enhanced,
  coltitle=black,
  top=0.15in,
  colframe=white,
  colback=gray!10!white,
  attach boxed title to top left=
  {xshift=0.5em,yshift=-\tcboxedtitleheight/2},
  boxed title style={size=small,colback=gray!40!white,colframe=gray!40!white}
}{summary}

\newcommand*\circled[1]{\tikz[baseline=(char.base)]{
            \node[shape=circle,draw,inner sep=0.6pt] (char) {#1};}}
\begin{document}

%%
%% The "title" command has an optional parameter,
%% allowing the author to define a "short title" to be used in page headers.
\title{Designing Adaptive User Interfaces for mHealth Applications Targeting Chronic Disease: A User-Centered Approach}

%%
%% The "author" command and its associated commands are used to define
%% the authors and their affiliations.
%% Of note is the shared affiliation of the first two authors, and the
%% "authornote" and "authornotemark" commands
%% used to denote shared contribution to the research.
\author{Wei Wang}
\email{wei.wang5@monash.edu}
\affiliation{%
  \institution{Monash University}
  \streetaddress{Wellington Rd}
  \city{Clayton}
  \state{Victoria}
  \country{Australia}
  \postcode{3031}
}

\author{John Grundy }
\affiliation{%
\institution{Monash University}
  \streetaddress{Wellington Rd}
  \city{Clayton}
  \state{Victoria}
  \country{Australia}
  \postcode{3031}
}
  \email{john.grundy@monash.edu}

\author{Hourieh Khalajzadeh}
\affiliation{%
  \institution{Deakin University}
  \streetaddress{221 Burwood Hwy}
  \city{Burwood}
  \country{Australia}
  \postcode{3125}}
\email{hkhalajzadeh@deakin.edu.au}

\author{Anuradha Madugalla}
\affiliation{%
\institution{Monash University}
  \streetaddress{Wellington Rd}
  \city{Clayton}
  \state{Victoria}
  \country{Australia}
  \postcode{3031}
}
  \email{anu.madugalla@monash.edu}

\author{Humphrey O. Obie}
\affiliation{%
 \institution{Monash University}
  \streetaddress{Wellington Rd}
  \city{Clayton}
  \state{Victoria}
  \country{Australia}
  \postcode{3031}
  }
  \email{humphrey.obie@monash.edu}

%%
%% By default, the full list of authors will be used in the page
%% headers. Often, this list is too long, and will overlap
%% other information printed in the page headers. This command allows
%% the author to define a more concise list
%% of authors' names for this purpose.
\renewcommand{\shortauthors}{Wang, et al.}

%%
%% The abstract is a short summary of the work to be presented in the
%% article.

\begin{abstract}

\textbf{Mobile Health (mHealth) }applications have demonstrated considerable potential in supporting chronic disease self-management; however, they remain underutilized due to low engagement, limited accessibility, and poor long-term adherence. These issues are particularly prominent among users with chronic disease, whose needs and capabilities vary widely. To address this, \textbf{Adaptive User Interfaces (AUIs)} offer a dynamic solution by tailoring interface features to users’ preferences, health status, and contexts. This paper presents a two-stage study to develop and validate actionable AUI design guidelines for mHealth applications. In \textit{stage one}, an AUI prototype was evaluated through focus groups, interviews, and a standalone survey, revealing key user challenges and preferences. These insights informed the creation of an initial set of guidelines. In \textit{stage two}, the guidelines were refined based on feedback from 20 end users and evaluated by 43 software practitioners through two surveys. This process resulted in nine finalized guidelines. To assess real-world relevance, a case study of four mHealth applications was conducted, with findings supported by user reviews highlighting the utility of the guidelines in identifying critical adaptation issues. This study offers actionable, evidence-based guidelines that help software practitioners design AUI in mHealth to better support individuals managing chronic diseases.
\end{abstract}

%%
%% The code below is generated by the tool at http://dl.acm.org/ccs.cfm.
%% Please copy and paste the code instead of the example below.
%%


%%
%% Keywords. The author(s) should pick words that accurately describe
%% the work being presented. Separate the keywords with commas.

\received{24 March 2024}
\received[revised]{02 Feb 2025}
\received[accepted]{17 April 2025}

%%
%% This command processes the author and affiliation and title
%% information and builds the first part of the formatted document.
\maketitle
\section{Introduction}
Chronic diseases, including conditions such as asthma, heart disease, and diabetes, present formidable challenges to healthcare systems around the world \cite{Who}. The management of these long-term health conditions transcends simple medical treatment, with an increasing emphasis on empowering patients to actively engage in \textbf{self-management} practices \cite{Who}. The use of \textbf{Mobile Health (mHealth)} applications has emerged as a promising avenue for promoting self-management by strengthening medication adherence and facilitating self-tracking capabilities \cite{hamine2015impact}. Despite their potential, studies indicate that many individuals who could benefit from mHealth technology do not fully engage with these tools, particularly in developing countries where non-adherence rates tend to be higher \cite{han2010professional, beaglehole2008improving}. Increasing mHealth adoption among chronic disease patients requires greater customization and flexibility \cite{choe2017semi,han2010professional}, yet several design and implementation challenges must first be addressed. Due to their inherent \textit{heterogeneity}, chronic diseases affect individuals in various ways and can co-occur with other medical or psychological disorders, making self-management more complex \cite{harvey2012future,audulv2013over,di2019chronic,audulv2013over}. In addition, chronic diseases typically persist throughout a person’s \textit{lifetime} \cite{Who, harvey2012future}, mHealth applications need to be designed to sustain user engagement and motivation over the long term.

One critical area of research that addresses barriers to sustained use of technology is accessibility. Traditionally, accessibility research has concentrated on meeting the needs of individuals with disabilities such as blindness, low vision, and physical impairment, dominating over half of the studies conducted in the last decade \cite{zhou2020making, mack2022chronically, mack2021we, depuy2023accessible}. Chronic diseases, which affect a significant portion of the population, remain underrepresented in accessibility research, highlighting the need for solutions to address the diverse challenges faced by individuals managing these conditions. \textbf{Adaptive User Interfaces (AUIs)} offer a promising solution to bridge this gap by dynamically tailoring the \textbf{User Interface (UI)} to accommodate the unique needs, goals, and contextual circumstances of each individual \cite{Norcio1989, firmenich2019user}. In this work, we focus on \textbf{\textit{adaptation}}, which broadly refers to interface modifications, encompassing both system-driven changes (adaptivity) and user-controlled customizations (adaptability) \citep{Norcio1989}. Additionally, we acknowledge mixed-initiative adaptation, where both the user and the system collaboratively share the responsibility of adapting the interface \cite{horvitz1999principles,mezhoudi2021toward}. Despite increasing interest in employing AUIs in chronic disease-related applications \cite{setiawan2019adaptive, talboom2018chronic}, they often overlook various user characteristics and interactions \cite{grua2020reference,mclean2011telehealthcare}. Most chronic disease-related applications view design adaptations as a \textit{complex task}, requiring expertise in psychology, physiology, human behavior, user experience analysis and interpretation with regard to the underlying behavior and health status \cite{grua2020reference,Vasilyeva2005}. Many software practitioners lack the expertise required to implement theories and models, making it challenging to access the necessary skills for mHealth-related projects \cite{chib2018theoretical}. Moreover, there is a lack of established resources and guidelines for AUI development \cite{zimmerman2020ux}. Existing studies on the development of AUI within these applications reveal a substantial deficiency in the literature concerning the foundational stages of AUI design \cite{wang2023adaptive}. This gap consequently limits the understanding of how AUIs are perceived and utilized by individuals with chronic diseases, potentially resulting in the underutilization of the benefits that adaptive systems can offer \cite{kornfield2020energy}. Bridging this gap is therefore essential for optimizing AUI design and improving user engagement in chronic disease-related applications.

In this work, we present a two-stage approach aimed at designing and validating mHealth adaptation guidelines for chronic disease-related applications. In \textbf{stage one}, we developed an AUI prototype and collected input from chronic disease patients through focus groups, interviews, and surveys, forming the initial design guidelines. In \textbf{stage two}, we refined these guidelines based on end-user and software practitioner feedback, and subsequently validated through real-world mHealth applications. We presented the initial findings of user study in stage one at the 2024 International Conference on Software Engineering (ICSE) \cite{wang2024adaptive}. In this work, we build on and significantly extend our earlier conference paper by advancing from stage one to stage two. Specifically, we: 1) enhance our user survey findings in stage one by including all collected data and conducting further analysis and interpretation; 2) develop a set of new guidelines for designing AUIs for mHealth application targeting chronic diseases; 3) refine these guidelines through an additional round of feedback from end-users and software practitioners; and 4) conduct a case study on existing mHealth applications to further validate the guidelines. This work offers five key research contributions:

\begin{enumerate}[left=0pt,itemsep=1pt, topsep=5pt]

    \item Advancement of the discussion on this topic by presenting design trade-offs in AUI when designing technology for users with chronic disease;
    \item Deeper understanding of how user preferences for different aspects of adaptations are influenced by different demographic factors, including cultural backgrounds, contextual circumstances, Health condition and age;
    \item Development of comprehensive actionable guidelines for researchers, practitioners, and designers for designing AUI in the chronic disease domain;
    \item Multi-stage evaluation and refinement of actionable guidelines through feedback collection from end-users and software practitioners;
    \item Validation of refined guidelines through a case study on real-world mHealth applications, demonstrating their practical applicability and effectiveness in addressing adaptation challenges in the chronic disease domain.
\end{enumerate}

The rest of the paper is organized as follows. Section \ref{sec:motiv} provides a summary of our study's motivation and summarizes key related work. In Section \ref{sec:method}, we present our research methodology. The results of the qualitative and quantitative analysis of stage one are detailed in Sections \ref{sec:stageoneinterview} and \ref{sec:stageonesurvey}, respectively. In stage two, Section \ref{sec:userfeedback} presents the evaluation results of the guidelines from an end-user feedback survey. Section \ref{sec:softfeedback} discusses the feedback from software practitioners on the refined guidelines. Furthermore, Section \ref{sec:case} evaluates the practical applicability of the guidelines through a case study involving real-world mHealth applications. The finalized set of guidelines is outlined in Section \ref{sec:guideline}. Lastly, Section \ref{sec:lim} discusses threats to validity of the study and Section \ref{sec:con} concludes the paper.

\section{Motivation and Related Work} \label{sec:motiv}

\subsection{Chronic Disease Self-Management} 
Responsible for 41 million deaths annually and representing 74\% of all global deaths \cite{Who, WHO2023}, chronic disease poses a major global health challenge. The prevalence of chronic diseases is steadily increasing, driven by the reclassification of previously fatal diseases as chronic diseases and the aging of the population \cite{beaglehole2008improving}. Traditional treatment paradigms do not address the multifaceted nature of chronic diseases \cite{Who,hird2016digital}, as treatment cannot be based solely on biological parameters, which require active engagement and self-management by patients \cite{deiss2006improved}. Self-management involves actively participating in self-care activities to improve behavior and well-being \cite{deiss2006improved}. Research has highlighted the efficacy of mHealth applications in supporting the self-management of chronic diseases \cite{pare2007systematic,hamine2015impact}. However, despite the potential benefits of such interventions, evidence suggests that those who stand to gain the most often exhibit lower levels of engagement and adoption \citep{han2010professional}. This disparity is especially concerning given that approximately 77\% of chronic disease-related fatalities occur in \textit{low- and middle-income countries}, where access to consistent and effective healthcare remains limited \cite{WHO2023}.

There are well-recognized challenges in designing mHealth technology for chronic disease self-management, one of which is the high heterogeneity of chronic diseases. These diseases affect patients differently in terms of triggers, symptoms, and severity \cite{harvey2012future,deiss2006improved,audulv2013over}, resulting in \textit{a wide range of self-management needs across individuals}. Secondly, the design of the UI must account for the evolving nature of chronic diseases over time \cite{di2019chronic,audulv2013over}. As many chronic diseases fluctuate, the corresponding adjustments to self-management strategies are required \cite{lorig2003self}. Furthermore, chronic diseases are often co-morbid with other medical or psychological disorders, resulting in a broader spectrum of user characteristics and functionality requirements \cite{di2019chronic,islam2014multimorbidity}. For example, diabetes can cause various complications, such as vision loss, amputation, neuropathy, end-stage renal disease, cardiovascular disease, infections, and cognitive impairment \cite{vijan2015type}. Thirdly, the vast majority of chronic diseases are \textit{long-lasting and generally lifelong} \cite{Who,harvey2012future}, generally manageable but not curable. Therefore, mHealth applications must sustain user engagement and motivation over the long term. In addition to the diverse nature of chronic diseases, patients also have a wide range of backgrounds, expertise, and demographic, psychological, and cognitive characteristics \cite{Vasilyeva2005}.
While adapting interfaces can improve user acceptance and motivation \cite{duplaga2015cross, han2010professional}, existing AUI solutions often have limited adaptation options, relying on predefined rules and overlooking unique characteristics of users \cite{grua2020reference, mclean2011telehealthcare}. Creating mHealth applications that offer access to knowledge and information is essential to prevent physical and social disparities \cite{han2010professional}, especially in developing countries where nonadherence to treatment remains a significant issue \cite{beaglehole2008improving}.

\subsubsection{Chronic disease-related applications and accessibility study}  While multidisciplinary efforts have contributed to chronic disease-related applications, many technological solutions remain anchored in \textbf{medicalized perspectives}, often viewing users solely as patients rather than individuals with various priorities and lifestyles \cite{burgess2019tricky,huh2012collaborative}. A considerable amount of research exists on the effect of these mHealth applications on treatment regimen adherence (e.g. \cite{hamine2015impact,gandapur2016role}), application design features (e.g. \cite{sockolow2021integrative,matthew2016designing}), and the evaluation of mHealth applications (e.g. \cite{yang2020intervention}). Some exceptions exist in exploring design strategies for individuals with chronic diseases (e.g. \cite{mack2022chronically,paymal2024good,epstein2016beyond}). These initiatives aim to explore the evolving nature of chronic diseases and how changes in users’ physical, cognitive, and emotional needs affect their interaction with technology. However, many self-management tools continue to generalize the patient experience and overlook the complexity of managing health in everyday contexts \cite{nunes2015self}. To address this gap, researchers have called for greater \textit{\textbf{customization and personalization}} in mHealth applications design, advocating systems that can adapt to users’ unique health trajectories and life contexts \cite{nunes2015self,ryu2023you,mclean2011telehealthcare, grua2020reference}. A critical domain that aligns with these goals is accessibility research, which traditionally focuses on improving system usability for individuals with disabilities \cite{radcliffe2021pilot,zhou2020making,alshayban2020accessibility,daihua2015accessibility}, including those with visual \cite{depuy2023accessible,alshayban2020accessibility,milne2014accessibility} and cognitive impairments \cite{yu2017accessibility}, or those of low socioeconomic backgrounds \cite{stowell2018designing}. Chronic diseases remain underrepresented in accessibility studies, despite the growing evidence of substantial accessibility challenges in self-management applications \cite{kim2019mobile,mack2021we}. Researchers argue that individuals with chronic diseases encounter nuanced challenges not fully captured by the general disability frameworks \cite{mack2022chronically}. In this context, AUI offers significant potential in mHealth applications by enhancing both accessibility and usability. By tailoring interfaces to the diverse capabilities and preferences of users, AUI ensures that these applications are not only functional but also inclusive \cite{wang2023adaptive}.

\begin{figure}[b]

     \begin{subfigure}[b]{0.25\textwidth}
         \centering
         \includegraphics[width=0.7\textwidth]{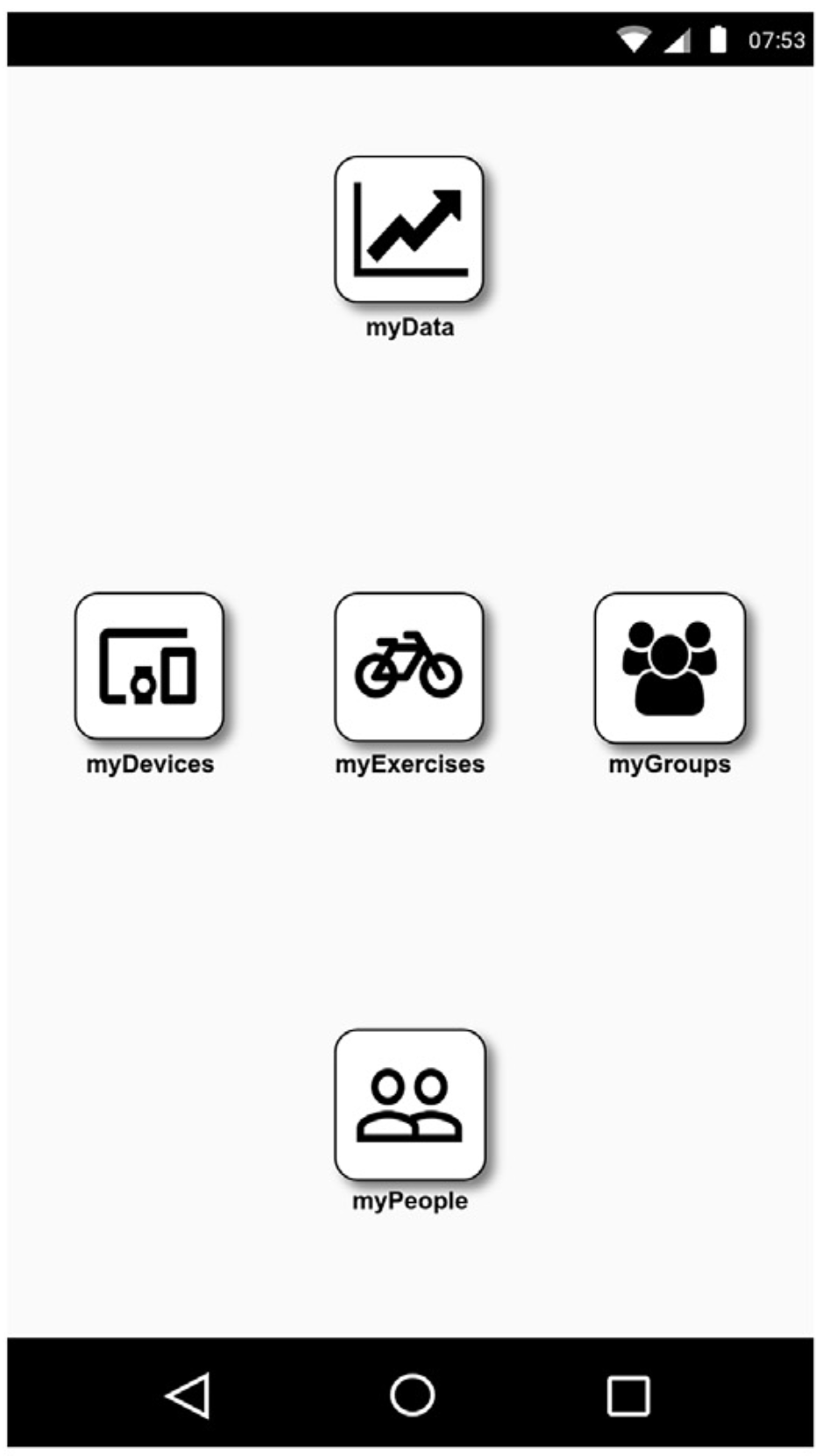}
         \caption{\scriptsize Normal page enabling regular application use \cite{pagiatakis2020intelligent}.}
         \label{sub:a}
     \end{subfigure}
     \hfill
      \begin{subfigure}[b]{0.25\textwidth}
         \centering
         \includegraphics[width=0.7\textwidth]{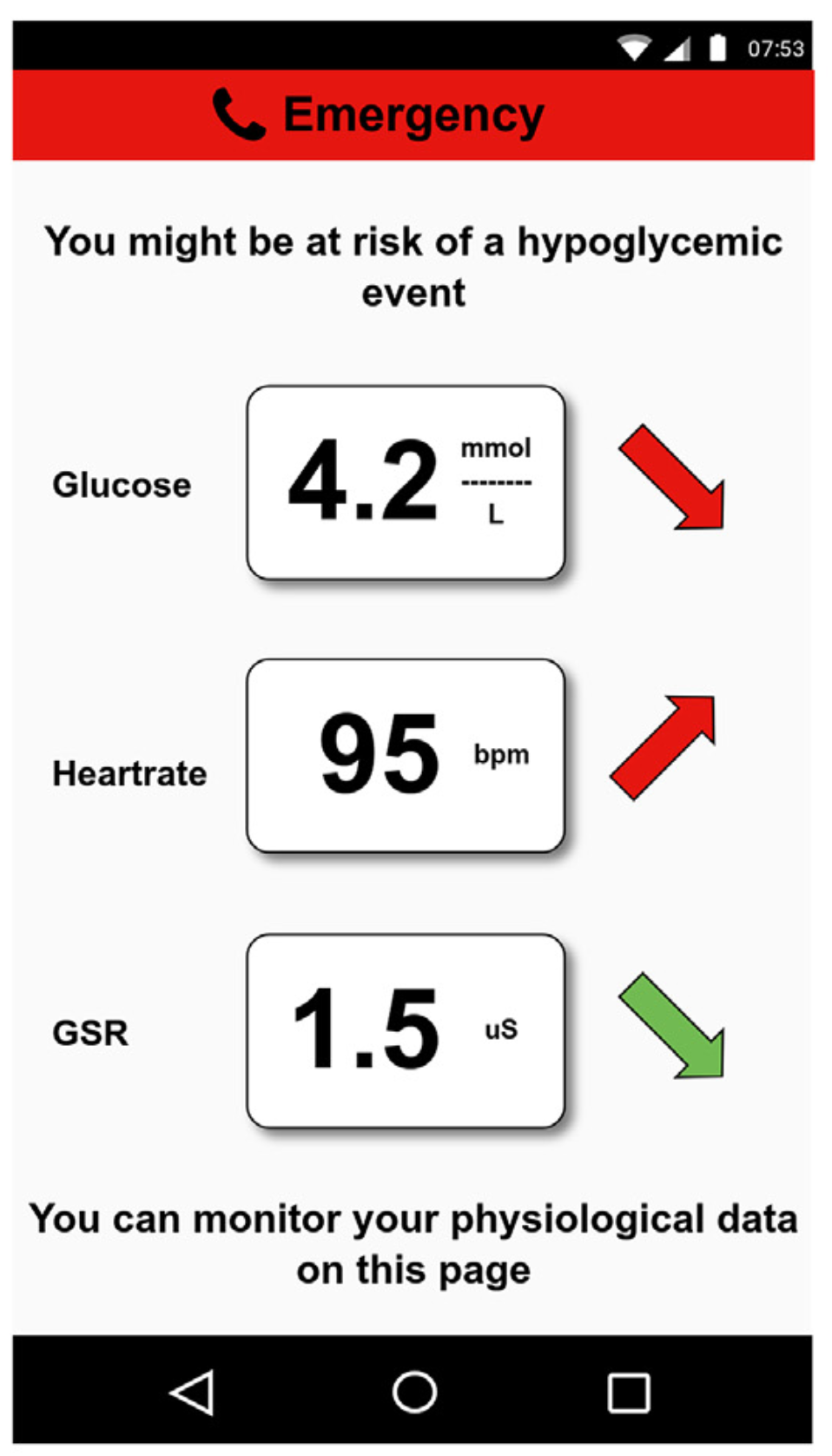}
         \caption{\scriptsize Page only presents relevant physiological information\cite{pagiatakis2020intelligent}.}
         \label{sub:b}
     \end{subfigure}
          \hfill
        \begin{subfigure}[b]{0.48\textwidth}
         \centering
         \includegraphics[width=\textwidth]{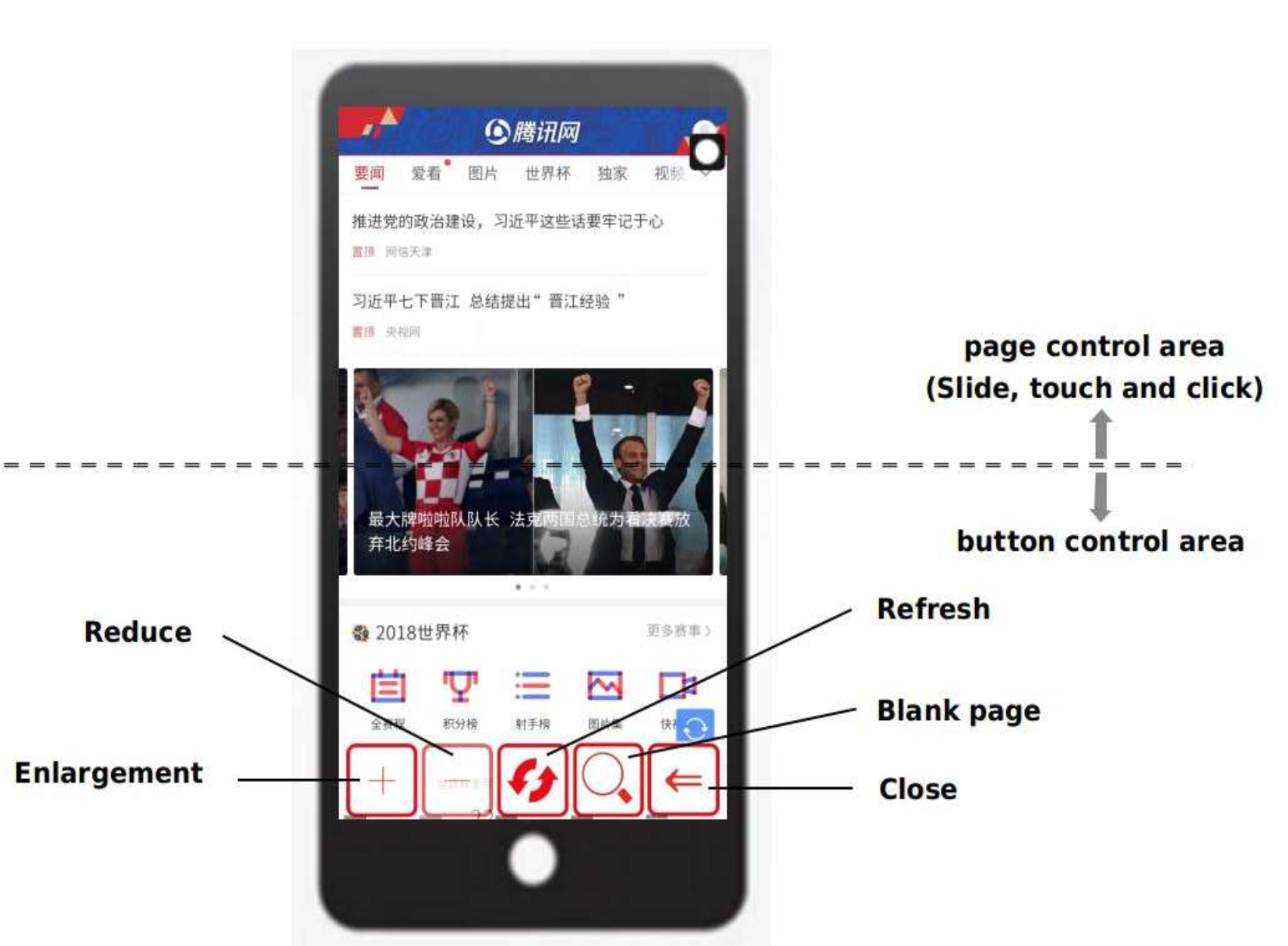}
         \caption{\scriptsize Adaptable web-browsing PD-helper\cite{jabeen2019improving}.}
         \label{sub:c}
     \end{subfigure}
    \caption{Examples of adaptive user interfaces. }

\end{figure}
\subsection{Adaptive User Interfaces}
Given the diversity of users and contexts, fixed-context UIs are often inadequate; AUIs, which dynamically adjust to user preferences and environments, offer a promising solution. \citet{mctear1993user} defines an AUI as \textit{``a software artefact that improves its ability to interact with a user by constructing a user model based on partial experience with that user”}. Recent research highlights the importance of adaptive mHealth applications in facilitating chronic disease self-management \cite{nunes2015self,ryu2023you,mclean2011telehealthcare, grua2020reference}. For example, in the treatment of diabetic hypoglycemia, an application developed by \citet{pagiatakis2020intelligent} adapts its navigation system during hypoglycemic events. Under normal conditions, the application displays a standard homepage for everyday use (see Figure \ref{sub:a}). However, during a hypoglycemic episode, as shown in Figure \ref{sub:b}, the application restricts access to non-essential sections and prominently features a quick access emergency contact button to ensure user safety. Similarly, \citet{jabeen2019improving} designed PD-Helper, AUIs tailored to support individuals with Parkinson’s disease (see Figure \ref{sub:c}). The application provides customization control, allowing users to adjust font size, refresh pages, open new tabs, and return to the main menu with a single tap, thus addressing the motor limitations commonly experienced by these patients.

\subsubsection{AUIs and healthcare}
Previous research has explored the application of AUIs in systems adapted for \textit{\textbf{healthcare professionals}} \cite{eslami2018user,greenwood2003agent,vogt2010adaptive}. \citet{eslami2018user} conducted interviews and observations to investigate user preferences regarding data entry, language and vocabulary, and information presentation, as well as providing help, warning, and feedback, with a primary focus on healthcare professionals. Similarly, \citet{vogt2010adaptive} examined AUI design issues that aim to simplify input and reduce the potential for errors, particularly in contexts such as smart hospitals. \citet{greenwood2003agent} introduced a novel approach using reactive agents for AUIs in diabetes treatment decision support, customizing data display according to clinician preferences. Despite the initial emphasis on healthcare professionals, there is an increasing recognition of the diverse user base of mHealth applications. Consequently, there has been an increasing number of studies describing various approaches, applications, and tools specifically tailored for \textit{\textbf{patient-focused}} AUIs. Existing studies on AUI frameworks often focus on particular adaptive components or specific aspects of patient management \cite{shakshuki2015adaptive, frohlich2009loca, yuan2012fuzzy}. \citet{shakshuki2015adaptive} proposed an AUI architecture for patient monitoring with a focus on health-related information adaptation. \citet{frohlich2009loca} explored the application of the LoCa (A Location and Context-aware eHealth Infrastructure) project, primarily to monitor physiological data and the activity status of patients within a digital home environment, facilitating context-aware adaptation of workflows. \citet{yuan2012fuzzy} designed a fuzzy-logic-based context model for personalized healthcare services in chronic diseases, prioritizing the prediction of health problems and preventive measures based on user data rather than UI adaptation to individual user needs and context. 

AUIs have been deployed across a spectrum of mHealth applications, ranging from stroke rehabilitation \cite{burke2009optimising}, diabetes \cite{pagiatakis2020intelligent}, cardiac disease \cite{mohan2008medinet}, dementia \cite{awada2018adaptive, gulla2019study}, and Parkinson's disease \cite{jabeen2019improving}. These applications exhibit different adaptations, ranging from the adjustment of the difficulty levels of exercise activity \cite{burke2009optimising} to the customization of health-related information \cite{awada2018adaptive, gulla2019study}, navigation adaptations \cite{pagiatakis2020intelligent, awada2018adaptive}, multimodal interfaces \cite{pagiatakis2020intelligent}, information architecture \cite{mohan2008medinet} and graphic design \cite{gulla2019study, mohan2008medinet, pagiatakis2020intelligent, jabeen2019improving}. Based on the findings of the previous \textbf{Systematic Literature Review (SLR)} by \citet{wang2023adaptive}, existing user models, which represent various user dimensions to support adaptation, predominantly utilize physical and physiological characteristics to generate AUI. Most existing studies acquire user data through methods such as \textit{user questionnaires} or by allowing users to \textit{manually adjust} settings and preferences during application usage. Regarding the adaptation mechanism, studies predominantly use \textit{rule-based} adaptation and predictive algorithm-based adaptation to adapt interfaces to user needs. However, most of these studies lack detailed explanations of their AUI development process, especially with regard to the initial stages that involve the collection of diverse end-user requirements. Furthermore, the typical evaluation approach for AUIs focuses on overall application effectiveness without proper comparisons to non-adaptive UIs, which complicates drawing specific conclusions about the impact of AUIs. Although research in other domains suggests that AUIs can improve user performance and satisfaction compared to non-adaptive baseline \cite{gajos2006exploring, paymans2004usability, Tsandilas2005AnEA, Li2014AdaptiveCA}, disruptive adaptations, which alter user accustomed interaction patterns or break conventions, can result in frustration or dissatisfaction \cite{Findlater2004ACO, rivera2005effect}. Despite these insights, our understanding of how individuals with chronic diseases use AUIs and how to design mHealth applications that integrate AUIs to maximize benefits while minimizing costs for this population remains limited.

\section{Methodology}\label{sec:method}
\begin{figure}[t]
\centering
\includegraphics[width=1\textwidth]{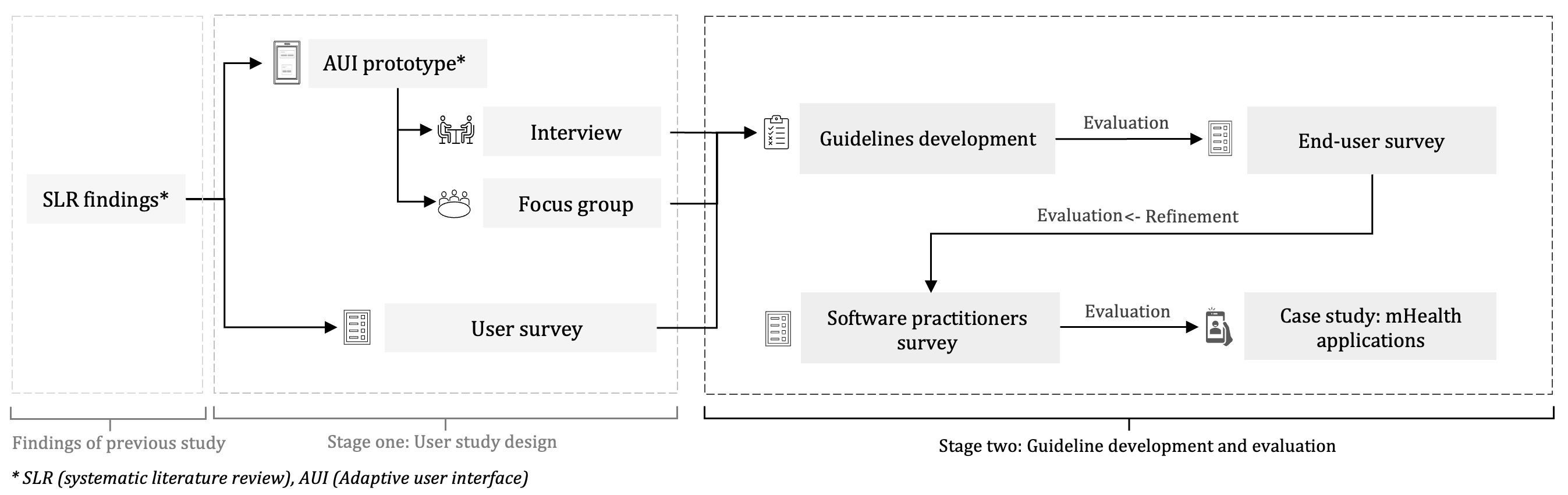}
\caption{Research methodology.}
\label{fig:studymethod}
\end{figure}

We conducted the study in two distinct stages to systematically develop and validate guidelines for AUI design in mHealth applications that target chronic disease management. In \textbf{\textit{stage one}}, we developed the AUI prototype tailored for chronic disease-related applications and conducted qualitative research through interviews and focus group studies with individuals managing chronic diseases. In parallel, a quantitative survey was administered to capture user preferences related to different aspects of adaptation. In \textbf{\textit{stage two}}, the analysis of both qualitative and quantitative findings, combined with insights synthesized from existing literature, led to the development of an initial set of guidelines for designing adaptive mHealth applications. The preliminary guidelines are further evaluated and refined through survey feedback from both end-users and software practitioners, resulting in the finalization of nine guidelines. The refined guidelines are then validated by applying them to real-world mHealth applications. We also compared our guidelines with existing mHealth usability guidelines and analyzed whether the issues identified through our approach aligned with those mentioned in user app review (see Figure \ref{fig:studymethod}).

\subsection{Stage One: User Study Design}
Building on the insights from an earlier SLR \cite{wang2023adaptive}, our research extends the adaptation categories and integrates them into a prototype that includes three primary types of adaptations: \textit{presentation adaptation, content adaptation, and behavior adaptation}. Details of the prototype can be found in \citet{wang2024adaptivep}. 
% The user study consists of two parallel investigations, as illustrated in Figure \ref{fig:studymethod}. We conducted a qualitative investigation through focus groups and interviews to examine how individuals experience AUI in the context of chronic diseases by using the AUI prototype. At the same time, a quantitative survey was administered to collect user preferences regarding different dimensions of adaptation.

\subsubsection{Focus group and interview studies}
Grounded in the SLR on contemporary developments in AUI within the field of chronic diseases \cite{wang2023adaptive}, we designed a focus group and interview protocol consisting of two sections. The \textbf{first} section collected detailed \textit{demographic information} about the participants, including their chronic disease and their use pattern of mHealth applications through a Qualtrics survey \footnote{https://www.qualtrics.com}. The \textbf{second} section collected the \textit{participant’s
views on the different adaptations we present in the AUI prototype}. The participants initially reviewed a brief adaptation video accompanied by audio explanations, providing detailed introductions to each type of adaptation. This visual aid was followed by hands-on interaction with the prototype. Instructions were provided in the accompanying slides, encouraging participants to actively engage with the prototype. In the event of any difficulties, participants could refer to the instructions for assistance, ensuring a seamless user experience. The researcher remained readily available to offer support, allowing participants to focus solely on the example adaptations. As a token of appreciation, all participants were offered an AU\$30 virtual gift voucher. Employing a \textbf{\textit{theoretical sampling}} approach to recruit participants entails choosing new individuals based on particular criteria, which include the advancement of data collection and analysis, as well as the development of categories and concepts \cite{hoda2021socio}. Qualitative data collection and analysis followed an iterative process in three different data collection iterations, with information on the recruiting iteration presented in \citet{wang2024adaptive}. To improve the reliability and validity of our findings through \textbf{\textit{methodological triangulation}} \cite{denzin2017research}, we employed both semi-structured interviews and focus groups. Focus groups provided valuable insight into collective attitudes and shared perceptions through dynamic group interactions \cite{acocella2012focus}, while interviews allowed a detailed exploration of individual experiences, allowing for a more nuanced understanding of personal perspectives \cite{WILSON201423}. By combining these methods, we captured both the breadth of group consensus and the depth of individual viewpoints.

\subsubsection{User survey} Participants in interviews and focus groups may provide socially desirable responses, introducing bias \cite{harris2019mixing}, which is further compounded by the subjectivity of inductive coding in qualitative analysis \citet{hoda2021socio}. To mitigate these biases, \textbf{\textit{between-method triangulation}} was employed, involving the triangulation of data using a combination of quantitative and qualitative techniques \cite{fusch2018denzin, denzin2017research, oppenheim2000questionnaire}. We conducted an anonymous online survey using Google Forms aimed at individuals with chronic diseases to gather quantitative feedback on their preferences on various aspects of adaptation. The survey design is informed by the SLR conducted in AUI in the context of chronic diseases \cite{wang2023adaptive}. The survey data collection went through three phases: 1) email ads and social media (58 responses), 2) physical posters distributed at Baker Institute Diabetes Clinic and Alfred Hospital (13 responses), and 3) advertisements through Dementia Australia, Stroke Foundation and Kidney Australia (19 responses). Participants who completed the survey were eligible for an AU\$20 virtual gift card draw. The user survey questions are provided in the Appendix \ref{app:A}.

\subsection{Stage Two: Guidelines Development and Evaluation}
The guideline development process follows and adapts the framework proposed by
\citet{hermawati2015user} and \citet{quinones2018methodology}. Data collected through surveys, interviews, and focus group sessions in stage one, along with findings from existing literature on guidelines, design considerations, and evaluation criteria, were synthesized to inform guideline creation. The process also included iterative refinement and validation based on participant feedback, ensuring the relevance and applicability of the guidelines.

\subsubsection{End-user guideline evaluation survey}
We carried out a preliminary evaluation of our guidelines with end-users, primarily concentrating on assessing their\textbf{ \textit{clarity}} and perceived \textbf{\textit{usefulness}}. Considering the potential scarcity of documentation on design considerations and guidelines, especially in emerging technology domains, it is imperative to \textit{maintain user involvement throughout the process}, and ensure that guidelines, though originally created for developers, designers, and researchers, hold significance for end-users \cite{hermawati2015user, zaphiris2007systematic, lechner2013chicken, zaphiris2009user}. Therefore, we strived to validate the efficacy and acceptance of these guidelines with the same cohort of users involved in our interview and focus group study (Stage One). We reached out to all participants via email, inviting them to complete a guideline evaluation survey. Participants were briefed on the findings of the user study and provided with an overview of the generated guidelines. To ensure anonymity and simplify the process, participants used a unique \textit{withdrawal code} from their initial registration to link their responses to the feedback survey and user study. Subsequently, participants were requested to evaluate each of the guidelines and offer suggestions regarding \textbf{\textit{additions, removals, or edits}} for each guideline. 

\subsubsection{Software practitioners guideline evaluation survey}
After end-user evaluation, another online survey was conducted among software practitioners and other relevant stakeholders involved in mHealth application development to evaluate the effectiveness of the proposed guidelines. In the survey, respondents were provided with definitions of mHealth applications and AUI, along with access to the proposed guidelines via publicly available links. Demographic information was gathered from respondents, along with their evaluations of the proposed guidelines. Participants were specifically asked to assess the \textit{applicability} of the guidelines in real-world practice, and to provide insights into their \textit{strengths, limitations, and potential areas for improvement}. The survey questions for evaluating the guidelines, which are adapted from \citet{shamsujjoha2024developer}, can be found in Appendix \ref{app:B}. Following ethics approval, we conducted a pilot study with representatives from the target population to evaluate the survey’s design, specifically the clarity of the questions and response options. Overall, participants expressed satisfaction with the survey and did not suggest changes to the questionnaire structure or answer formats. However, they recommended the inclusion of \textbf{comprehension checks} to ensure that the respondents had adequately understood the guidelines, especially in cases where the guidelines may not be thoroughly reviewed prior to answering related questions. The survey was revised to include two comprehension check questions, allowing participants two attempts to complete them, with \textit{\textbf{exclusion}} from the study if both attempts were failed. Moreover, participants who do not meet the screening criteria, which focused on their \textit{expertise in developing health-related applications, particularly those targeting chronic diseases}, were also excluded. Traditional attention check questions were not included, as the study emphasized qualitative responses and prioritized meaningful engagement and understanding of the proposed guidelines. An initial set of 9 responses was collected through personal networks. Due to the limited sample size, survey distribution was subsequently expanded via Prolific\footnote{https://www.prolific.com/}, resulting in 34 additional responses and a total of 43 participants. During recruitment on Prolific, a customized screening tool was employed to ensure that participants were professionals working in the technology sector. To maintain the integrity of the sample, individuals whose responses did not align with their prescreening information on Prolific were excluded from the study.

\subsubsection{Case study}\label{sec:reviewanalysis}
Case studies are an effective tool for validating guidelines, complementing expert reviews \cite{goundar2024development}, and providing insights into their effectiveness by analyzing existing mHealth applications. To ensure meaningful comparisons across the selected applications, all were selected in the domain of \textbf{diabetes} management, providing a consistent context for evaluation. The selection of applications was guided by three key criteria: 1) consistently high user ratings in both the iOS App Store and Google Play, 2) a large number of downloads and installations across the iOS and Android platforms, and 3) availability as free applications with optional in-app purchases on both operating systems \cite{radcliffe2021pilot}. The case study aimed to assess whether widely used mHealth applications incorporate adaptive features, evaluate the effectiveness of the proposed guidelines in identifying critical design issues, and compare these findings with those derived from a control guideline. To further validate the evaluation outcome, user reviews were analyzed to determine whether the problems flagged by the evaluators were also reflected in the end-user feedback. The analysis was limited to reviews in English. Prior to analysis, the review texts were pre-processed using the NLTK library \cite{NLTK}, including tokenization, stemming, spelling correction, case normalization, and noise word removal \cite{dkabrowski2022analysing}. Then a keyword search was performed using a refined list developed from the proposed guidelines, the control guideline, and existing accessibility standards. This list was iteratively updated to align with emerging themes related to adaptation, accessibility, and usability.

% \begin{figure}[h!]
%     \centering
%     \includegraphics[width=1\linewidth]{Figure/Guideline_evaluation.png}
%     \caption{Guideline evaluation process}
%     \label{fig:guideline}
% \end{figure}

\subsection{Data Analysis}

\subsubsection{Stage One: Qualitative data analysis} 

We used the \textit{data analysis procedures} of \textbf{Socio-Technical Grounded Theory (STGT)} \cite{hoda2021socio} to analyze data from the focus group and the interview study. This decision was primarily driven by the close alignment between the focus of our study and the principles of the socio-technical research framework that STGT is built upon, as our investigation revolves around AUIs in applications related to chronic diseases, a socio-technical phenomenon that encompasses both human and social aspects, as well as technical aspects. STGT allows for selective application by integrating its core data analysis procedures of open coding, constant comparison, and memoing, while traditional grounded theory methods such as Glaserian \cite{glaser1992basics} and Strauss-Corbinian \cite{strauss1994grounded} are developed as standalone methodologies for theory development. We obtained consent from the participants to transcribe the audio recordings, and subsequently stored and analyzed the data using NVivo\footnote{https://lumivero.com/products/nvivo/}.

\begin{figure}[h]
        \centering
        \includegraphics[width=0.98\textwidth]{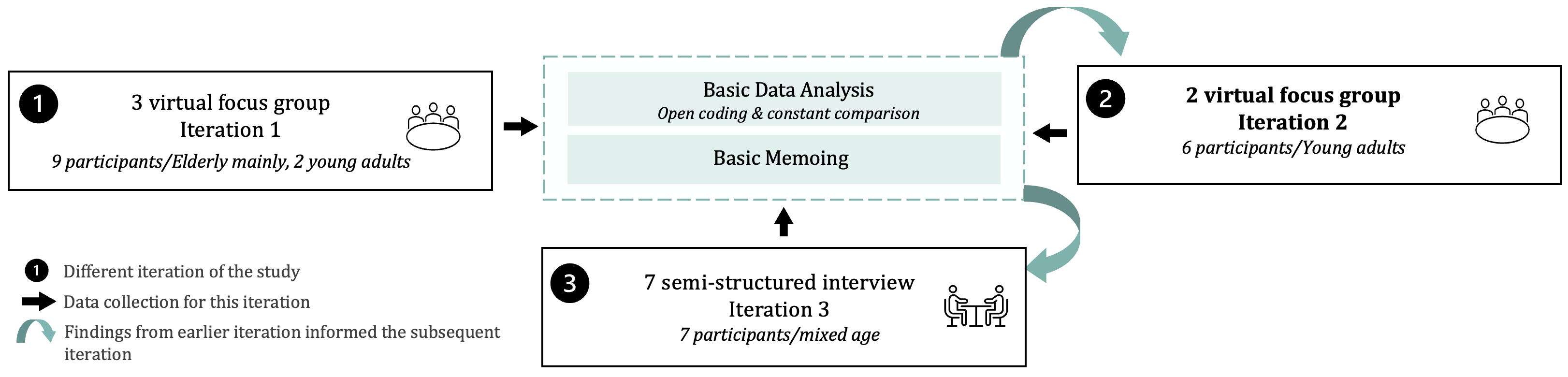}
        \caption{Process of applying STGT for data analysis in the focus group and interview study.}
        \label{fig:STGT}
    \end{figure}

\begin{figure}[b]
        \centering
        \includegraphics[width=0.95\linewidth]{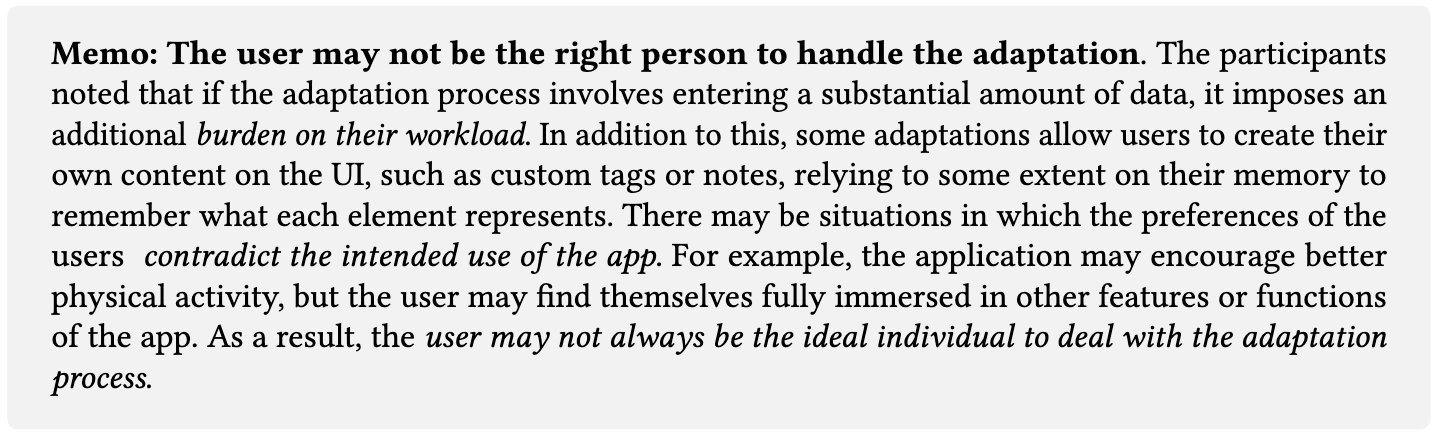}
        \caption{Memo example}
        \label{fig:memo}
    \end{figure}

The data collection and analysis process followed an \textit{iterative and interleaved} approach (Figure \ref{fig:STGT}). \textit{Saturation}, which indicates the point at which no new categories or concepts properties emerged, was achieved in the third iteration of the user study. The qualitative data was analyzed by the first author and was subsequently shared with the remaining authors to encourage collaborative discussion at each stage of the process. In Section \ref{sec:stageoneinterview}, we present the key concepts and categories derived from the STGT analysis. An example of a process for applying STGT for data analysis is provided below.
\begin{enumerate}[left=0pt,itemsep=5pt, topsep=5pt]
    \item \textbf{Open Coding and Constant Comparison:} We analyzed the audio transcripts and extracted various codes from the raw data. \\
    \textbf{Raw Quote 1:} \textit{ "If you want to set it, then you don't want to change it. Likewise, you don't want to have to go through all those settings again. I think that would be complicated. ”} \\\hcode{\textbf{Code 1:} mental workload for adaptation}\\
    \textbf{Raw Quote 2}: \textit{“I am not going to log my daily blood sugar levels in the application because I am too lazy to do it sometimes. Will the adaptation hide this important function because I don't use it a lot?” }\\\hcode{\textbf{Code 2:} user preferences contradict the app's intended usage}\\
    
    \noindent The two code examples given above suggest: \hconcep{\textbf{Concept:} The user may not be the right person to handle the adaptation}. Drawing from the \textit{memos} generated during the coding process (see Figure \ref{fig:memo}), along with the identified codes and concepts, it illustrates the \hcate{\textbf{Category:} Who should take charge of the adaptation process.} 
    
    \item \textbf{Theoretical Sampling:} Initial data collection began with \textit{convenience sampling}, primarily involving older adults in focus group sessions. After we conceptualized concepts from the focus group data, \textit{theoretical sampling} was employed to further refine and elaborate on the emerging themes. For example, most of the focus group participants in \textit{\textbf{Iteration 1}}, primarily older adults, express difficulty engaging in the adaptation process and \textit{prefer it to occur without their direct involvement}, as the multi-step nature of the process can be confusing to them. In contrast, a participant of a much younger age has a contrasting view, showing interest in adapting the features or functions. As a result, the concept \hconcep{the user may not be the right person to handle the adaptation} needs to be refined, prompting us to recruit more participants from a younger age group in \textbf{\textit{Iteration 2} }(see Figure \ref{fig:STGT}).
    
    \item \textbf{Memoing} played a crucial role in our approach, allowing us to explore emerging concepts and potential relationships between them, as described by \citet{hoda2021socio}. These memos served as invaluable tools for capturing \textit{key insights and reflections} gleaned from our open-coding efforts, which are further detailed in Section \ref{meomoreflect}. An example memo is provided in Figure \ref{fig:memo}.
\\
    \end{enumerate}

% \memo{\textbf{Memo: The user may not be the right person to handle the adaptation}. The participants noted that if the adaptation process involves entering a substantial amount of data, it imposes an additional \textit{burden on their workload}. In addition to this, some adaptations allow users to create their own content on the UI, such as custom tags or notes, relying to some extent on their memory to remember what each element represents. There may be situations in which the preferences of the users \textit{ contradict the intended use of the app}. For example, the application may encourage better physical activity, but the user may find themselves fully immersed in other features or functions of the app. As a result, the \textit{user may not always be the ideal individual to deal with the adaptation process}. } 

\subsubsection{Stage Two: Qualitative data analysis} Given the relatively small amount of unstructured data collected in stage two survey studies, which contrasts with the richer data from stage one focus group and interview studies, we do not anticipate discovering multi-layered findings \cite{hoda2024qualitative}.  Consequently,\textit{ thematic analysis} is employed in stage two to delve deeper into the qualitative aspects of the survey responses, allowing the discovery of common themes that surpassed the survey responses \cite{braun2012thematic,braun2006using}. The qualitative data was analyzed by the first author and was subsequently shared with the remaining authors to encourage collaborative discussion. 

\subsubsection{Quantitative data analysis}
 In \textbf{stage two}, as the surveys targeting end-users and software practitioners primarily comprised open-ended questions, the analysis predominantly relied on descriptive statistics to summarize the responses. For the user survey in \textbf{stage one}, quantitative data were analyzed using R\footnote{https://www.r-project.org}. Descriptive statistics were employed to summarize respondent characteristics and preferences, providing insights into distribution patterns and emerging trends within the dataset. \textit{Chi-square tests} were used to determine whether there was a significant association between the preferences of users with respect to various aspects of adaptations and their demographic characteristics, such as age, gender, nationality, education, and chronic diseases. To ensure that the data distribution meets the prerequisites for the Chi-square independence test, related variables were grouped into categorical variables beforehand. Age was categorized into two groups: 18-45 and 45-74. The level of education was classified into three groups: Less than Bachelor's degree, Bachelor's degree, and Postgraduate (Master's and Doctoral degrees). Chronic disease conditions were categorized as detailed in Section \ref{sec:surveypart}. If a significant association is found, \textit{binary logistic regression or multinomial logistic regression} will be subsequently employed to model the relationship between these variables. Understanding how these demographic factors influence user preferences provides valuable information on how to tailor AUIs to accommodate various user needs and preferences. The common significance level of \begin{math}\alpha= 0.05\end{math} is chosen for statistical analysis.

\section{Stage One: Findings of interview and focus group studies}\label{sec:stageoneinterview}

This section presents key findings on the user’s perspectives toward the AUI prototype in the context of mHealth applications. We identified four overarching challenges that participants faced while interacting with the AUI prototype. As participants describe the challenges they encountered, they also offer recommendations for improving the adaptation design. For more comprehensive insights into each challenge and recommendation, refer to \citet{wang2024adaptive}. In addition, a further detailed analysis explores why certain recommendations are more effective for some users than others, identifying three \textbf{contextual factors} that influence how individual user characteristics shape adaptation preferences and outcomes (Section \ref{sec:context}). Drawing on STGT data analysis and the existing literature \cite{wang2023adaptive, abrahao2021model}, we grouped the identified challenges into four distinct categories: \textit{What to adapt}, which focuses on pinpointing specific UI components that require adaptation and recognizing the associated implementation challenges; \textit{Who should initiate adaptations}, which explores the assignment of responsibilities between users and systems in triggering adaptations; \textit{How to adapt}, which examines the strategies and mechanisms used to carry out adaptations effectively; and \textit{When to adapt}, which considers the appropriate timing and contextual conditions for initiating adaptations. Participants participated in discussions about recommendations related to \textit{controllability, user support, and alignment} of adaptation design. The priorities of the participants placed on various recommendations are diverse and were shaped by their desired degree of \textit{involvement} with the system, their familiarity with \textit{mHealth applications}, as well as their personal \textit{health conditions} (contextual factors).

% \begin{figure*}[!ht]
% \centering
% \includegraphics[width=1\textwidth]{Figure/Strategies_number.pdf}
% \caption{Strategies for adaptations suggested by participants}
% \label{fig:strategies}
% \end{figure*}

\subsection{Contextual Factors of Recommendations}\label{sec:context}
\subsubsection{User involvement}
The activeness level of users significantly influences recommendations related to user participation, such as user support and controllability. In many human-computer interaction models, users are typically categorized as \textit{"active process operators"} or \textit{"passive process operators"}  \cite{persson2001passive}. This distinction is common in shared technologies such as health technologies used by physicians, nurses, and patients \cite{xu2014different}. Users can be active, having direct control, or passive, interacting without control \cite{montague2012understanding, inbar2009feature}. Our research categorizes users as active or passive based on their willingness to engage with the system, regardless of the shared technology \cite{eiband2021support, gomez2024human}. \textbf{ Active user involvement} is exemplified by participants who proactively experiment with various data sources to understand how they impact the system's output, desire for active participation, and a sense of control over the adaptation process. They are active in the adaptation process, willing to explore and approve various adaptation suggestions \cite{Gajos2017TheIO}. On the contrary, \textbf{ passive user involvement} characterizes individuals who are more inclined to seek information about how the adaptation works but do not actively provide feedback or corrections to the system. Passive participants may stop exploring once they believe the current UI meets their minimum requirements \cite{eiband2021support, parra2015user}. Participants with a passive involvement with the application, with some admitting that they have not fully explored what the application can do, others expressing a lack of concern about the system's adaptation process, some indicating tolerance for most generated adaptations, and some preferring minimal interaction with the application. Participants can actively experiment with the way that different adaptation settings influence system output if the software provides switches and configurable options within the UI. For example, they can access a dashboard for the adaptation types and configure the settings for different adaptations. In contrast, users with a negative perception of the software show less enthusiasm for this level of participation. They tend to prefer less or even no interaction when it comes to experimenting with adaptation settings. Participants also highlighted the importance of including \textit{\textbf{family members or caregivers}} in the use of the system, as caregivers often play an active role in medical care decisions \cite{xie2009older, Cassileth1980}. Allowing others to handle the adaptation process can alleviate cognitive effort for older users \cite{schiavo2022trade}, but could potentially reduce their independent use of the technology \cite{munteanu2017multimodal}.

\subsubsection{User experience with mHealth applications} 
Prior technology experience significantly affects users' perceptions and interactions, influencing effort expectancy and their intention to use technologies \cite{parameswaran2015within, venkatesh2012consumer}. Users are more likely to adopt mHealth applications if they are easy to use, as this reduces the effort expectancy \cite{sun2013understanding}. Familiarity with the system reduces the cognitive load, prompting experienced users to explore adaptation options \cite{albers1997cognitive}. Inexperienced mHealth users often find it difficult to understand adaptations. As experience improves memory access \cite{fazio1978attitudinal}, these inexperienced users generally expect systems to offer substantial assistance with minimal cognitive effort \cite{taylor1995assessing}. Participants expressed concerns about adaptations affecting their \textit{technologically inexperienced parents}, who struggle with technology and may hesitate to explore adaptations due to fear of making mistakes.

\subsubsection{Health condition}
Chronic diseases exert varying physical, psychological, and mental impacts on participants \cite{harvey2012future, di2019chronic, lorig2003self, Who}. Our findings indicate that extensive adaptation control and support might not be appropriate for individuals who struggle with decision making, particularly when faced with numerous lifestyle and treatment options \cite{scherer2021patient}. This is especially relevant for those with serious health conditions or recent diagnoses of chronic diseases \cite{Pizzoli2019FromLT,zhang2021designing}. Certain participants who consider themselves experts in the management of their health condition tend to become their own primary caregivers and demonstrate a greater willingness to explore various adaptation possibilities. Individuals experiencing more severe symptoms and facing greater challenges in managing their health express a stronger preference for simplicity and greater system assistance.  

\subsection{Mapping of Challenges to Recommendations}\label{meomoreflect}

\begin{figure}[b]
\centering
\includegraphics[width=\textwidth]{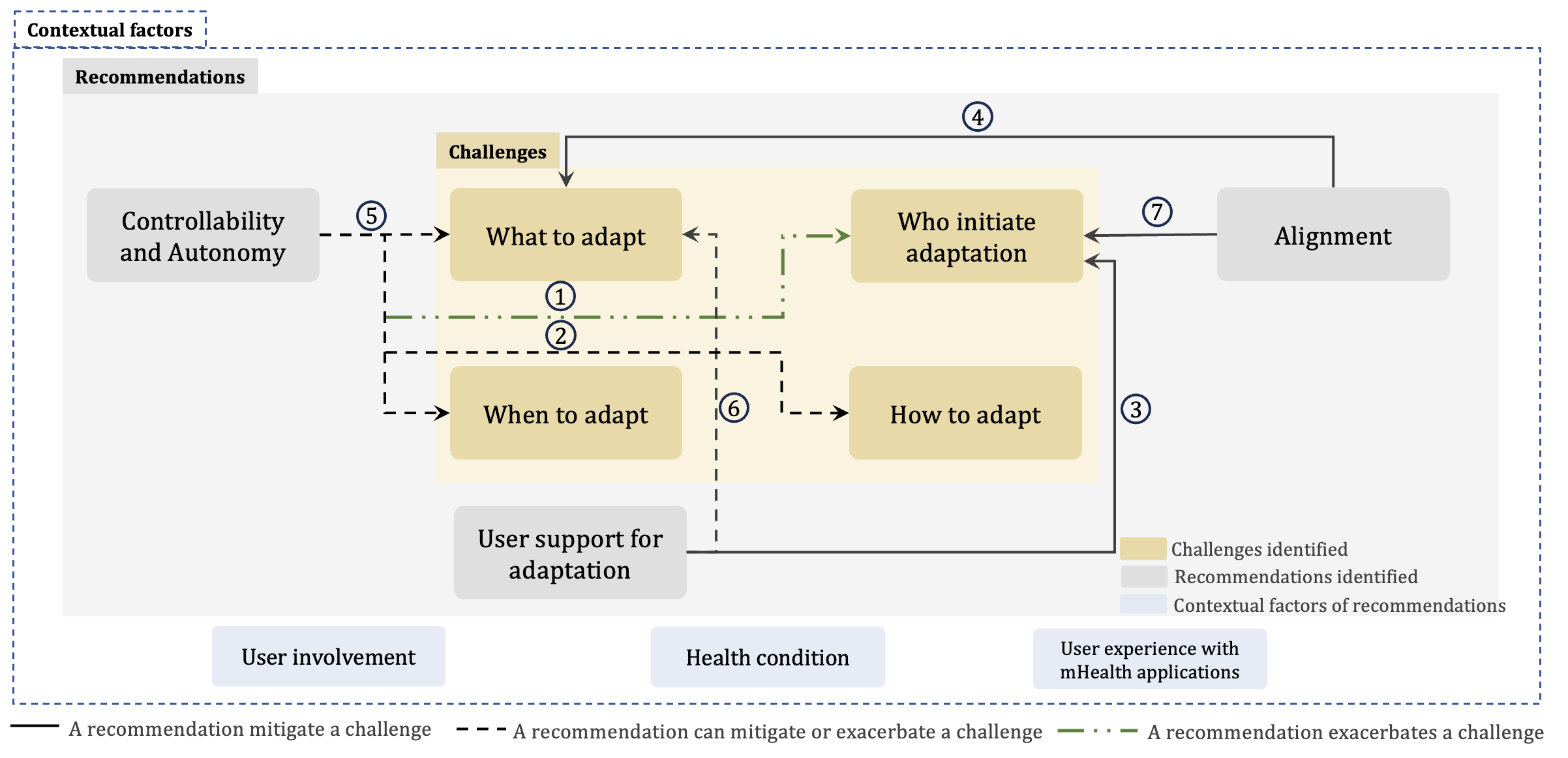}
\caption{An overview of challenges and recommendations for designing adaptation.}
\label{fig:mapping}
\end{figure}

Drawing from our data analysis and the memos recorded while following the STGT method, we have found some insights spanning various categories. Figure \ref{fig:mapping} illustrates the correlation between the \textit{challenges} identified and the \textit{recommendations} outlined.

\subsubsection{Trade-off between user burden, user support and controllability} (\circled{1}, \circled{2}, \circled{3})\label{sec:control}
Users have consistently highlighted the importance of having various options to control and support adaptations. However, effectively managing the complexity of providing users with control and support remains a challenge \cite{paymans2004usability,bunt2004role, eiband2021support}. User control over adaptation offers significant benefits, empowering users to customize their experiences to better meet their needs and preferences \cite{parra2015user, kay2001learner}. It also encourages a sense of ownership and agency, boosting engagement and motivation for effective system use \cite{sundar2010personalization,sundar2008self}. Leaving excessive control can lead to distraction and inefficiency, particularly among users who lack the necessary knowledge or interest to make informed decisions \cite{jameson2002pros} (\circled{1}). Furthermore, as discussed in \citet{kay2001learner}, the user preferences for control can vary significantly. Variations in user preferences for controlling adaptations and interface elements can be explained by Hofstede’s cultural dimensions model \cite{hofstede2001culture}, which suggests that users of individualistic cultures often prefer personal control, while those from collectivist cultures can rely on preset options \cite{bunt2004role}. Too many control options can lead to distraction and fatigue (\circled{2}). Furthermore, differences in user perspectives on controllability may also be related to specific tasks the user performs, as highlighted by \citet{Gajos2017TheIO}. This underscores the importance of considering these factors, as users often engage in a mix of tasks when using mHealth applications, and some tasks are more frequent or demanding than others \cite{bunt2004role}. In essential and frequent tasks, user controllability becomes less important, and users may prefer the system to manage these tasks automatically \cite{peissner2013user}. Therefore, delegating control to users requires considering various factors, including cultural influence and the nature of the task. In addition to controllability, research has attempted to support the adaptation process. For example, some studies have explored the use of animated transitions to demonstrate the adaptation process to users \cite{dessart2011showing}, while others have investigated the effectiveness of providing detailed explanations \cite{kuhme1993user}. However, the provision of support materials may not necessarily reduce cognitive demands, as \textit{additional support itself can impose an additional burden on users }\cite{kulesza2013too}. Users may not always value support materials such as explanations, especially when they lack control over the adaptation process, perceive the system as effective, or find the effort to understand the information not worth the benefits \cite{bunt2012explanations}. Therefore, while improvements in controllability and support are beneficial, it is crucial to carefully consider their implementation and ensure that they meet user needs (\circled{3}).

\subsubsection{Trade-off for usability issues} (\circled{4},\circled{5},\circled{6})
AUIs are increasingly seen as a solution to cope with the growing diversity of usage contexts, devices, and users \cite{browne1990build, hook1998tutorial,Norcio1989}. They offer solutions to various usability issues in mobile applications, including improving accuracy, efficiency, and user learning, as well as addressing information overload and helping in the use of complex systems \cite{browne1990build, hook1998tutorial}. However, previous research on AUIs has also shown a trade-off between adaptive mechanisms and usability \cite{paymans2004usability, jameson2007adaptive, mezhoudi2015toward, deuschel2018influence, hook2000steps, glass2008toward, bunt2004role, gajos2006exploring, gajos2008predictability, turk2019exploring}. We found several usability challenges associated with AUI, including \textit{ privacy concerns, predictability issues, comprehensibility difficulties, and UI obtrusiveness} in our user study. Interestingly, our study reveals that certain recommendations proposed by participants to improve the adaptation design can alleviate or exacerbate some existing usability challenges. For example, aligning visual elements and icons with user preferences can reduce the obtrusiveness of the UI (\circled{4}). However, introducing lifelike character icons, such as a doctor or ambulance, can unrealistically increase expectations of the system's competence in adapting to the user \cite{jameson2007adaptive, sproull1996interface}. Setting accurate user expectations is key for effective design, as misaligned expectations can lower trust and reduce the likelihood of reusing the interface, emphasizing the importance of avoiding both overestimation and underestimation of the system competence \cite{hook2000steps, glass2008toward, pu2006trust}. Recommendations such as improving adaptation control or privacy can require additional user interactions, feedback input, or system notifications, potentially disrupting and distracting users \cite{jameson2007adaptive, Deuschel2016OnTI} (\circled{5}). We have observed that while certain recommendations can mitigate specific usability challenges, they can inadvertently exacerbate others \cite{jameson2007adaptive}. Meanwhile, \textit{the effectiveness of a recommendation depends on its appropriate application}, as it can have varying impacts on usability. For example, aligning with chronic disease and providing explanatory materials can improve adaptation comprehensibility (\circled{4},\circled{6}). However, excessive provision of support content can become obtrusive, difficult to grasp, and alter user learnability \cite{eiband2021support,paymans2004usability}. Similarly, step-by-step adaptation at regular intervals could improve UI predictability and could also increase the intrusiveness of the adaptation process (\circled{5}). However, constantly adapting the UI can be obtrusive and disrupt user workflow \cite{wesson2010can}. 

Mitigating specific usability challenges in AUI design often involves trade-offs with other usability goals. Our findings highlight two key considerations for dealing with conflicting demands. 

\textit{\textbf{1) User priorities.}} Users often have varying priorities regarding conflicting usability goals \cite{rich2004expected}. For example, in our user study, we observed that some participants preferred interfaces with consistent layouts and design elements across various screens, as it fosters familiarity and ease of use. In contrast, others prioritize adaptations and enjoy exploring new interface designs and features, even if it means sacrificing some consistency. Similarly, users may have different privacy priorities. Although some people prioritize privacy and are cautious about sharing personal information, others are more inclined to offer data in return for a customized UI. These examples highlight the critical importance of offering alternative solutions that \textit{accommodate the diverse usability priorities} of users \cite{bunt2004role, gajos2008predictability}. This approach assumes that different users may be open to trade-offs among different usability aspects. \citet{harper2015putting} found that users reported usability problems with a system but were overall satisfied because the benefits of having control options outweighed the inconveniences encountered. 

\textit{\textbf{2) Granularity.}} Usability objectives such as predictability, comprehensibility, and controllability can be achieved at different levels of granularity \cite{bunt2004role, jameson2007adaptive}. Our observations revealed that making minor adjustments during the adaptation process could not significantly impact the overall user experience. Furthermore, low-level granularity adaptations have a minimal effect on comprehensibility, as they only affect small parts of the UI \cite{abrahao2021model}. The adaptation of high-level granularity involves modifying multiple aspects of the UI simultaneously \cite{cristea2003three, zeidler2013evaluation}, such as altering the overall color scheme, font styles, navigation menus, and placement of key features at the same time. It may introduce more usability challenges, as it could be perceived as a completely new application from the user's perspective. High-level granularity adaptations are infrequent and may happen just once to customize the application for the user, helping to address some usability issues. This approach is particularly advantageous when users first engage with the system, as they do not have preconceived expectations of its appearance or functionality \cite{bunt2004role}. However, it may pose challenges for users who are less familiar with such applications and lack insight into which design and functionality would best suit their needs \cite{gajos2006exploring}. Some usability issues introduced by AUI can be mitigated without necessarily improving the user's mental model of the adaptive system \cite{paymans2004usability,Rosman2014OnUB}. Systems that help users overcome these problems foster trust and understanding, encouraging the continued use of mHealth applications \cite{martin2019exploring}. As users recognize the system's adaptive benefits, they appreciate its role in enhancing their experience and well-being.

\subsubsection{Trade-off between independence and assistance}(\circled{3}, \circled{7})
User support typically focuses on facilitating the execution of application functions, but this type of support may hinder the user learning process and diminish their overall experience with the system \cite{lanier1995agents}. Learning is the process of acquiring skills, knowledge, and competencies in a specific domain, allowing greater independence. However, this comes with costs such as time and effort, and can be prone to errors and time-consuming (\circled{3}). Mitigating the cognitive effort to adapt the system can be achieved by having caregivers or health professionals manage the process (\circled{7}). Users should decide whether to invest time in learning or delegate tasks to the system or caregivers. Providing options for independent and assisted application use can be advantageous. Support features should align with user goals and context, accommodating their preferences and comfort with challenges \cite{fischer2006distributed}.

The inevitability of trade-offs complicates decision making in UI adaptation, as users frequently differ in their prioritization of conflicting goals, level of competence, objectives for using the app, and the specific usage scenario. It is imperative to provide alternative solutions that cater to users with varying priorities \cite{billsus2007adaptive}.

\section{Stage One: Findings of the user survey}\label{sec:stageonesurvey}
\subsection{User Survey Participants} \label{sec:surveypart}
\begin{table}[b]
\centering
\renewcommand{\arraystretch}{1}
\caption{Survey participants demographics information (n=90)}
\label{tab:surveydemographics}
    \resizebox{0.98\textwidth}{!}{%
\begin{tabular}{p{87mm}p{115mm}}
\hline
\toprule
    \begin{tabular}[t]{p{45mm}p{2mm}l}
    \textbf{Demographics} & \textbf{\#}  & \textbf{\% of Participants}  \\ \hline
    \multicolumn{3}{l}{\textit{\textbf{Age$^1$}}}\\ \hline
        18-24 & 17 & \mybarhhigh{5.7}{19\%} \\
        25-34 & 30 & \mybarhhigh{10.0}{33\%} \\
        35-44 & 19 & \mybarhhigh{6.3}{21\%} \\
        45-54 & 10 & \mybarhhigh{3.3}{11\%} \\
        55-64 & 11 & \mybarhhigh{3.7}{12\%} \\
        65-74 & 3  & \mybarhhigh{1.0}{3\%} \\
    \multicolumn{3}{l}{\textit{\textbf{Gender$^2$}}}\\ \hline
        Female & 38 & \mybarhhigh{12.7}{42\%} \\
        Male & 50 & \mybarhhigh{16.7}{56\%} \\
        Prefer not to say & 2 & \mybarhhigh{1.0}{3\%} \\
    \multicolumn{3}{l}{\textit{\textbf{Education$^3$}}}\\ \hline
        Less than Bachelor’s degree & 31 & \mybarhhigh{10.3}{34\%} \\
        Bachelor’s degree & 40 & \mybarhhigh{13.3}{44\%} \\
        Postgraduate & 19 & \mybarhhigh{6.3}{21\%} \\
    \multicolumn{3}{l}{\textit{\textbf{Chronic Disease Categories$^4$}}}\\ \hline
        Cardiometabolic & 47 & \mybarhhigh{15.7}{52\%} \\
        Immune-related & 31 & \mybarhhigh{10.3}{34\%} \\
        Mental health conditions & 7 & \mybarhhigh{2.3}{8\%} \\
        Respiratory & 11 & \mybarhhigh{3.7}{12\%} \\
    \end{tabular}
&
\begin{tabular}[t]{p{70mm}p{2mm}l}
    \textbf{Demographics} & \textbf{\#}  & \textbf{\% of Participants}  \\ \hline
    \multicolumn{3}{l}{\textit{\textbf{Country of Residence}}}\\ \hline
        Australia & 44 & \mybarhhigh{14.7}{49\%} \\
        China & 21 & \mybarhhigh{7.0}{23\%} \\
        USA & 12 & \mybarhhigh{4.0}{13\%} \\
        UK & 6 & \mybarhhigh{2.0}{7\%} \\
        Other (Nigeria, Canada, Korea, Spain, Sri Lanka) & 7 & \mybarhhigh{2.3}{8\%} \\
    \multicolumn{3}{l}{\textit{\textbf{Main reason to use the app$^5$}}}\\ \hline
        To monitor my chronic disease symptoms&52&\mybarhhigh{17.3333333333333}{58\%}\\
        Increase my physicial activity levels&49&\mybarhhigh{16.3333333333333}{54\%}\\
        To track what I eat&42&\mybarhhigh{14}{47\%}\\
        To get education about my chronic disease&34&\mybarhhigh{11.3333333333333}{38\%}\\
        Help with weight loss&28&\mybarhhigh{9.33333333333333}{32\%}\\
        Manage my medications&26&\mybarhhigh{8.66666666666667}{29\%}\\
    \multicolumn{3}{l}{\textit{\textbf{Type of mHealth Application Used$^5$}}}\\ \hline
        Health promoting and self-monitoring & 76 & \mybarhhigh{25.3}{84\%} \\
        Informative application & 41 & \mybarhhigh{13.7}{46\%} \\
        Assistive application & 29 & \mybarhhigh{9.7}{32\%} \\
        Communication application & 24 & \mybarhhigh{8.0}{27\%} \\
        Health game & 16 & \mybarhhigh{5.3}{18\%} \\
        Rehabilitation application & 16 & \mybarhhigh{5.3}{18\%} \\
        Exercise application & 7 & \mybarhhigh{2.3}{8\%} \\
    \end{tabular} \\
        \hline  \multicolumn{2}{p{200mm}}{\textit{* 1,2,3: The percentage does not strictly add up to 100\% due to rounding. 4: The percentage does not strictly add up to 100\% due
to multimorbidity. 5: The percentage does not strictly add up to 100\% due to the use of multiple applications/purpose.}}\\
\bottomrule
\end{tabular}}
\end{table}

Concurrently with focus group and interview studies, we surveyed 90 participants with chronic diseases, all of whom have experience using mHealth applications. Their detailed demographic information is summarized in Table \ref{tab:surveydemographics}. Most of the participants identified as men (56\%), between the ages of 18 and 74. In particular, the 25-34 age group constituted the largest group, comprising about 33\% of all participants. There was representation from older age groups, with 16\% of participants over 55 years of age and 21\% between 35 and 44 years of age. Geographically, our survey captured responses from a diverse set of countries, with approximately half of the responses coming from Australia. China and the USA are also significant contributors, accounting for 23\% and 13\% of the responses, respectively. In terms of educational background, a bachelor's degree is the most common attainment (44\%), followed by 34\% with education levels below a bachelor's degree and 21\% with higher degrees, such as master's or doctorate degrees. 

The chronic disease reported by the participants were systematically categorized into four groups: cardiometabolic (e.g., diabetes, high blood pressure, obesity, and heart disease), respiratory (e.g., allergies, asthma, and chronic lung disease), immune-related (e.g., cancer, Parkinson's disease, and compromised immune system), and mental health conditions \cite{camacho2020attitudes}. Cardiometabolic diseases were the most prevalent, reported by 52\% of the participants. It should be noted that some participants reported multiple chronic diseases, such as having diabetes and high blood pressure, or a combination of cardiometabolic disease and another type of disease, such as asthma (a respiratory disease). Indicates a phenomenon known as \textbf{multimorbidity} \cite{islam2014multimorbidity}, where individuals may experience two or more chronic diseases simultaneously, a finding that is also consistent with other studies \cite{anderson2016mobile}. Participants were asked to indicate their familiarity with the mHealth applications. As shown in Table \ref{tab:surveydemographics}, respondents reported using various types of mHealth applications, with health-promoting and self-monitoring applications being the most prevalent, utilized by 84\% of participants. These applications typically target functions related to fitness, medication, and diet, which is consistent with findings from previous research \cite{kayyali2017awareness}. This prevalence can be attributed to the ongoing need for individuals with chronic diseases to monitor their health status and follow prescribed treatment regimens \cite{hamine2015impact}. The primary motivations cited for using mHealth applications include symptom monitoring (58\%), promoting physical activity (54\%), and management of dietary intake (47\%). Participants were also asked about the frequency of using the mHealth app.  In our survey study, 54\% of the participants reported using health applications daily, while 33\% reported weekly usage. Most users spend between 1 to 20 minutes per session, aligning with findings from previous research \cite{robbins2017health, kc2021types}. Users who access the application two or more times per day typically allocate a shorter duration per session. In contrast, people who use it less frequently (less than once a month or less than once a week) tend to engage in longer sessions. The former group may use the application for quick, frequent interactions or brief check-ins throughout the day, while the latter may delve into more comprehensive interactions during less frequent usage, possibly for specific tasks or content consumption.

\begin{table}[b]
\centering
\caption{Survey participants' perspective towards different aspects of adaptations.}
\label{tab:surveyaspects}
    \resizebox{0.95\textwidth}{!}{%
\begin{tabular}{llllll}
\hline
\toprule \textbf{Aspects of adaptations} & \textbf{\#}  & \textbf{\% of Participants}  &\textbf{Aspects of adaptations} & \textbf{\#}  & \textbf{\% of Participants}  \\
\hline
\textbf{\textit{Different types of adaptation*}} & & &\textit{\textbf{Data source of adaptation*}}&&\\
\hline

\textbf{P:} Graphic design\cellcolor{greya} &52\cellcolor{greya}&\mybarhhigh{13}{58\%}\cellcolor{greya}&\textbf{UC:} Physiological characteristics&53&\mybarhhigh{13.25}{59\%}\\
\textbf{P:} Information architecture\cellcolor{greya}&30\cellcolor{greya}&\mybarhhigh{7.5}{33\%}\cellcolor{greya}&\textbf{UC:} Physical characteristics&52&\mybarhhigh{13}{58\%}\\
\textbf{P:} Sound effect\cellcolor{greya}&25\cellcolor{greya}&\mybarhhigh{6.25}{28\%}\cellcolor{greya}&\textbf{UC:} Preference&48&\mybarhhigh{12}{53\%}\\
\textbf{C:} Content complexity&53&\mybarhhigh{13.25}{59\%}&\textbf{UC:} Psychological characteristics&47&\mybarhhigh{11.75}{52\%}\\
\textbf{C:} Interface elements rearrangement&42&\mybarhhigh{10.5}{47\%}&\textbf{UC:} Demographics&37&\mybarhhigh{9.25}{41\%}\\
\textbf{B:} Multimodal interaction\cellcolor{greya}&41\cellcolor{greya}&\mybarhhigh{10.25}{46\%}\cellcolor{greya}&\textbf{UC:} Social activity&29&\mybarhhigh{7.25}{32\%}\\
\textbf{B:} Difficulty level\cellcolor{greya}&39\cellcolor{greya}&\mybarhhigh{9.75}{43\%}\cellcolor{greya}&\textbf{IR:} Feedback\cellcolor{greya}&41\cellcolor{greya}&\mybarhhigh{10.25}{46\%}\cellcolor{greya}\\
\textbf{B:} Add on functions\cellcolor{greya}&36\cellcolor{greya}&\mybarhhigh{9}{40\%}\cellcolor{greya}&\textbf{IR:} Interaction with the interface\cellcolor{greya}&34\cellcolor{greya}&\mybarhhigh{8.5}{38\%}\cellcolor{greya}\\
\textbf{B:} Navigation adaptation\cellcolor{greya}&31\cellcolor{greya}&\mybarhhigh{7.75}{34\%}\cellcolor{greya}&\textbf{IR:} Emotions\cellcolor{greya}&34\cellcolor{greya}&\mybarhhigh{8.5}{38\%}\cellcolor{greya}\\
\textbf{B:} Different persuasive strategy\cellcolor{greya}&31\cellcolor{greya}&\mybarhhigh{7.75}{34\%}\cellcolor{greya}&\textbf{IR:} Performance in game\cellcolor{greya}&29\cellcolor{greya}&\mybarhhigh{7.25}{32\%}\cellcolor{greya}\\
&&&\textbf{TS:} Goals&40&\mybarhhigh{10}{44\%}\\
\multicolumn{3}{p{110mm}}{\textbf{\textit{P:}} \textit{Presentation adaptation}, \textbf{\textit{C:}} \textit{Content presentation}, \textbf{\textit{B:}} \textit{Behaviour adaptation}}&\textbf{TS:} Motivation&35&\mybarhhigh{8.75}{39\%}\\
\multicolumn{3}{p{110mm}}{\textbf{\textit{UC:}} \textit{User
characteristics}, \textbf{\textit{IR:}} \textit{Interaction
related}, \textbf{\textit{TS:}} \textit{Task specific}} &\textbf{TS:} Role&24&\mybarhhigh{6}{27\%}\\

\hline
\textbf{\textit{Different data collection method*}} & & &\multicolumn{3}{p{80mm}}{\textit{\textbf{Preferred level of involvement in the adaptation}}}\\
\hline

Smartphone sensor&61&\mybarhhigh{15.25}{68\%}&Semi-automatic&49&\mybarhhigh{12.25}{54\%}\\
Wearable sensor&58&\mybarhhigh{14.5}{64\%}&Automatic&34&\mybarhhigh{8.5}{38\%}\\
User input through the application&49&\mybarhhigh{12.25}{54\%}&Manual&7&\mybarhhigh{1.75}{8\%}\\
Analysis of user behaviour through the application&41&\mybarhhigh{10.25}{46\%}&&& \\
Analysis of activities with keyboard&27&\mybarhhigh{6.75}{30\%}&&& \\

\\
        \hline  \multicolumn{6}{p{200mm}}{\textit{*: Because respondents could select multiple options, the percentages do not sum precisely to 100\%.}}\\

\bottomrule
\end{tabular}
}
\end{table}

\subsection{Different Types of Adaptations} 
In the earlier SLR on AUIs for chronic disease-related applications \cite{wang2023adaptive} compiled a comprehensive taxonomy of adaptation implemented by researchers to support users in managing chronic diseases. Our aim with this survey is to discern the users' perspectives regarding the significance and value attributed to these adaptations. Our survey revealed a diversity of preferences among the respondents, with no singular adaptation type dominating over others. Content complexity (59\%) and graphic design (58\%) emerged as the most prevalent, closely followed by the rearrangement of the interface elements (47\%) and multimodal interaction (46\%). In contrast, the sound effects exhibited the lowest frequency among the listed adaptations (28\%) (see Table \ref{tab:surveyaspects}). This finding is not consistent with the dominant trends in existing AUI studies in the domain of chronic diseases \cite{wang2023adaptive}, where the SLR emphasized graphic design (presentation adaptation), as the predominant adaptation type; however, the survey findings highlighted the significant emphasis of users on content adaptation and some types of behavior adaptation. It highlights the significance of synchronizing researchers' initiatives with user needs and priorities to guarantee the efficient design and application of AUIs in chronic disease management tools.

\begin{table}[t]
\centering
\caption{Binary logistic regression results of the adaptation types and demographic aspects (\vcenteredincludep{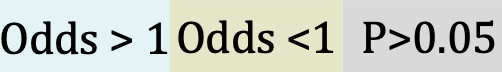})}
\label{tab:BLR}
{\scriptsize
\begin{tabular}{p{20mm}p{25mm}p{17mm}p{17mm}p{17mm}p{17mm}}
\hline
\toprule
\textbf{Variables}&\textbf{Categories}&\textbf{CC*}&\textbf{AD*}&\textbf{DP*}&\textbf{MI*}\\
\hline
\textbf{Age group}	&	45-74	&	3.212/(0.065)\cellcolor{greyb}&	3.764/(0.02)\cellcolor{signif}	&	0.637/(0.43)\cellcolor{greyb}&	5.824/(0.002)\cellcolor{signif}	\\
\hline
\multirow[c]{4}{4em}{\textbf{Nationality}} &	UK	&	0.066/(0.16)\cellcolor{greyb}	&	0.825/(0.843)	\cellcolor{greyb}&	0.889/(0.908)	\cellcolor{greyb}&	0.133/(0.095)\cellcolor{greyb}\\
&	USA	&	0.276/(0.107)	\cellcolor{greyb}&	0.299/(0.178)	\cellcolor{greyb}&	1.504/(0.567)	\cellcolor{greyb}&	0.346/(0.193)\cellcolor{greyb}\\ &	China	&	0.125/(0.003)\cellcolor{signifl}		&	0.215/(0.025)\cellcolor{signifl}		&	1.179/(0.796)	\cellcolor{greyb}&	1.081/(0.896)	\cellcolor{greyb}\\
&	Other	&	0.783/(0.803)	\cellcolor{greyb}&	0.341/(0.264)	\cellcolor{greyb}&	1.2/(0.845)\cellcolor{greyb}&	0.533/(0.492)\cellcolor{greyb}\\
\hline
\multirow[c]{3}{8em}{\textbf{Chronic diseases}} &	Mental health	&	4.173/(0.033)\cellcolor{signif}		&	0.358/(0.093)	\cellcolor{greyb}&	2.126/(0.181)	\cellcolor{greyb}&	0.661/(0.475)	\cellcolor{greyb}\\
&Cardiometabolic&	12.215/(0.013)\cellcolor{signif}		&	8.093/(0.029)\cellcolor{signif}		&	11.264/(0.006)\cellcolor{signif}		&	1.782/(0.471)	\cellcolor{greyb}\\
&	Immune-related &	0.536/(0.554)	\cellcolor{greyb}&	3.823/(0.168)	\cellcolor{greyb}&	2.293/(0.339)	\cellcolor{greyb}&	1.463/(0.664)	\cellcolor{greyb}\\
\hline
\multicolumn{6}{p{120mm}}{{\tiny This table shows how demographic factors affect preferences for mHealth adaptations using binary logistic regression.
\begin{itemize}[left=5pt]
    \item Each cell shows the odds ratio (OR) for preference likelihood and the p-value for statistical significance.
    \item An OR $>1$ indicates higher preference, OR $<1$ indicates lower preference, and $p \leq 0.05$ means statistical significance.
\end{itemize}
 }}\vspace{-8pt}\\
 
\multicolumn{6}{p{120mm}}{{\tiny \textit{* CC=Content complexity, AD=Add on functions, DP=Different persuasive strategy, MI=Multimodal interaction}}}\\

\bottomrule
\end{tabular}}
\end{table}

\subsubsection{Relationship between users' preferences for adaptations and their demographic characteristics}
Conducting a Chi-square independence test, we found significant associations between age, nationality, chronic diseases, and preferences for adaptations such as content complexity, add-on functions, persuasive strategy, and multimodal interaction. Subsequently, a binary logistic regression analysis was employed to examine the relationships between various adaptations and significant demographic factors identified in the Chi-square test (see Table \ref{tab:BLR}). It is important to note that statistical models such as binary logistic regression are built on certain assumptions about the data, including independent observations and non-perfect multicollinearity \cite{harris2021primer}, which are met in our dataset. Compared to participants from Australia, individuals from China are less likely to prefer adaptations related to content complexity (OR = 0.125) and add-on functions (OR = 0.215). Individuals from China may have different expectations or preferences regarding the complexity of content or additional functions in mHealth applications, leading to a lower likelihood of preferring content complexity adaptations compared to participants from Australia. Participants with cardiometabolic diseases showed a significantly higher inclination toward adaptations such as different persuasive strategies, additional functions, and content complexity compared to those with respiratory diseases. Similarly, participants with mental health conditions also demonstrated a notable preference for content complexity. These preferences may arise from the self-management nature of cardiometabolic and mental health conditions, which often necessitate individuals to actively monitor their health and follow treatment plans \cite{cruz2022mhealth, chow2016mhealth, torous2015realizing}. Furthermore, participants over 45 years of age exhibited a higher tendency to seek multimodal interaction (OR = 5.824) and additional function (OR = 3.764) adaptations compared to their younger counterparts. This trend among older users may be due to a preference to minimize interaction efforts and an increased need for additional assistance with application navigation \cite{trajkova2020alexa, betts2019there, martin2019exploring}. As individuals age, they may face challenges related to vision, dexterity, or cognitive abilities, which makes features such as multimodal interaction and additional functions particularly appealing to facilitate their interaction with mHealth applications.

\subsection{Data Sources for Adaptation} 
Our SLR identified various sources of data used as a basis for adaptation \cite{wang2023adaptive}. Based on these findings, the present survey aimed to investigate end-user preferences on the types of data they consider most appropriate to inform adaptation in mHealth applications. We found that the participants exhibited various preferences about the source of the adaptation data and almost half expressed the desire for the application to adapt according to physiological, physical, and psychological characteristics, user preferences, and user goals (see Table \ref{tab:surveyaspects}). The user characteristics data were the most popular compared to the interaction-related data and the task-specific data. This finding is consistent with existing research trends \cite{wang2023adaptive}. 

\begin{table}[t]
\centering
\caption{Binary logistic regression results of the data types and demographic aspects (\vcenteredincludep{Figure/icon/Odds.png})}
\label{tab:BLRData}
{\scriptsize
\begin{tabular}{p{10mm}p{18mm}p{17mm}p{17mm}p{12mm}p{12mm}p{12mm}p{12mm}}
\hline
\toprule
\textbf{Variable}&\textbf{Category}&\textbf{Physiological characteristics}&\textbf{Physical characteristics}&\textbf{Preference}&\textbf{Feedback}&\textbf{Goals}&\textbf{Motivation}\\
\hline
% \textbf{Age group}	&	45-74	&	3.212/(0.065)	&	3.764/(0.02)\cellcolor{signif}	&	0.637/(0.43)	&	5.824/(0.002)\cellcolor{signif}	\\
% \hline
\multirow[c]{4}{4em}{\textbf{Nationality}} &	UK	&	0.093/(0.053)\cellcolor{greyb}&	0.042/(0.019)\cellcolor{signifl}	&	0.047/(0.018)\cellcolor{signifl}	&	1.618/(0.633)	\cellcolor{greyb}&0.222/(0.899)	\cellcolor{greyb}&	0.727/(0.767)	\cellcolor{greyb}\\
&	USA	&	0.121/(0.009)\cellcolor{signifl}	&	0.032/(0.002)\cellcolor{signifl}	&	0.138/(0.018)\cellcolor{signifl}	&	0.205/(0.07)	\cellcolor{greyb}&0.11/(0.015)\cellcolor{signifl}	&	0.168/(0.052)	\cellcolor{greyb}\\
&	China	&	0.303/(0.066)	\cellcolor{greyb}&	0.485/(0.283)	\cellcolor{greyb}&	0.127/(0.003)\cellcolor{signifl}	&	0.216/(0.016)\cellcolor{signifl}	&	0.129/(0.005)	\cellcolor{greyb}&	0.048/(<.001)	\cellcolor{signifl}\\
&	other	&	0.224/(0.092)	\cellcolor{greyb}&	0.736/(0.747)	\cellcolor{greyb}&	0.624/(0.647)	\cellcolor{greyb}&	0.565/(0.512)	\cellcolor{greyb}&	0.238/(0.122)	\cellcolor{greyb}&	0.101/(0.057)\cellcolor{greyb}	\\
\hline
\multirow[c]{3}{4em}{\textbf{Chronic diseases}} &	Mental health	&	1.055/(0.929)	\cellcolor{greyb}&	0.833/(0.783)	\cellcolor{greyb}&	0.202/(0.013)\cellcolor{signifl}	&	0.471/(0.193)	\cellcolor{greyb}&	1.682/(0.412)	\cellcolor{greyb}&	0.844/(0.801)	\cellcolor{greyb}\\
&	Cardiometabolic	&	136.518/(0.999)	\cellcolor{greyb}&	6.172/(0.179)	\cellcolor{greyb}&	0.885/(0.895)	\cellcolor{greyb}&	4.012/(0.138)	\cellcolor{greyb}&	13.948/(0.034)\cellcolor{signif}	&	3.352/(0.208)	\cellcolor{greyb}\\
&	Immune-related	&	0.354/(0.269)	\cellcolor{greyb}&	0.057/(0.017)\cellcolor{signifl}	&	1.244/(0.835)	\cellcolor{greyb}&	0.477/(0.456)	\cellcolor{greyb}&	2.954/(0.287)	\cellcolor{greyb}&	0.167/(0.17)	\cellcolor{greyb}\\
\hline
\multirow[c]{2}{5em}{\textbf{Education level}} &	Bachelor’s degree	&	1.023/(0.97)	\cellcolor{greyb}&	1.09/(0.896)	\cellcolor{greyb}&	0.952/(0.937)	\cellcolor{greyb}&	1.428/(0.544)	\cellcolor{greyb}&	1.1/(0.881)	\cellcolor{greyb}&	7.835/(0.006)	\cellcolor{signif}\\
&	Postgraduate	&	2.116/(0.355)	\cellcolor{greyb}&	1.568/(0.602)	\cellcolor{greyb}&	1.71/(0.513)	\cellcolor{greyb}&	1.511/(0.569)	\cellcolor{greyb}&	1.634/(0.52)	\cellcolor{greyb}&	6.798/(0.035)\cellcolor{signif}	\\
\hline
\multicolumn{8}{p{130mm}}{\tiny This table shows how demographic factors affect preferences for using of different data source using binary logistic regression.
\begin{itemize}[left=5pt]
    \item Each cell shows the odds ratio (OR) for preference likelihood and the p-value for statistical significance.
    \item An OR $>1$ indicates higher preference, OR $<1$ indicates lower preference, and $p \leq 0.05$ means statistical significance. 
\end{itemize}}\vspace{-8pt}\\ 
\multicolumn{8}{p{120mm}}{\tiny \textit{* Odds Ratios/(P-value)}}\\
\bottomrule
\end{tabular}}
\end{table}
\subsubsection{Relationship between users' preferences for data sources for adaptations and their demographic characteristics}
We used a Chi-square independence test to explore how demographic variables influence users' preferences for data sources for adaptation. The findings revealed significant associations between nationality, chronic diseases, education level, and various types of data (see Table \ref{tab:BLRData}). Subsequently, we performed a binary logistic regression analysis to examine the relationships between different data sources and the significant results obtained from the Chi-square test. Participants from different countries exhibit varying preferences for data adaptation based on physiological, physical, preference, feedback, goals, and motivation factors. Given the relatively small number of participants from the UK and USA in our survey, it is essential to acknowledge the potential impact on the statistical power of our analysis. With a reduced sample size, there is a risk of diminishing the ability to detect genuine effects or associations accurately. Consequently, the observed relationships between nationality and data adaptation preferences may be less reliable, introducing uncertainty into our findings \cite{nemes2009bias}. Individuals with higher levels of education exhibit a stronger inclination toward the adaptation of their motivation to use the application (Bachelor: OR= 7.835, Postgraduate: OR= 6.798). Individuals with higher levels of education often have a greater awareness of the benefits of using technology for health management and may also have a greater understanding of the importance of motivation in achieving health-related goals. Therefore, people with higher educational levels can show a greater tendency to adapt their motivation to use the application due to their improved understanding of the role of motivation in achieving health outcomes \cite{woldaregay2018motivational}. Individuals with cardiometabolic diseases are more inclined to desire adaptations that align with their goals (OR=13.948). This preference could be attributed to the treatment of cardiometabolic diseases, which often involves monitoring the levels of diet and physical activity and striving to achieve specific goals \cite{cruz2022mhealth, chow2016mhealth, torous2015realizing}.

Our survey reveals that the predominant methods of data collection include smartphone sensors (68\%) and wearable sensors (64\%) (see Table \ref{tab:surveyaspects}). The existing literature shows a pronounced focus on wearable sensors over smartphone sensors in mHealth applications targeting chronic diseases \cite{wang2023adaptive}. With the widespread adoption of smartphones, an increasing number of users prefer to use their smartphones for data collection rather than rely on other devices. In particular, no significant correlations were found between the data collection method and demographic variables.

\subsection{Preferred Level of Involvement in the Adaptation}
The SLR found varied ways of the involvement of the user in the adaptation process \cite{wang2023adaptive}. In the survey, we discovered that participants generally preferred a mixed-initiative adaptation approach (54\%), which involves collaboration between the system and the end-users to achieve adaptation \citep{mukhiya2020adaptive, abrahao2021model}. In contrast, only a small minority of the participants (8\%) expressed a preference for a fully manual system, where users have complete control over the modification of specific UI elements to suit their needs \citep{Akiki2014}. The limited preference for fully manual systems suggests that users may find manual adaptation processes cumbersome or time consuming, particularly those with limited technological literacy or cognitive abilities \cite{Akiki2014}. The SLR also indicated a lower preference for manual systems, which aligns with our survey findings. Although significant research efforts have been dedicated to automatic systems, there has been a recent surge in interest in mixed-initiative adaptation \cite{wang2023adaptive}. No significant correlations are found between the level of involvement and demographic factors.

\subsection{Key Findings from the User Survey}
From our analysis, we have identified four \textbf{key findings (KFs)} that help unravel the intricacies of user preferences, complementing and contextualizing the qualitative insights from the interview and focus group studies (Figure \ref{fig:surveykeyfinding}).

\begin{enumerate}[left=8pt,itemsep=5pt, topsep=10pt]
    \item [\textbf{KF1}] \textbf{The multimorbidity nature of chronic diseases.} The prevalence of \textit{multimorbidity} among participants highlights the  common experience of managing multiple chronic diseases simultaneously, a phenomenon consistently reported and highlighted in other research \cite{islam2014multimorbidity, anderson2016mobile}. Despite the widespread occurrence of multimorbidity, much of the existing exploration in the realm of mHealth applications for chronic disease management has primarily \textit{focused on addressing a single chronic disease} (e.g., hypertension \cite{hallberg2016supporting} and asthma \cite{tinschert2017potential}). Similarly, in the context of AUIs, research has focused mainly on investigating their efficacy in managing single chronic diseases \cite{burke2009optimising, pagiatakis2020intelligent}. This phenomenon links to what could be adapted in the UI and users' preference for controllability. Users may feel overwhelmed by managing multiple diseases, leading to a lack of motivation and interest in actively intervening or taking responsibility themselves (see Figure \ref{fig:surveykeyfinding}).

\begin{figure}[b]
\centering
\includegraphics[width=\textwidth]{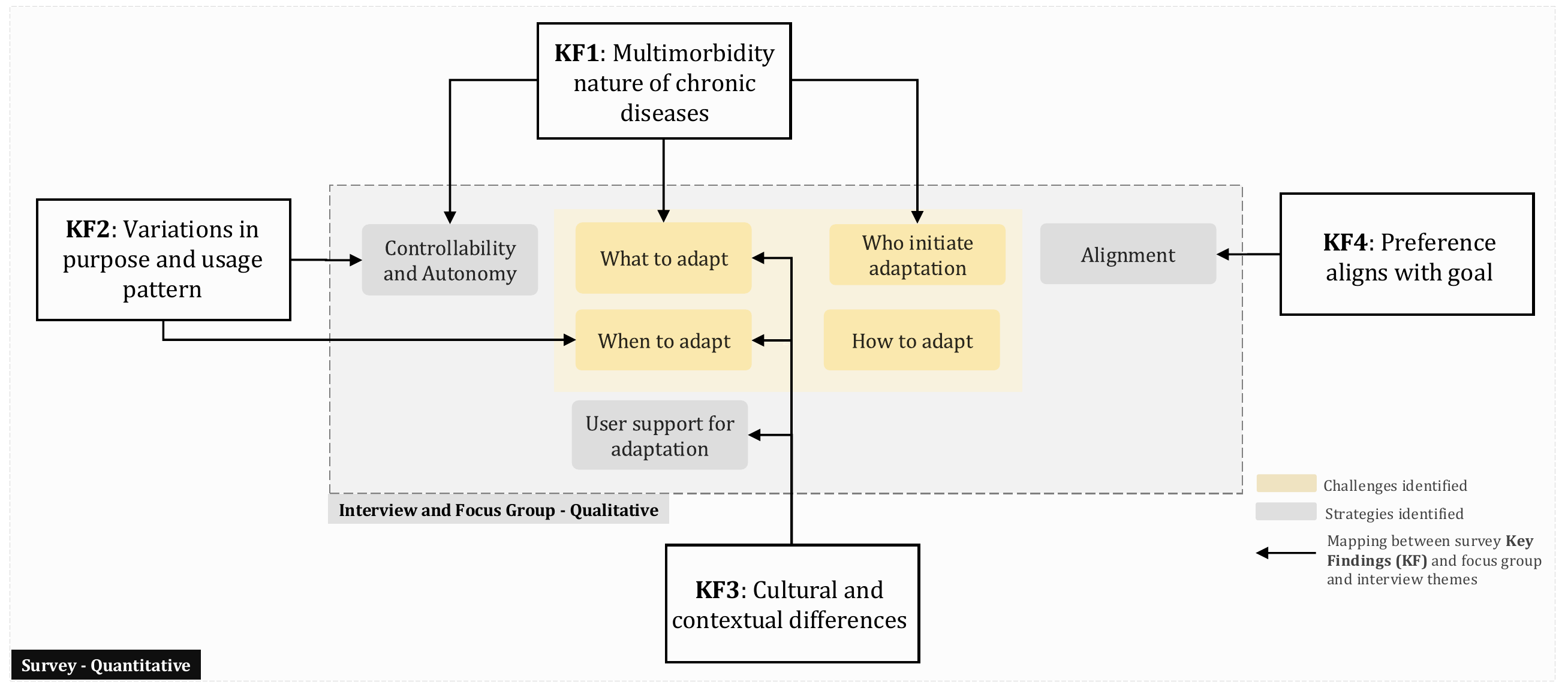}
\caption{Mapping between survey key findings and user study categories.}
\label{fig:surveykeyfinding}
\end{figure}
    \item [\textbf{KF2}] \textbf{Variations in purpose and usage pattern.} The extensive use of mHealth applications for various purposes, ranging from symptom monitoring to dietary management, highlights their vital role in supporting the self-management of individuals with chronic diseases. However, even within the same functionality, users may exhibit fluctuating usage patterns, with no fixed frequency or duration. Research indicates that individuals engage with physical activity tracking applications in \textit{intermittent intervals}, characterized by periods of consistent use followed by breaks and subsequent reengagement \cite{lin2018ll, meyer2016exploring}. These task-related characteristics have implications for the design of AUIs and their performance, especially concerning users' preferences for controllability over adaptations and the timing and frequency of adaptation \cite{lavie2010benefits, peissner2013user} (see Figure \ref{fig:surveykeyfinding}). For example, tasks that require minimal effort and are performed daily, users might prefer the system to handle these tasks automatically without requiring explicit confirmation and \textit{not desire frequent changes} to occur each time they log in.

    \item [\textbf{KF3}] \textbf{Cultural and contextual differences.} The analysis identified significant correlations between \textit{demographic factors such as age, nationality, and chronic diseases}, and preferences for specific types of adaptation. For example, Chinese individuals exhibited a lower inclination toward content complexity adaptations compared to Australian participants, suggesting cultural and contextual differences in adaptation preferences. This observation could be explained by \textit{Hofstede’s cultural dimensions model }\cite{hofstede2001culture}, which is widely used to examine human-computer interaction and cross-cultural challenges in UI design. Specifically, Hofstede's Uncertainty Avoidance dimension, which is defined as the degree to which individuals in a culture perceive ambiguity as a threat and seek to reduce it, offers an explanation. In societies with high uncertainty avoidance, individuals tend to favor predefined UI over experimenting with new adaptations, whereas a society with lower uncertainty avoidance may exhibit greater openness to try new adaptations \cite{alsswey2021role}.  Previous research highlights the importance of culturally specific design preferences in influencing system usefulness in different cultural contexts \cite{evers1997role}. For example, \textit{political orientation} and \textit{social structure} may affect users' perception of the hierarchy and complexity of information presentation \cite{schmid2000language}. Moreover, individuals with higher levels of education demonstrated a stronger tendency to adapt their motivation to use the application, likely due to their deeper understanding of motivation's role in health outcomes. Individuals with different levels of education may require varying levels of support and cues from the systems \cite{reinecke2013knowing} (see Figure \ref{fig:surveykeyfinding}). Cultural and contextual factors significantly influence how well an adaptation aligns with user expectations, and misalignment may result in cultural mismatches that hinder usability and acceptance.

    \item [\textbf{KF4}]\textbf{Preference aligns with goal.} Participants with cardiometabolic diseases showed a greater inclination toward adaptations that align with their usage goals. This preference likely arises from the urgent need for self-management inherent in cardiometabolic diseases. The availability of various applications for the management of chronic diseases or multimorbidities in the App Store underscores the importance of adapting to user primary usage goals and adding value to their overall application experience \cite{bricca2022quality}. This resonates with the overarching category observed in the user study section (see Figure \ref{fig:surveykeyfinding}), emphasizing the essential value brought by adaptation and its alignment with users' primary usage goals \cite{trajkova2020alexa}. Selecting the perfect solution from such diverse options is challenging, as it depends on individual user experiences and the type of application and its adaptations, as suggested by the study conducted by \citet{jameson2002pros}, while ongoing research continues to try to offer recommendations and guidance to users, helping them select the most appropriate options based on the data available within the system \cite{reinecke2013knowing}.
    \end{enumerate}

\begin{Summary}{Summary of Stage One}{Stage one} Stage one employs a mixed-method approach to collect user feedback through interviews, focus groups, and a survey study. We identified four key challenges in how users perceive the adaptation process, with participants highlighting the importance of user control, support, and alignment. These recommendations, shaped by user involvement, experience with mHealth applications, and health condition, involve trade-offs between user burden, support, controllability, usability, and balancing independence with assistance. Our user survey revealed diverse adaptation preferences influenced by different demographic factors. The survey data analysis yielded four key findings, covering the prevalence of multimorbidity in chronic disease, varied usage patterns among users, cultural and contextual differences, and the need for alignment with user goals and preferences. These four key findings offer contextual information or complement the qualitative findings derived from interview and focus group investigations.
\end{Summary}

\section{Stage Two: Guidelines for designing AUIs in mHealth applications targeting chronic diseases}
Our guideline development approach draws on and integrates the methodologies proposed by \citet{hermawati2015user} and \citet{quinones2018methodology}. While \citet{quinones2018methodology} outlines structured stages for guideline specification and validation, their framework offers limited emphasis on implementation procedures and user involvement. In contrast, \citet{hermawati2015user} adopts a user-centric approach specifically tailored to develop domain-specific heuristics. By combining the strengths of both methodologies, we designed a hybrid process that incorporates iterative validation, domain relevance, and active end-user participation throughout the development of AUI guidelines.

\begin{enumerate}[left=5pt,itemsep=5pt, topsep=5pt]
    \item \textbf{Exploratory phase}: Our review of the literature offered critical insights that informed the development of AUI design guidelines for mHealth applications. These guidelines are grounded in existing frameworks and design considerations identified in prior research. A detailed analysis supporting this development is presented in Section \ref{sec:liter}.
    \item \textbf{Experimental phase}: User studies using interviews, focus groups, and a survey explored user experience, preference, and challenges with AUI in mHealth applications. The analysis of both qualitative and quantitative data from this study provided valuable insights that informed the development of the design guidelines.
    \item \textbf{Descriptive phase}: To formalize the primary guidelines, key categories and concepts were derived from the qualitative and quantitative data gathered during the experimental phase. The user needs were systematically grouped and aligned with the guidelines and design recommendations established during the exploratory phase, following the approach proposed by \citet{hermawati2015user}. For insights that did not correspond to existing recommendations, new guidelines were abstracted, accompanied by precise definitions of the application domain and illustrative examples to improve clarity, relevance, and practical applicability.
    \item \textbf{Validation phase}: Given the critical importance of this phase, which is deeply embedded within the specific application domain, particular emphasis was placed on collecting feedback from end-users to ensure the relevance of the guidelines \cite{hermawati2015user}. Building on this foundation, input was also obtained from software practitioners responsible for implementing these guidelines, allowing for further refinement based on their practical experience. Finally, to assess the applicability and effectiveness of the developed guidelines, a case study was conducted in which real-world mHealth applications were evaluated using the proposed guideline set.
\end{enumerate}

\subsection{Literature Review over Existing Guidelines or Design Recommendations}\label{sec:liter}

A review of general design guidelines for mHealth applications was conducted to establish a foundational understanding of best practices. In 2019, the Xcertia guidelines were introduced to specifically address usability concerns in the mHealth application domain \cite{Xcertia_2020}. The guidelines consist of ten distinct sections that address key usability issues for mHealth applications, as summarized below: 1) Visual design, 2) Readability, 3) App navigation, 4) Onboarding, 5) App feedback, 6) Notifications, 7) Alerts and alarms, 8) Historical data, 9) Ongoing application evaluation, 10) Help resources and troubleshooting. Additionally, \citet{appevaluation} developed nine rubrics to help users evaluate the relevance, quality, functionality,
and security of medical applications. Complementing these efforts, \citet{stoyanov2015mobile} proposed the Mobile App Rating Scale, a framework for evaluating mHealth applications. The scale includes four objective quality dimensions: engagement, functionality, aesthetics, and information quality, as well as a \textit{subjective quality} dimension. The World Wide Web Consortium (W3C) has established widely recognized accessibility standards, which provide comprehensive guidelines for digital and mobile accessibility \cite{W3C}. WCAG 2.0, released in 2008, introduced foundational accessibility guidelines focused primarily on \textit{web accessibility}. WCAG 2.1, introduced in 2018, added 17 new criteria to address accessibility challenges in \textit{mobile applications}, low vision support, and cognitive disabilities. WCAG 2.2 (2023) introduced additional criteria targeting further inclusivity. Despite increasing research on accessibility within the realm of mobile applications \cite{alshayban2020accessibility,ross2017epidemiology}, and some initiatives that address mHealth applications \cite{milne2014accessibility, daihua2015accessibility}, significant accessibility challenges persist in these applications \cite{alshayban2020accessibility, ross2017epidemiology}. In particular, there is still a lack of comprehensive guidelines to improve the accessibility of mHealth applications. Current initiatives emphasize three critical dimensions in improving accessibility for mHealth applications: 1) \textbf{\textit{Accessible content presentation} }ensures that content and information are effectively delivered to all users, regardless of their abilities \cite{radcliffe2021pilot,zhou2020making,stowell2018designing}. 2) \textbf{\textit{Inclusive interaction}} focuses on enabling users to seamlessly interact with the application, regardless of physical or cognitive limitations \cite{radcliffe2021pilot,zhou2020making}. 3) \textbf{\textit{Reliable and assistive functionality}} ensures that users can depend on the application to provide accurate and error-free information through assistive technologies \cite{radcliffe2021pilot,zhou2020making}. 

Chronic diseases remain significantly underrepresented in accessibility research \cite{mack2021we, mack2022chronically}. There is a lack of standardized guidelines for tailoring applications related to various chronic diseases for different usage contexts. However, three recurring themes have emerged as critical for the design of AUI in this domain: 1) \textbf{\textit{Transparency}} emerged as a dominant theme in multiple studies. The researchers consistently underscored the need for transparency in various aspects of AUI, including data management, utilization, and specific decision-making processes \cite{martin2019exploring, lage2014role, mezhoudi2015toward}. This emphasis on transparency aimed to empower users by providing them with comprehensible explanations regarding the adaptations made by the system. 2) \textbf{\textit{Autonomy}} is another salient topic that has attracted considerable attention. Numerous studies have highlighted the importance of providing users with a sense of autonomy within AUI, particularly in terms of customizing system content, interaction modalities, and data management procedures \cite{zhang2021designing, sanders2019exploring, martin2019exploring}. This autonomy-centric approach aimed to empower users and cater to their individual preferences and needs. 3) \textbf{\textit{Learnability}} emerged as a crucial consideration to ensure the accessibility and usability of AUI. Several studies focused on facilitating a smooth and intuitive learning curve for users, especially those with limited technical proficiency. Strategies to improve learnability included the provision of user-friendly interfaces, intuitive navigation systems, and customized information presentations adapted to existing user knowledge and familiarity with relevant concepts \cite{martin2019exploring, eslami2018user}. 

\subsection{Evaluation of Guidelines with End-users}\label{sec:userfeedback}
\begin{figure}[b]
    \centering
    \includegraphics[width=1\linewidth]{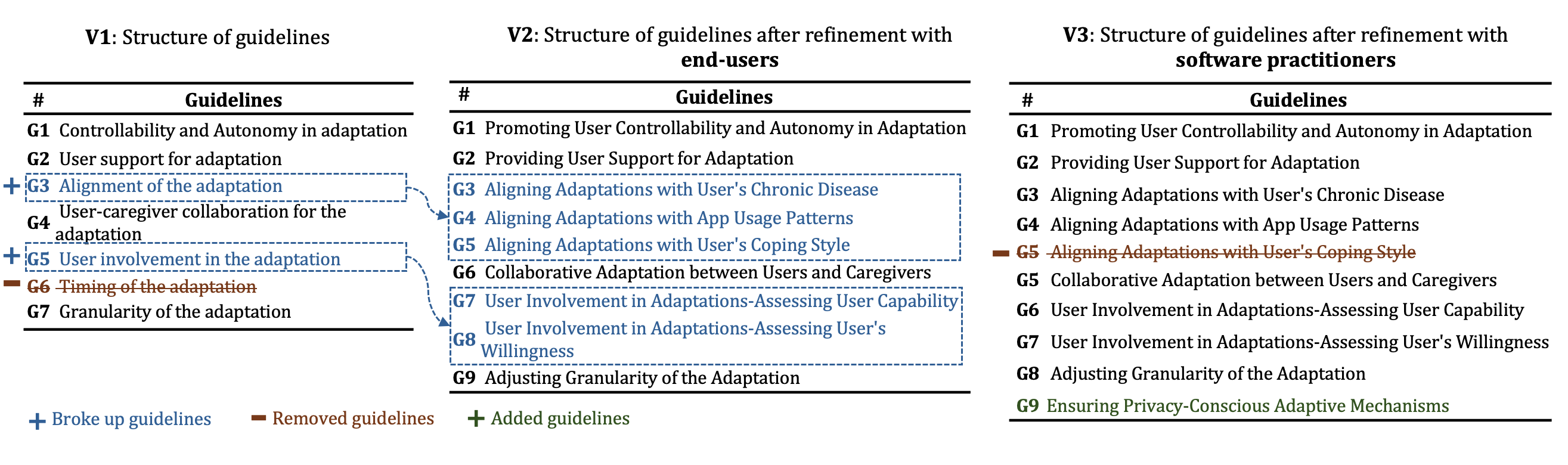}
    \caption{Evolution of the structure of the guidelines(Version One->Version Two->Version Three)}
    \label{fig:guidelinechangev1}
\end{figure}

A total of 20 participants completed the guideline evaluation survey. Based on their feedback, the original set of guidelines (Version One) was refined to produce a more comprehensive and actionable set (Version Two). As illustrated in Figure \ref{fig:guidelinechangev1}, two guidelines (G3\textsubscript{v1} and G5\textsubscript{v1}) were expanded into more specific subguidelines to improve clarity and usability. One guideline (G6\textsubscript{v1}) was removed due to its limited contribution and conceptual overlap with others. These revisions led to a final structure consisting of nine distinct guidelines in Version Two, as shown in Figure \ref{fig:guidelinechangev1}.

% \begin{figure}[ht!]
% \centering
% \includegraphics[width=0.97\textwidth]{Figure/Guideline_change.pdf}
% \caption{Evolution of the structure of the guidelines following the feedback survey.}
% \label{fig:guidelinechange}
% \end{figure}

\subsubsection{Clarity of the guidelines}
Although most of the participants found the guideline understandable, a few encountered difficulties with certain technical terms, particularly \textit{"Granularity"} (G7\textsubscript{v1}) and \textit{"Autonomy"} (G1\textsubscript{v1}). These terms may require further clarification or simplification to ensure greater comprehension among users. Meanwhile, a critique regarding the \textbf{\textit{difficulty in relating to the examples}} provided for each guideline underlines the need for clearer and more context-specific illustrations. Participants expressed confusion about how the examples were applied to their own experiences and situations, particularly in the case of the \textit{alignment of the adaptation (G3\textsubscript{v1})} guideline. The challenge of creating relevant and accessible examples is a common issue in guideline development, as noted in the literature \cite{ning2019addressing}. It is essential to recognize that the guidelines were developed primarily for software practitioners, whereas end-users may interpret and prioritize aspects of these guidelines differently due to their distinct experiences, needs, and contextual understanding.

Some guidelines like G3\textsubscript{v1} (\textit{Alignment of the Adaptation}) and G5\textsubscript{v1} (\textit{User Involvement in the Adaptation}) have attracted confusion because they can be too broad and difficult to interpret. This was further supported by the clarity ratings, where both G3\textsubscript{v1} and G5\textsubscript{v1} received the lowest levels of strong agreement on clarity, each at only 40\%. To address this issue, G3\textsubscript{v1} was subdivided into more focused and granular guidelines: G3\textsubscript{v2} (\textit{Aligning Adaptations with User's Chronic Disease}), G4\textsubscript{v2} (\textit{Aligning Adaptations with App Usage Patterns}) and G5\textsubscript{v2} (\textit{Aligning Adaptations with User’s Coping Style}). G5\textsubscript{v1} is broken down into: G7\textsubscript{v2} (\textit{User Involvement in Adaptations-Assessing User Capability}) and G8\textsubscript{v2} (\textit{User Involvement in Adaptations-Assessing User’s Willingness}) (see Figure \ref{fig:guidelinechangev1}). Upon further examination, G6\textsubscript{v1} (\textit{Timing of the adaptation}) has been \textbf{removed} from our guidelines because participants perceive it as too general, affecting its clarity, and it overlaps with G4\textsubscript{v1}.

\subsubsection{Usefulness of the guidelines}
When evaluating the guidelines, participants expressed difficulty in relating to some of them and struggled to understand their purpose. It is frequently observed in the development of guidelines that they tend to state actions without explaining their rationale or offering implementation advice \cite{carter1999incorporating}. We have supplemented each guideline with its \textbf{purpose }to offer context and aid in understanding. Furthermore, participants with experience in application design noted the potential\textit{ overlap} between the proposed guidelines and existing general guidelines for designing mHealth applications. This overlap in guidelines aligns with similar findings from other studies \cite{srivastava2021actionable}. In addition, the participants in the evaluation study emphasized the need to \textbf{prioritize} certain guidelines under different situations. This emphasizes the challenge of designing for the hypothetical \textit{"general"} user \cite{carter1999incorporating}. Feedback from the survey revealed that users engage with chronic disease-related applications in a variety of ways. For instance, younger users often utilize these applications to independently monitor health metrics, which can render guidelines focused on user-caregiver collaboration, such as G4\textsubscript{v1}, less relevant for this subgroup. The achievement of all design goals for a computer-based product or service often involves \textbf{trade-offs} \cite{johnson2020designing}, as the guidelines proposed in this study, discussed in Section \ref{meomoreflect}, are not exempt from such tensions, with certain recommendations potentially conflicting depending on user needs and system priorities.

\subsection{Evaluation of Guidelines with Software Practitioners}\label{sec:softfeedback}
A total of 43 software practitioners completed our evaluation survey. Each participant was assigned a unique identifier (e.g., [S1]) to reference their input throughout the analysis. Based on their feedback, guidelines have been refined to better correspond to the mHealth application design, with actionable tips added to help developers apply the guidelines in diverse design contexts. One new guideline is \textit{introduced} (G9\textsubscript{v3}), while another is \textit{removed} due to redundancy and limited applicability (G5\textsubscript{v2}). The resulting nine guidelines (Version Three) are illustrated in Figure \ref{fig:guidelinechangev1}, with detailed descriptions for each guideline provided in Section \ref{sec:guideline}.

\begin{table}[b]
\centering
\renewcommand{\arraystretch}{1}
\caption{Survey software practitioners participants demographics information (n=43)}
\label{tab:softwaredemographics}
    \resizebox{0.92\textwidth}{!}{%
\begin{tabular}{p{69mm}p{130mm}}
\hline
\toprule
    \begin{tabular}[t]{p{30mm}p{2mm}l}
    \textbf{Demographics} & \textbf{\#}  & \textbf{\% of Participants}  \\ \hline
    \multicolumn{3}{l}{\textit{\textbf{Gender}}}\\ \hline
         Female&  13& \mybarhhigh{4.3}{30\%}\\
        Male & 29 &\mybarhhigh{10}{70\%} \\ 
        \multicolumn{3}{l}{\textit{\textbf{Age}}}\\ \hline
        18 - 24&18& \mybarhhigh{6}{42\%}\\
        25 - 34&17& \mybarhhigh{5.6}{40\%}\\
        35 - 44&5& \mybarhhigh{1.7}{12\%}\\
        45 - 54&2& \mybarhhigh{0.6}{5\%}\\
        55 - 64&1& \mybarhhigh{0.3}{2\%}\\
    \multicolumn{3}{l}{\textit{\textbf{Company size}}}\\ \hline
    Less than 10 & 11 & \mybarhhigh{3.67}{21\%}\\
    11-50 & 4 & \mybarhhigh{1.33}{8\%}\\
    51-100 & 9 & \mybarhhigh{3.00}{17\%}\\
    101-500 & 4 & \mybarhhigh{1.33}{8\%}\\
    501-1000 & 5 & \mybarhhigh{1.67}{9\%}\\
    More than 1000 & 8 & \mybarhhigh{2.67}{15\%}\\
    Prefer not to say & 2 & \mybarhhigh{0.67}{4\%}\\
    \multicolumn{3}{l}{\textit{\textbf{Years of working experience}}}\\ \hline
        0-2 years & 17 & \mybarhhigh{5.67}{40\%}\\
        3-5 years & 11 & \mybarhhigh{3.67}{26\%}\\
        6-10 years & 11 & \mybarhhigh{3.67}{26\%}\\ 
        11+ years & 4 & \mybarhhigh{1.33}{9\%}\\
    \multicolumn{3}{l}{\textit{\textbf{Ethnicity simplified}}}\\ \hline    
    White & 20 & \mybarhhigh{6.67}{47\%} \\
    Asian & 14 & \mybarhhigh{4.67}{33\%} \\
    Black & 4 & \mybarhhigh{1.33}{9\%} \\
    Other & 3 & \mybarhhigh{1.00}{7\%} \\
    Mixed & 2 & \mybarhhigh{0.67}{5\%} \\

    \end{tabular}
&
\begin{tabular}[t]{p{93mm}p{2mm}l}
    \textbf{Demographics} & \textbf{\#}  & \textbf{\% of Participants}  \\ \hline
    \multicolumn{3}{l}{\textit{\textbf{Country of residence}}}\\ \hline
        Australia & 11 & \mybarhhigh{3.7}{26\%}\\
        Canada & 7 & \mybarhhigh{2.3}{16\%}\\
        Portugal & 6 & \mybarhhigh{2.0}{14\%}\\
        Poland & 4 & \mybarhhigh{1.3}{9\%}\\
        Mexico & 3 & \mybarhhigh{1.0}{7\%}\\
        Chile & 2 & \mybarhhigh{0.7}{5\%}\\
        India & 2 & \mybarhhigh{0.7}{5\%}\\
        Italy & 2 & \mybarhhigh{0.7}{5\%}\\
        United States of America & 2 & \mybarhhigh{0.7}{5\%}\\
        \multicolumn{3}{p{115mm}}{Greece, Netherlands, New Zealand, United Kingdom of Great Britain and Northern Ireland 1\% Each}\\

    \multicolumn{3}{l}{\textit{\textbf{Roles in the team *}}}\\ \hline
    Programmer & 21 & \mybarhhigh{7.0}{49\%} \\
    User interface or Graphical user interface designer & 16 & \mybarhhigh{5.33}{37\%} \\
    Software architect & 15 & \mybarhhigh{5.0}{35\%} \\
    Tester & 10 & \mybarhhigh{3.33}{23\%} \\
    Project manager & 8 & \mybarhhigh{2.67}{19\%} \\
    App animator or operations developer/engineer & 4 & \mybarhhigh{1.33}{9\%} \\
    QA engineer & 3 & \mybarhhigh{1.0}{7\%} \\
    Requirements analyst & 2 & \mybarhhigh{0.67}{5\%} \\ 
    \multicolumn{3}{p{115mm}}{Business consultant/Marketing manager/Sales personnel, Technical Lead, Researcher 2\% Each}\\
      \multicolumn{3}{l}{\textit{\textbf{Experience in health-related applications}}}\\ \hline
    Chronic disease management (e.g., diabetes tracking apps) & 14 & \mybarhhigh{4.67}{33\%} \\
    General health and wellness Tools (e.g., diet apps)& 14 & \mybarhhigh{4.67}{33\%} \\
    Specialized applications (e.g., tumor detection)& 12 & \mybarhhigh{4.00}{28\%} \\
    Experience in related fields (e.g., testing for health apps)& 3 & \mybarhhigh{1.00}{7\%} \\
    
    \end{tabular} \\
        \hline  \multicolumn{2}{p{200mm}}{\textit{* This does not added up to 100\%, because some participants took several roles. Other categories of demographic data may not sum to 100\% due to rounding.}}\\
\bottomrule
\end{tabular}}
\end{table}
\subsubsection{Demographic of software practitioners}
Table \ref{tab:softwaredemographics} provides a summary of the demographic details of the participants. Participants are predominantly men (70\%) and relatively young, with 42\% aged 18-24 and 40\% aged 25-34, indicating a sample largely composed of early career professionals. Most of the respondents had 0-2 years (40\%) or 3-10 years (52\%) of experience, and only 9\% had more than 11 years of experience. The largest groups by geography hail from Australia (26\%), Canada (16\%), and Portugal (14\%), with smaller groups from various countries in Europe, North America, and Asia. The participants are employed in organizations of varying sizes, with 21\% working in small companies (with fewer than 10 employees) and others in medium to large organizations. In terms of professional roles, the largest group is identified as programmers (49\%), followed by UI designers (37\%), software architects (35\%), and testers (23\%). These demographics highlight the diversity of professional backgrounds and responsibilities of survey participants within the software development sector.

The responses reveal a diverse range of experiences in developing health-related applications, highlighting significant contributions to chronic disease management (33\%), general health tools (33\%), specialized applications (28\%) and other experiences in health-related fields (7\%). Many survey respondents reported working on applications for managing chronic diseases, incorporating features such as tracking vital signs, medication adherence, and personalized health recommendations. Others contributed to general health and wellness applications, including applications promoting healthy routines, dietary management, and senior-focused medication management systems. Specialized projects ranged from tumor detection applications and mental health solutions to orthodontics-related tools and the digitization of healthcare facilities like nursing homes and hospitals. Participants also played key roles in testing, UI design, and  information technology (IT) support.

\begin{figure}[b]
    \centering
    \includegraphics[width=1\linewidth]{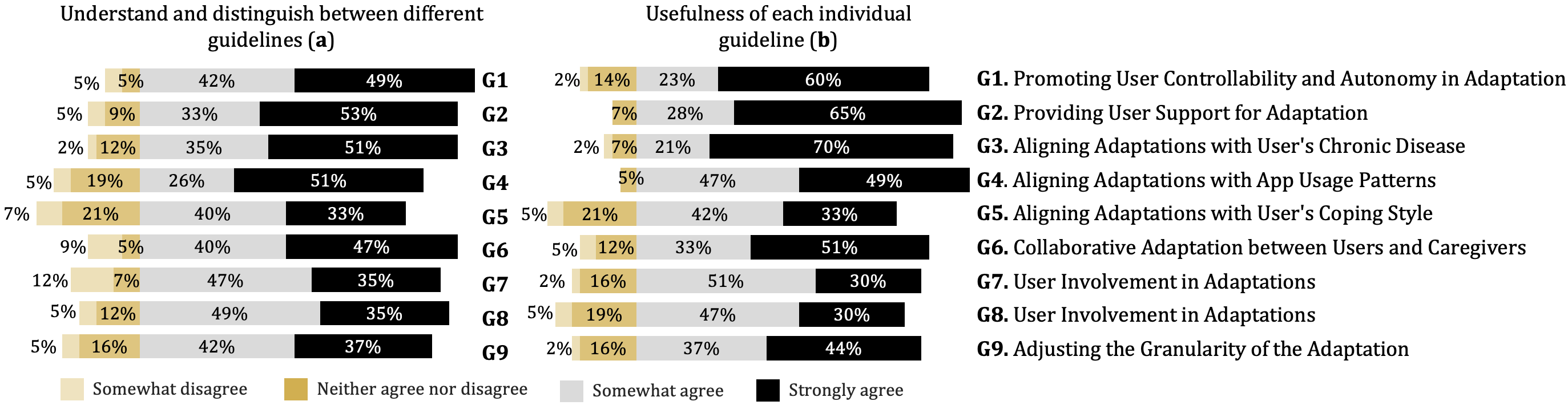}
    \caption{Distribution of participants’ rating on the proposed guidelines (\textit{Version Two}) for – (a) Understand and distinguish between different guidelines (b) Usefulness of each individual guideline }
    \label{fig:perception}
\end{figure}

\subsubsection{Understandability and usefulness of guidelines}

Figure \ref{fig:perception} provides insight into participants’ perceptions of the guidelines related to user adaptation in applications, evaluating both their understanding and their usefulness. Most of the participants reported strong agreement with the clarity of G1\textsubscript{v2} (49\%) and G2\textsubscript{v2} (53\%), indicating that these two guidelines were among the most easily understood and distinguishable. Similarly, G3\textsubscript{v2} and G4\textsubscript{v2} received strong agreement from 51\% of the respondents, indicating that they were generally well understood. In contrast, G9\textsubscript{v2} and G5\textsubscript{v2} had lower strong agreement ratings (37\% and 33\%, respectively), suggesting that these guidelines were perceived as more ambiguous and may require further clarification. In terms of perceived usefulness, G3\textsubscript{v2} stood out as the guideline most positively rated, with 70\% of the participants strongly agreeing on its utility, followed by G2\textsubscript{v2} (65\%) and G1\textsubscript{v2} (60\%). Guidelines such as G6\textsubscript{v2} and G9\textsubscript{v2} also received considerable support, with 51\% and 44\% in strong agreement, respectively. The survey participants overwhelmingly demonstrated a strong preference (91\%) for using the proposed guidelines over existing standalone guidelines to design mHealth applications, although a small minority (9\%) expressed skepticism. In general, while most guidelines were perceived as useful and understandable, such as G9\textsubscript{v2} and G5\textsubscript{v2}, may require refinement to improve clarity and effectiveness. The positive reception of user-centered guidelines highlights their value in improving usability in applications.

\subsubsection{Strengths of the guidelines} 
\begin{figure}[b]
    \centering
    \includegraphics[width=0.8\linewidth]{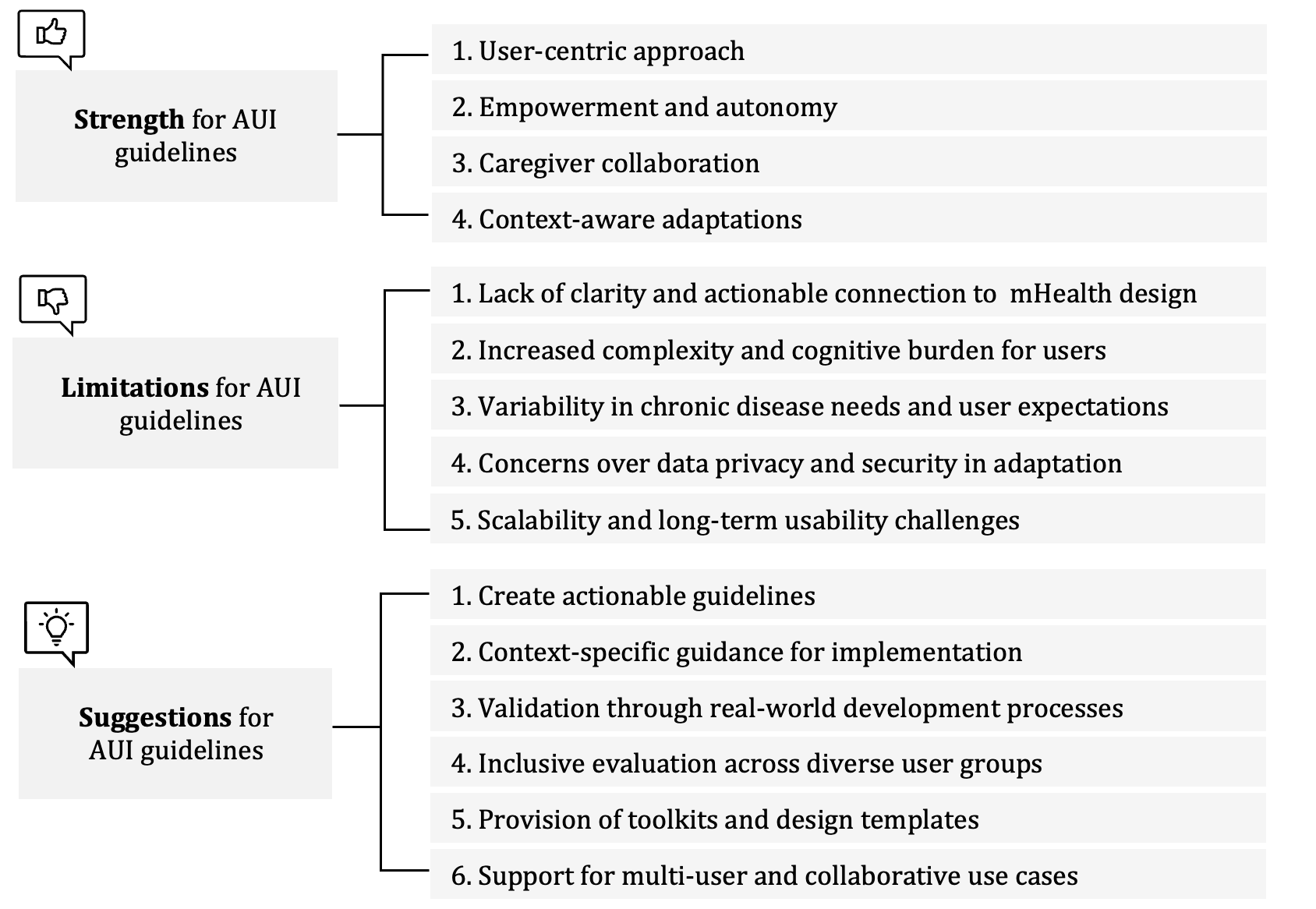}
    \caption{Opinions of software practitioners regarding the strengths, limitations, and suggestions for the guidelines}
    \label{fig:strengthlimitation}
\end{figure}

Software practitioners also provided several comments that highlighted additional strengths and limitations of the proposed guidelines (see Figure \ref{fig:strengthlimitation}). The guidelines demonstrate significant strengths in creating \textbf{user-centric}, personalized applications, particularly for individuals who manage chronic diseases. Personalization emerged as the theme most frequently endorsed, referenced in 78\% of participant responses, with many commending the guidelines for enabling a personalized application experience. In particular, G3\textsubscript{v2} (\textit{Aligning Adaptations with Chronic Diseases}) and G1\textsubscript{v2} (\textit{Promoting User Controllability and Autonomy}) were specifically praised for empowering users to manage their health while maintaining the flexibility and user-friendliness of the system. Another key strength is the emphasis on\textbf{ empowerment and autonomy}, identified in 62\% of the responses. The participants appreciated that the guidelines allow users to control application adaptation, reduce frustration, and encourage sustained use. G1\textsubscript{v2} and G9\textsubscript{v2} were frequently cited as ensuring user control without overwhelming them. \textbf{Caregiver collaboration}, addressed in G6\textsubscript{v2}, was mentioned in 38\% of the responses, emphasizing its critical role in shared health management. Including caregivers not only supports users who need additional help, but also improves adherence to health routines through clear communication and shared decision making. A focus on \textbf{context-aware adaptations} was observed in 85\% of the responses, highlighting the importance of aligning the application features with usage patterns, coping styles, and real-time user behavior and chronic disease management. Guidelines such as G5\textsubscript{v2} (\textit{Aligning Adaptations with Coping Styles}) and G4\textsubscript{v2} (\textit{Aligning Adaptations with App Usage Patterns}) are considered critical to maintaining the usability and relevance of the application. 
% The participants also appreciated how the guidelines addressed both the emotional and functional needs of individuals with chronic diseases, allowing routine maintenance and self-management. For example, several respondents noted that these guidelines avoid a one-size-fits-all approach, giving users and developers flexibility to adapt to unique challenges.
\subsubsection{Limitations of the guidelines} 
While the proposed guidelines demonstrate substantial strengths, the evaluation also uncovered several limitations that may hinder their practical implementation. One key challenge lies in the difficulty of concretely \textbf{linking} certain guideline elements to specific \textit{mHealth application design} decisions, with some components perceived as inconsistently applicable. For example, [S4] highlighted the challenge of clearly defining the concept of user involvement and its applicability across various types of mHealth applications. One common concern is the risk of overwhelming \textbf{complexity}, as participants emphasized the challenge of maintaining an appropriate balance between the level of granularity, user autonomy, and overall interface simplicity. [S15] noted: 

\boxquote{Excessive customization, mental overload, and caregiver dependency could affect user experience. Collaborative adaptation requires sharing personal health data between the user and their caregivers. Both parties need to have access to sensitive health information, but managing who has access to what data, like diagnosis, medication and treatment progress, can add more complexity [to the adaptation].} 

\noindent Guidelines such as G1\textsubscript{v2} (\textit{Promoting User Controllability and Autonomy in Adaptation}) and G9\textsubscript{v2} (\textit{Adjusting the Granularity of the Adaptation}) exemplify this tension, as they must navigate the fine line between offering users sufficient control and customization without introducing excessive complexity that could hinder usability or reduce engagement. An additional limitation identified was the challenge in accommodating the \textbf{variation} associated with chronic disease requirements. Guidelines such as G3\textsubscript{v2} (\textit{Aligning Adaptations with User's Chronic Disease}) must accommodate fluctuating health condition, which may require frequent updates, risking user fatigue or misaligned adaptations. [S23] noted that the \textit{willingness }of users to modify the application can vary depending on the context. For example, users might initially resist making changes but become more open to adaptation as their familiarity with the applications or their health condition evolves over time. \textbf{Privacy and security} concerns are also prominent, especially when sharing sensitive health data with caregivers. [S16] highlighted the constant risk of fraud and the possibility of unauthorized access to patient personal medical information. \textbf{Scalability} and \textbf{long-term usability} challenges emphasize the importance of continuous testing and refinement to ensure that the system remains relevant and effective over time. Furthermore, [S6] pointed out a potential drawback of customization.

\boxquote{They [the system] might modify crucial settings that could negatively impact them in the future without their[the user's] awareness. These designs must ensure that the guidelines are put into practice effectively without harming the user experience.} 

\noindent [S18] noted that if users are not in a suitable physical or mental state, prolonged use of the application can decrease, reducing its effectiveness. Furthermore, [S9] raised concerns about excessive \textit{dependency} on the application, especially during urgent or emergency situations, suggesting that such features should be limited and paired with clearer usage instructions to prevent misuse. Several participants emphasized that certain guidelines warrant particular attention, with G5\textsubscript{v2} (\textit{Aligning Adaptations with User’s Coping Style}) being especially challenging. Developing an application that effectively reflects the user's coping style necessitates continuous fine-grained data collection over time, which raises concerns about privacy, security, and the potential for user fatigue from repeated prompts or data input. [S31] illustrated an intriguing situation in which the importance of different guidelines can fluctuate, highlighting the \textbf{\textit{interconnected}} nature of the guidelines. This participant also noted that specific guidelines can \textbf{\textit{interact with}} others, implying that changes to one may require adjustments to another.

\boxquote{G3\textsubscript{v2} addresses the fluctuation in the severity of chronic diseases, an initial configuration would be more beneficial for those with stable chronic diseases. An additional scenario is when a user has complete mobility at first and does not need any special adaptations initially. However, their mobility may decrease with time, which could make them less able (or willing) [G8\textsubscript{v2}] to change settings later [S31].}

\textit{Response to the limitations}. Some limitations noted earlier have been thoroughly examined in the AUI literature, including concerns such as privacy, learnability, high complexity, and usability challenges \cite{paymans2004usability, gajos2008predictability, wesson2010can, lavie2010benefits, peissner2013user, wang2024adaptive}. Our aim is to highlight these issues in the guidelines while offering strategies for designing AUIs that effectively address or offset these drawbacks to meet users' needs. Our attention will be directed towards improving the clarification of the guidelines, connecting this with the design of the mHealth application, and refining the guidelines associated with the management of chronic diseases.

\subsubsection{Suggestions for the guidelines}
Participants also offered valuable suggestions to improve the practicality and impact of the guidelines. Several respondents emphasized the importance of translating the guidelines into clear, \textbf{actionable }steps for development teams. Guidelines such as G1\textsubscript{v2}, G3\textsubscript{v2}, G5\textsubscript{v2}, G6\textsubscript{v2}, and G9\textsubscript{v2} were identified as needing more concrete implementation strategies, particularly in areas such as adaptive learning pathways and data privacy protections. For example, [S1] suggested improving G1\textsubscript{v2} (\textit{Promoting User Controllability and Autonomy in Adaptation}) by allowing users to gradually learn about controls through tutorials or default settings. Furthermore, [S28] emphasized the incorporation of explicit consent mechanisms for caregiver access under G6\textsubscript{v2} (\textit{Collaborative Adaptation between Users and Caregivers}), as well as the provision of secure sharing options, such as temporary access tokens or granular data permissions for collaborative features. The participants called for clarity on the \textbf{context-specific guidance for implementation}, with a strong focus on understanding the \textit{interaction} between different guidelines. For example, [S11] highlighted the importance of examining how guidelines interact in scenarios that involve collaborative adaptations with caregivers. Further recommendations included developing a \textbf{holistic onboarding process} to assess the physical and mental capacities of users. This process should allow periodic reassessments to ensure that users are not permanently classified based on their initial evaluations. Participants emphasized the need for more explicit and \textbf{context-specific} guidance to support the effective implementation of the guidelines. [S11] highlighted the importance of examining how guidelines interact in scenarios involving collaborative adaptations with caregivers. Several participants stressed the importance of conducting a continuous and iterative evaluation of the guidelines throughout the \textbf{actual development process}. Such practical evaluations would help identify gaps between the theoretical guidance and its usability in real development environments. In addition, several participants advised to perform extensive user research covering \textbf{various demographics}, such as age, socioeconomic status, and cultural background, to ensure that the guidelines are applicable and inclusive across the board. Some participants also advised to consider offering \textbf{toolkits or design templates} to developers, particularly for complex aspects such as G3\textsubscript{v2} (\textit{Aligning Adaptations with the User’s Chronic Disease}) and G9\textsubscript{v2} (\textit{Adjusting Granularity of the Adaptation}). These could serve as an initial guide for developers who might lack experience in dealing with AUIs. In addition, it could be advantageous to include examples or suggestions for the implementation of the guidelines in \textbf{multi-user scenarios}, such as family-oriented applications or shared interfaces between caregivers and patients. 

\textit{Response to the suggestions}. The guidelines have been refined in response to the specificity of the suggestions in the context of mHealth application design. A new guideline, G9\textsubscript {v3} (\textit{Ensuring Privacy-Conscious Adaptive Mechanisms}), has been introduced to improve practical applicability, while G5\textsubscript {v2} (\textit{Aligning Adaptations with User’s Coping Style}) has been removed. To assist software practitioners, actionable tips have been provided to apply the guidelines in various scenarios. Although some suggestions hold significant promise, they could be further explored and refined in future work.

\subsection{Guidelines}\label{sec:guideline}
\begin{table}[b!]
    \centering
        \caption{Importance ratings of guidelines (Version Two)}
        \resizebox{0.9\textwidth}{!}{%
    \begin{tabular}{cp{100mm}p{35mm}|p{35mm}}
        \hline
        \textbf{ID} & \textbf{Guidelines} & \textbf{End user} & \textbf{Software practitioner} \\ \hline
        G1\textsubscript{v2} & Promoting User Controllability and Autonomy in Adaptation & \cellcolor{signifl} Important & \cellcolor{signifl} Important \\ 
        G2\textsubscript{v2} & Providing User Support for Adaptation & \cellcolor{signif} Critical & \cellcolor{signif} \cellcolor{signif} Critical \\ 
        G3\textsubscript{v2} & Aligning Adaptations with User's Chronic Disease & \cellcolor{greyb} Helpful & \cellcolor{signif} Critical \\ 
        G4\textsubscript{v2}\ & Aligning Adaptations with App Usage Patterns & \cellcolor{greyb} Helpful & \cellcolor{signif} Critical \\ 
        G5\textsubscript{v2}   & Aligning Adaptations with User's Coping Style & \cellcolor{greyb} Helpful & \cellcolor{greyb} Helpful \\ 
        G6\textsubscript{v2}& Collaborative Adaptation between Users and Caregivers & \cellcolor{signif} Critical & \cellcolor{signifl} Important \\ 
        G7\textsubscript{v2}  & User Involvement in Adaptations- Assessing User Capability & \cellcolor{greyb} Helpful& \cellcolor{greyb} Helpful \\
        G8\textsubscript{v2}  & User Involvement in Adaptations- Assessing User's Willingness & \cellcolor{greyb} Helpful& \cellcolor{greyb} Helpful \\ 
        G9\textsubscript{v2}  & Adjusting Granularity of the Adaptation & \cellcolor{signif} Critical & \cellcolor{signifl} Important \\ \hline
        \multicolumn{4}{p{150mm}}{\textit{* Critical >Important >Helpful}}\\
        \hline
    \end{tabular}}

    \label{tab:guidelines}
\end{table}
Table \ref{tab:guidelines} highlights the varying importance of guidelines for adaptation in mHealth applications from the perspective of\textit{ end-users} and \textit{software practitioners}, categorized into critical, important, and helpful tiers. It is important to note that end-user evaluations were based on Version One of the guidelines, which included only seven guidelines (see Figure \ref{fig:guidelinechangev1}). Broader guidelines like \textit{Alignment} and \textit{User Involvement} had not yet been subdivided at that stage. This context explains why alignment- and involvement-related guidelines are largely rated as only \textit{“helpful”} by end-users (see Table \ref{tab:guidelines}). Several guidelines, particularly G2\textsubscript {v2}, G6\textsubscript {v2}, and G9\textsubscript {v2}, emerge as universally important from both the end-user and the software practitioner's perspectives. In addition, examining the relationships between these guidelines reveals key interdependence. For example, G1\textsubscript{v2}, and G9\textsubscript{v2} are closely \textbf{interconnected}, as a high degree of granularity typically requires strong user support mechanisms and effective user control. Similarly, G6\textsubscript{v2} can improve the implementation of G7\textsubscript{v2} and G8\textsubscript{v2} by leveraging caregiver participation to better address primary user capability and willingness to engage in the adaptation process. In light of these insight, the \textit{Version Three} guidelines have been categorized into four groups, each reflecting a distinct design focus. 

\begin{enumerate}[left=0pt]
    \item \textbf{User Support and Interaction:} Users often have varying levels of familiarity with digital platforms, which can create barriers to effective interaction. By providing clear guidance and support, this category of guidelines ensures that all users, regardless of their technical skills or physical abilities, can navigate and utilize the application seamlessly.\\
    \textbf{Associated guidelines}: \textbf{G2:} Providing User Support for Adaptation, and \textbf{G6:} User Involvement in Adaptations-Assessing User Capability.
    \item \textbf{Context-Aware Adaptations:} Chronic disease management often involves varying needs based on the user's health condition and application usage patterns. The different purposes for using mHealth applications result in different usage patterns, with differences in the frequency and duration of each session. Context-aware adaptations ensure that the application remains relevant and effective by aligning its functionalities with the user’s goal of using it. \\ 
    \textbf{Associated guidelines}: \textbf{G3:} Aligning Adaptations with User's Chronic Disease, \textbf{G4:} Aligning Adaptations with App Usage Patterns, and \textbf{G9:} Ensuring Privacy-Conscious Adaptive Mechanisms.
    \item \textbf{Caregiver Collaboration and Adaptation:} Carers are key in overseeing application management and modifications, particularly for users who have limited ability or engagement. Shared usage scenarios require mechanisms that support collaboration between the user and their caregiver. Adaptations should be designed to empower both parties without introducing unnecessary complexity or privacy risks.\\
    \textbf{Associated guidelines}: \textbf{G5:} Collaborative Adaptation between Users and Caregivers, \textbf{G7:} User Involvement in Adaptations-Assessing User's Willingness, and \textbf{G9:} Ensuring Privacy-Conscious Adaptive Mechanisms.
    \item \textbf{Empowerment and Autonomy:} Granting users comprehensive control and autonomy regarding the app's adaptive functionalities is key to some users. It highlights the significance of allowing users to customize their application experience according to their individual needs and preferences while staying informed about the application's operations. This category ensures that the user retains ownership of their health management journey by providing them with meaningful choices and control mechanisms.
    \\
    \textbf{Associated guidelines}: \textbf{G1:} Promoting User Controllability and Autonomy in Adaptation, \textbf{ G8:} Modifying Granularity and \textbf{G9:} Ensuring Privacy-Conscious Adaptive Mechanisms.
\end{enumerate}
\begin{figure}[b]
    \centering
    \includegraphics[width=1\linewidth]{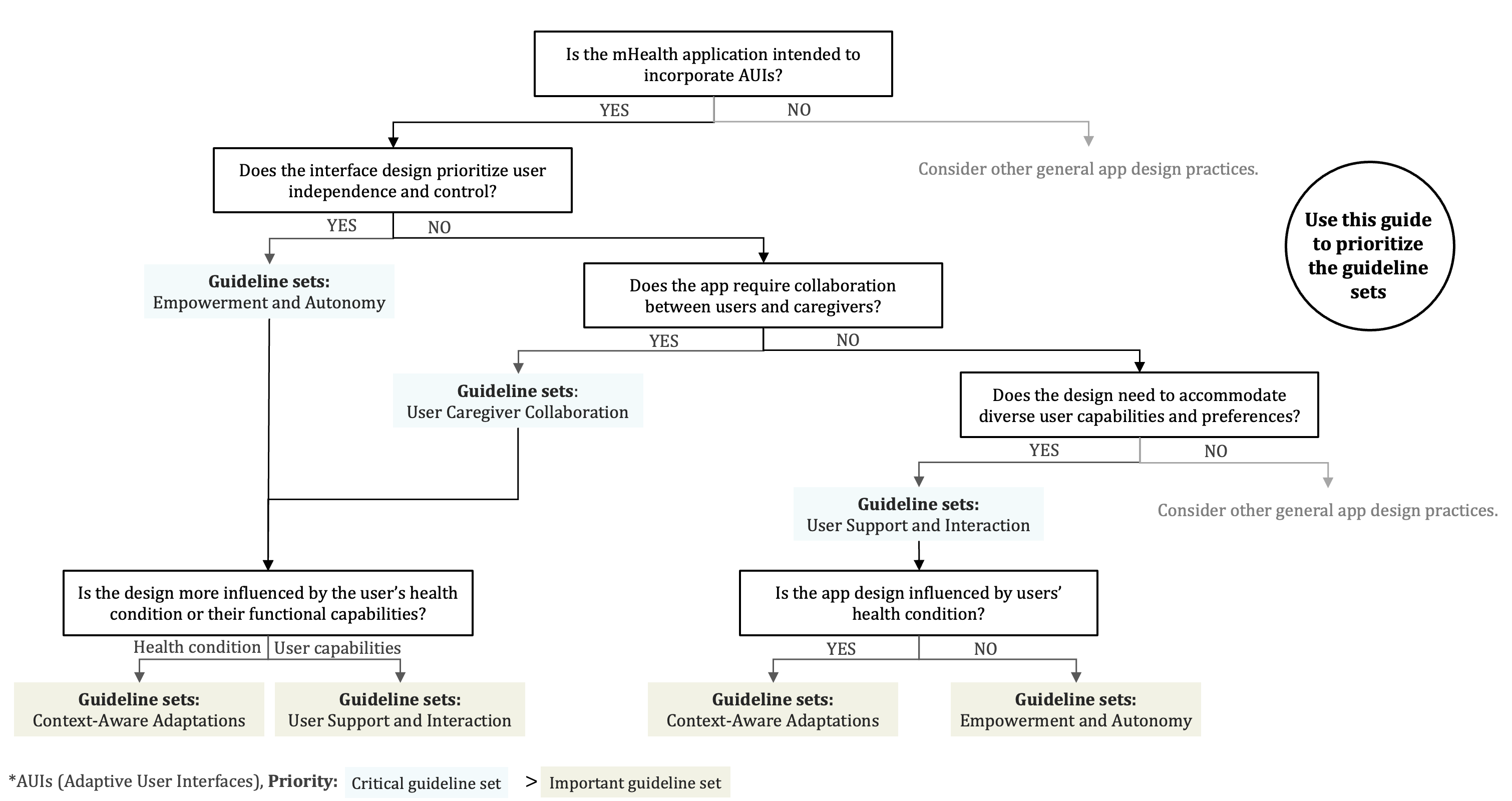}
    \caption{Guideline sets prioritization guide}
    \label{fig:guideflow}
\end{figure}

This set of guidelines may hold with different levels of importance depending on the scenario, as the software practitioners of the evaluation study recommend clearer prioritization steps for a specific context. This approach aligns with studies on accessibility in user review, advocating for a severity-based priority system to address critical needs \cite{reyes2022accessibility}. Figure \ref{fig:guideflow} provides a structured guide for prioritizing guideline sets when designing AUIs for mHealth applications that target chronic diseases. The selection process begins by determining whether the application is designed to empower the user \textit{independence} and control, in which case the \hconcep{Empowerment and Autonomy} guideline set is recommended. If not, the next step assesses whether the design requires collaboration between \textit{users and caregivers}, leading to the \hconcep{User Caregiver Collaboration} guideline set. If the application does not involve caregiver collaboration, the focus shifts to accommodating \textit{diverse user capabilities} and preferences, for which the \hconcep{User Support and Interaction} guideline set is applicable. Ultimately, the design's impact on users' \textit{health conditions} serves to further refine the prioritization between the \hconcep{Context-Aware Adaptations} or \hconcep{Empowerment and Autonomy} guideline sets. The following content presents Version Three of the guidelines, which represents the final refined set developed through iterative feedback and evaluation.

\linkphd{}{\textbf{G1:} Promoting User Controllability and Autonomy in Adaptation}

\noindent G1 involves empowering users to manage the adaptation process by balancing user control and system automation. It emphasizes the importance of offering users the flexibility to personalize their interactions, which ultimately leads to an improved user experience.

\begin{itemize}[left=16pt, topsep=5pt, itemsep=5pt]
    \item [\textbf{G1.a}] mHealth applications for chronic disease management can feature an \textit{“extra-UI”}, a dedicated adaptation dashboard that allows users to personalize their interface according to their specific needs and preferences \cite{melchior2012comparative}. For example, a user managing diabetes might configure the dashboard to prioritize glucose monitoring tools on the home screen while minimizing less relevant features, such as exercise tracking, to streamline their daily interactions with the application. An example of such a dashboard interface is illustrated in Figure \ref{fig:app1}.
    \item [\textbf{G1.b}] The mHealth application could offer a \textbf{step-by-step adaptation process}, allowing users to progressively experiment with different levels of adaptation. This approach aligns with the findings that people often prefer tasks of moderate complexity, which fosters a greater sense of competence and engagement \cite{deci2013intrinsic}. Moreover, gradual adaptation strategies have been shown to \textit{outperform both non-adaptive and fully adaptive systems} in terms of usability and user satisfaction \cite{todi2021adapting}. In the context of end-user development, several studies advocate starting with a minimal application and enabling users to iteratively solve tasks, each unlocking or adjusting new features based on prior interactions \cite{johnsson2020towards,castelli2017happened}. For example, \citet{castelli2017happened} demonstrated how users could customize smart home data visualizations through a guided incremental process, while \citet{schobel2016end} supported physicians in intuitively developing customized applications using similar step-by-step techniques.
    
    \item [\textbf{G1.c}] \textbf{Opt-in and opt-out features} can empower users by allowing them to selectively enable or disable specific adaptation functions based on their preferences and routines. In addition, the inclusion of a scheduling mechanism for adaptations enables users to define temporal boundaries, activate adaptive features during particular periods, such as work hours or active health management phases, and deactivate them during rest or downtime. 
    
    \item [\textbf{G1.d}] \textbf{Centralized adaptation} refers to the provision of a dedicated section within the system where users can configure all adaptive features. This approach reduces disruptions to the primary UI and preserves \textit{spatial consistency}. This design strategy aligns with previous findings that emphasize users’ preference for spatial stability, as frequent interface changes can increase cognitive load and hinder usability \cite{gajos2006exploring, Li2014AdaptiveCA, Deuschel2016OnTI}. As noted in \citet{deuschel2018influence}, maintaining spatial stability supports better user orientation and interaction efficiency, particularly in health-related applications where \textit{reliability} is critical.
\end{itemize}

\linkphd{}{\textbf{G2:} Providing User Support for Adaptation}

\noindent G2 focuses on ensuring that users can navigate and utilize AUIs effectively by providing adequate assistance and clear guidance. This guideline highlights the importance of helping users understand how adaptive features work, what they can expect from these features, and how to interact with them efficiently. By delivering streamlined support, users with varying levels of digital literacy can confidently engage with adaptive features. This user-centric approach ensures that the benefits of adaptation are fully realized without introducing confusion or cognitive overload.

\begin{itemize}[left=16pt, topsep=5pt, itemsep=5pt]
    \item [\textbf{G2.a}] mHealth applications for chronic disease management could incorporate \textbf{quick-access shortcuts} to streamline interaction with frequently used adaptive features, \textit{minimizing the workload} for users to navigate through layered menus. For example, a person who manages diabetes might find value in a one-tap shortcut on the home screen that allows immediate adjustment of notification preferences for glucose monitoring or dietary alerts. To help users engage with these adaptive features, \textit{onboarding tutorials} can be introduced during the initial setup process. These tutorials would provide guidance on how to configure and utilize adaptive options, ensuring that users understand the benefits and functionalities from the outset.

    \item [\textbf{G2.b}] \textbf{Access to relevant adaptation suggestions} is essential to support users who may face difficulties in customizing technology due to their health conditions. As highlighted in previous research \cite{zhang2021designing}, users experiencing significant mental health challenges often struggle with personalization tasks, making adaptive support critical. To accommodate this, mHealth applications could offer preset configurations, such as a \textit{“low-energy mode”} that simplifies the interface, minimizes notifications, and reduces visual clutter, to ease interaction and reduce cognitive load. These suggestions should offer immediate support yet be flexible, enabling users to adjust settings over time as their preferences and needs evolve.
    
    \item [\textbf{G2.c}]  Providing \textbf{contextual explanations for adaptations} within the application is essential to improve user understanding and ensure transparency in the adaptation process \cite{hook2000steps}. Users benefit from being able to interpret changes made to the interface. For example, \citet{teevan2009changing} demonstrated how \textit{highlighted adapted sections} on web pages helped users track content changes. Similarly, \citet{dessart2011showing} introduced \textit{animated transitions} to visualize the progression from a pre-adaptation interface to its adapted form. 
\end{itemize}

\linkphd{}{\textbf{G3:} Aligning Adaptations with User's Chronic Disease}
\noindent G3 emphasizes tailoring adaptive features to accommodate user-specific health conditions, including cases of multimorbidity, varying levels of disease understanding, and progression of chronic disease. If users do not perceive clear and practical relevance in the adaptation, they are unlikely to remain engaged with the application over time \cite{o2008user, short2018measuring}. Reflecting the nuanced needs of users with complex or evolving conditions, the system fosters a sense of support and empowerment, contributing to better long-term health outcomes.

\begin{itemize}[left=16pt, topsep=5pt, itemsep=5pt]
    \item [\textbf{G3.a}] Adaptations could include customized dashboards for users managing \textbf{multiple chronic diseases}, such as diabetes and hypertension. These dashboards can visually differentiate disease-specific information through intuitive icons, such as a syringe that represents insulin tracking or a heart symbol that indicates blood pressure monitoring, allowing users to quickly identify and navigate to the relevant sections. 
    
    \item [\textbf{G3.b}] Adaptive UI features can dynamically tailor the layout and content of the interface according to the user’s \textbf{health condition}. For example, a user with advanced diabetes might see a streamlined interface that prioritizes quick access to blood glucose tracking, insulin dose logs, and emergency contacts. In contrast, a user in the early stages of diabetes might have an interface that emphasizes educational tools to build awareness and encourage healthy habits. To support this approach, \citet{pagiatakis2020intelligent} presents a system that adapts its navigation structure during hypoglycemic events, restricts access to non-critical functions and prominently enables emergency contact features, highlighting how condition-sensitive adaptation improves usability and safety. An example of such an adaptation is shown in Figure \ref{sub:b}).
    % \item \textbf{G3.c:} Practical implementation might involve \textbf{disease-specific educational resources} within the app, dynamically adjusted based on user input. For example, if a user reports new symptoms or worsening conditions, the application could adapt by surfacing relevant tutorials, connecting users to support groups, or suggesting a consultation with a healthcare provider. 
    \item [\textbf{G3.c}] Adaptation strategies should account for users’ \textbf{attitudes toward their chronic disease}, particularly the \textit{coping mechanisms} they employ in response to health-related stress \cite{csikszentmihalyi2014positive}. Some users adopt approach-based coping styles and may seek continuous, detailed feedback on their health status. For these users, the application could provide regular data visualizations, trend alerts, and actionable recommendations. In contrast, users with avoidance-based coping tendencies might find frequent feedback overwhelming or demotivating. In such cases, the application could offer minimalistic summaries with customizable options to access more detailed information on demand, while still ensuring that critical alerts are delivered in a less intrusive, emotionally sensitive manner. The study by \citet{sefidgar2024migrainetracker} highlights how patients’ differing goals, such as monitoring, learning, or anticipating symptoms, influence their expectations of health data and applications, highlighting that individual attitudes significantly shape the engagement with adaptive health technologies.
    
    \item [\textbf{G3.d}] The adaptation process plays a critical role in supporting individuals with chronic diseases by \textbf{aligning} the application’s interface and features with their specific \textbf{health management needs}. For example, in a diabetes management app, adaptive UI components can highlight priority tasks such as blood glucose monitoring, medication reminders, or dietary tracking based on the user’s current health status and routines. By streamlining access to relevant functions and minimizing irrelevant content, adaptive systems can enhance usability, promote user confidence, and maintain long-term engagement, factors consistently highlighted as crucial in the literature on mHealth application adoption \cite{trajkova2020alexa, melenhorst2001use}.
\end{itemize}

\linkphd{}{\textbf{G4:} Aligning Adaptations with App Usage Patterns}

\noindent G4 emphasizes the alignment of adaptive features with users’ actual \textbf{usage patterns}, including how frequently, how long, and with what level of effort they engage with different functionalities in the application. The goal is to support a seamless user experience by integrating adaptations that feel intuitive, avoiding disruptions to users’ established routines. This behavioral alignment not only preserves workflow efficiency, but also fosters continued user engagement by delivering personalized, context-aware support that adapts to evolving usage habits.
\begin{itemize}[left=16pt, topsep=5pt, itemsep=5pt]
    \item [\textbf{G4.a}] Research indicates that the balance between routine and non-routine tasks, along with the effort involved in task execution, directly influences the effectiveness of AUIs \cite{lavie2010benefits, peissner2013user}. Therefore, tasks that are performed \textbf{frequently and require minimal cognitive or physical effort} are suitable for automation. In mHealth applications, this could be operationalized through smart automation for repetitive behaviors. For example, if a user habitually logs water intake after meals, the system could offer prefilled values based on historical patterns, requiring only user confirmation or minor edits. In contrast, tasks that are infrequent and more complex, such as setting or adjusting long-term health goals, may be best managed through user-driven interactions. In such cases, the application might provide guided instructions or suggestions to help users review and update their goals, ensuring that the process remains user-driven while offering support as needed. 
    
    \item [\textbf{G4.b}] The \textbf{timing of adaptations} should be aligned with individual user interaction patterns to maximize usability and minimize disruption in chronic disease management applications. For example, in a diabetes management application, a user who accesses the application only occasionally may benefit from immediate prompts upon login, such as a quick setup panel to adjust display preferences. In contrast, users who interact with the application more regularly might receive adaptation prompts, such as suggestions for adjusting lifestyle goals or daily activity targets, later in their session when they are more engaged and receptive to change.

\end{itemize}

\linkphd{}{\textbf{G5:} Collaborative Adaptation between Users and Caregivers}

\noindent G5 emphasizes a collaborative adaptation model, in which both end-users and caregivers jointly contribute to customizing and optimizing the mHealth application. This approach addresses the cognitive and logistical challenges users may face when managing adaptations independently, as the mental effort involved in the oversight of interface changes can offset the efficiency gains promised by automation \cite{gajos2006exploring, Harman2014}. In addition, users might unintentionally steer the adaptation process toward personal preferences that diverge from clinical or functional priorities, potentially affecting the intended purpose of the app. Given that caregivers often play an important role in medical decision-making \cite{xie2009older, Cassileth1980}, their involvement ensures that adaptations are practical and aligned with user needs. This collaboration recognizes the caregiver's role in helping with the usage of the application, ensuring that the adaptations meet the needs of end-users. This collaboration fosters a shared sense of responsibility and makes the application more effective in managing chronic diseases, particularly for users who rely heavily on caregiver assistance.

\begin{itemize}[left=16pt, topsep=5pt, itemsep=5pt]
    \item [\textbf{G5.a}] \textbf{Adaptation Lock} enables caregivers to securely access and adjust specific adaptive features within the mHealth application. Through an \textit{access code or caregiver authentication}, the system grants temporary control over interface configurations such as activating high-contrast display modes, simplifying navigation by hiding non-essential features, or reordering dashboard elements to better reflect the priorities of the user under the caregiver’s supervision. Once the caregiver completes these modifications, the system re-locks the settings, preventing accidental or unauthorized changes. This mechanism facilitates collaborative customization while reducing cognitive overload for individuals with limited digital literacy or age-related impairments.
    
    \item [\textbf{G5.b}] \textbf{Role-based customization} enables distinct user roles (e.g., patient, caregiver and healthcare provider) to access distinct interfaces \textit{tailored to their specific tasks and responsibilities}. For example, caregivers might be granted permissions to modify system settings, manage medication schedules, or monitor key health indicators over time, while patients maintain control over personal health data and interact with an interface focused on daily self-management tasks, such as tracking physical activity or dietary logging. This design ensures usability and security by aligning the interface with the contextual needs of each user role.
    
    \item [\textbf{G5.c}] A clearly maintained \textbf{audit trail} can track all adaptations and changes made by caregivers and users, enhancing transparency and accountability. This is especially important in multi-user settings where \textit{conflicting preferences} may arise, such as disagreements over which features should be prioritized or modified. Without oversight, such conflicts can result in miscommunication, data misinterpretation, or inappropriate use of the application \cite{aljedaani2021challenges, lewis2014mhealth}. Figure \ref{fig:app3} shows such an example, where the application supports communication channels between patients and caregivers, and notifications document patient-related activities. To ensure traceability, caregiver actions could also be logged in a similar way, allowing both parties to reference adaptation histories.

\end{itemize}
\linkphd{}{\textbf{G6:} User Involvement in Adaptations-Assessing User Capability}

\noindent G6 highlights the need to assess whether users possess the physical and mental capacity to handle the added responsibilities introduced by adaptive features. This guideline protects against overwhelming users with cognitive or interaction demands that may exceed their abilities. By tailoring the adaptation process to the user's capabilities, the application can accommodate a diverse range of users, from those who are highly tech-savvy and comfortable with extensive customization to those who require a simpler, more guided experience to effectively interact with the system. 

\begin{itemize}[left=16pt, topsep=5pt, itemsep=5pt]
    \item [\textbf{G6.a}] Implementing an adaptive \textbf{onboarding process to assess user capability} in the initial stage of the application interaction can help to tailor adaptations. For example, a brief questionnaire or interactive tutorial can assess a user's digital literacy, confidence in health management, and comfort with interface customization. Based on the responses, the application could recommend a suitable level of adaptation while enabling more granular control for users with higher confidence and technical proficiency. For users with limited physical or cognitive capacity, the application could provide \textit{pre-set adaptation options} instead of requiring manual adjustments. 
    
    \item [\textbf{G6.b}] Based on the evaluation of the user's capabilities in \textbf{G6.a}, they can be offered several predefined options. These predefined options could be: 1) \textit{Vision-friendly AUIs}: This mode improves visual accessibility by increasing font size, increasing contrast between text and background, eliminating distracting background images \cite{nedopil2013knowledge}, and reducing the dependence on peripheral vision. It also optimizes the display settings for low light environments to ensure that text and icons remain visible under various conditions \cite{campbell2015designing}. 2) \textit{Motor-friendly AUIs}: Given the high prevalence of motor impairments among people with chronic diseases \cite{di2019chronic,islam2014multimorbidity}, this mode groups related buttons in logical sequences with adequate spacing to prevent accidental input. It simplifies interactions by minimizing the use of gestures, scrolling, and double taps, replacing them with single-touch commands to improve usability. 3) \textit{Cognitive-friendly AUIs:} This mode could aid in simplifying tasks and reduce cognitive load. The application might also offer adaptive feedback, such as highlighting the next action to take and ensuring that all necessary information is displayed clearly without clutter. Additionally, the application could limit the display of parallel information and reduce the number of steps in any process, ensuring that the interface remains intuitive and task-oriented.
    
\end{itemize}

\linkphd{}{\textbf{G7:} User Involvement in Adaptations-Assessing User's Willingness}

\noindent G7 emphasizes that users differ in their willingness to engage with the adaptation process, influenced by factors such as personality, cultural background, and contextual preferences. This guideline advocates for empowering users by offering them the flexibility to actively participate in shaping the adaptation or passively accept predefined system configurations. This approach ensures that both proactive and passive users can interact comfortably with the application. 

\begin{itemize}[left=16pt, topsep=5pt, itemsep=5pt]
    \item [\textbf{G7.a}] The application could introduce several \textbf{involvement modes} during onboarding to accommodate different user preferences for adaptation. An \textit{active mode} would enable users to take full control over adaptive settings, allowing them to explore and personalize the interface based on their needs and preferences. Conversely, a \textit{passive mode} would apply default configurations with minimal user input, while still offering opportunities for basic UI adjustments if desired. This dual-mode approach ensures inclusivity by supporting users who prefer hands-on control and those who opt for a more guided, effortless experience.
    
    \item [\textbf{G7.b}] Integrate short \textbf{personality or cultural assessments} to better tailor the adaptation process to user preferences. For example, users from cultures characterized by high uncertainty avoidance may prefer simplified and clearly structured interfaces that minimize ambiguity and reduce perceived risk. In such cases, the system could default to passive adaptation modes with intuitive icons, consistent navigation, and minimal customization requirements, ensuring a more comfortable and culturally aligned user experience \cite{alsswey2021role}.
    
\end{itemize}

\linkphd{}{\textbf{G8:} Adjusting the Granularity of the Adaptation}

\noindent G8 highlights the importance of managing the degree or scope of interface adaptations, emphasizing a balanced approach that avoids overwhelming users while still allowing meaningful customization. Excessive changes can lead to steep learning curves and poor usability, while overly limited adaptations can restrict user engagement and satisfaction \cite{cristea2003three, zeidler2013evaluation}. To address this, the guideline promotes a tiered system of adaptation granularity, where users can begin with fundamental adjustments to the interface and gradually access more advanced customization options. For example, a health monitoring application could offer three levels of granularity: 1) \textit{Basic tier} focuses on incremental adjustments that improve accessibility and usability without significantly altering the interface. Users can make essential changes, such as adjusting font size, enabling high contrast mode, or changing button spacing; 2) \textit{Intermediate tier} allow users modify the dashboard layout, reorder widgets (e.g., glucose tracker or exercise log), or switch between simplified and detailed data views; and 3) \textit{Advanced tier} empowers users to implement extensive, system-wide changes, granting full control over the interface's behavior and functionality. 

\linkphd{}{\textbf{G9:} Ensuring Privacy-Conscious Adaptive Mechanisms}

\noindent G9 emphasizes the implementation of robust privacy mechanisms designed to protect sensitive health information during the adaptation process. The goal is to maintain a careful balance between delivering personalized user experiences and addressing valid privacy concerns. Previous studies have highlighted the importance of maintaining user privacy in adaptive systems \cite{findlater2007evaluating}, particularly in the context of mHealth technologies where transparency about data use and system behavior is essential \cite{glass2008toward, martin2019exploring}.  This guideline advocates for privacy-sensitive adaptation strategies that clearly communicate how user data is collected, processed, and applied.

    \begin{itemize}[left=16pt, topsep=5pt, itemsep=5pt]
        \item [\textbf{G9.a}] Clearly \textbf{communicate the rationale behind adaptive changes or interface customizations} to improve transparency. For example, the system could inform users that the dashboard has been reorganized to highlight frequently used features based on their recent interaction patterns, while explicitly assuring users that their personal data remains secure and private. However, research indicates that users tend to lose interest in such explanations when they are not given sufficient control over adaptations \cite{bunt2012explanations}. Providing both rationality and user-controlled adaptation fosters greater engagement and trust.
        
        \item [\textbf{G9.b}] Implement adaptive systems that operate on \textbf{minimal data input}, collecting only \textit{essential information} for specific adaptive features. For example, the application could adjust the placement and prominence of frequently used UI elements, based solely on the user’s navigation patterns, without collecting unnecessary data such as search history or inactive screen interactions.

        \item [\textbf{G9.c}] For applications involving multiple users, such as caregiver-patient scenarios, enable end-users to retain control over caregiver access through \textbf{easy-to-configure privacy settings}. The application should support fine-grained permissions that allow users to specify what information caregivers can view or modify. For example, a caregiver can manage medication schedules, but is restricted from accessing sensitive data such as personal notes or detailed health trends unless explicitly authorized.
    \end{itemize}

\subsection{Evaluation of Guidelines Against Real mHealth Applications} \label{sec:case}
\begin{figure}[h]
    \centering
    \includegraphics[width=\linewidth]{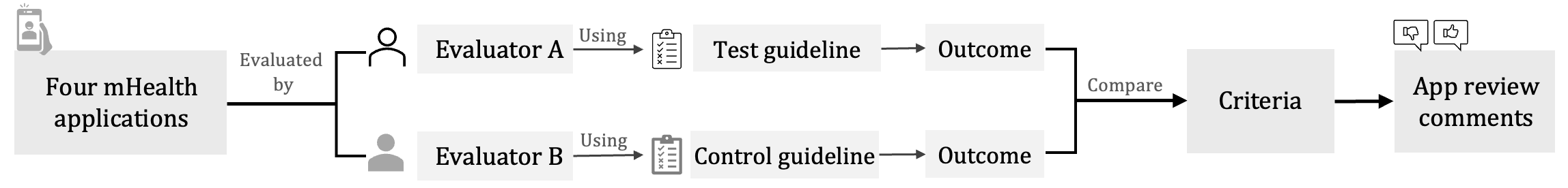}
    \caption{Evaluation process of AUI guidelines through case study analysis}
    \label{fig:comparing}
\end{figure}
The evaluation was conducted in accordance with validation strategy proposed by \citet{quinones2018methodology}, with the detailed evaluation process illustrated in Figure \ref{fig:comparing} and the application details summarized in Table \ref{tab:diabetes_apps}. The selected applications are evaluated by human experts against the \textit{tested guideline} and another set of \textit{control guideline}, with the latter serving as the basis for comparing the results obtained during the evaluation process. The \textit{Xcertia Usability Guidelines} \cite{Xcertia_2020} are adopted as the control guideline in this study, as they offer a comprehensive and widely recognized framework addressing key usability aspects specific to mHealth applications. The control guideline provides a solid standard for comparison, which includes areas including: \textit{ 1) Visual design, 2) Readability, 3) Application navigation, 4) Onboarding, 5) Application feedback, 6) Notifications, alerts, and alarms, 7) Help resources and troubleshooting, 8) Historical data, 9) Accessibility, and 10) Continuous application evaluation.} The chosen applications are evaluated by two evaluators who possess comparable experience in UI design, with both evaluators reviewing the same applications. \textbf{Evaluator A} relies exclusively on the set of test guidelines, while \textbf{Evaluator B} focuses solely on the control guideline, and subsequently, the issues of each application identified by two evaluators are compared \cite{langevin2021heuristic}. In the following sections of the article, we will refer to our \textbf{\textit{test guidelines as T1 through T9}}, corresponding to the \textbf{\textit{Version Three guidelines G1 through G9}}. To evaluate the effectiveness of the test guidelines, the issues identified using the test guidelines were compared with those identified using the control guideline, based on two criteria adapted from prior studies \cite{quinones2018methodology, chen2024case, miniukovich2019guideline}: 1) the number of incorrectly assigned problems to the guideline, and 2) the number of identified problems deemed to be of higher severity.

\begin{table}[b]
    \centering
    \caption{Number of issues and average severity rating found by the experts for both test guideline and control guideline}
        \resizebox{0.98\textwidth}{!}{%
    \begin{tabular}{lllllllll}
    \hline
    \textbf{App ID} &\textbf{App Name} & \textbf{Rate/Downloads} &\textbf{Review}&\textbf{Flaged review}&\textbf{Test(Num)\textsuperscript{5}} &\textbf{Control(Num)\textsuperscript{5}} &\textbf{Test(Sev)\textsuperscript{6}} &\textbf{Control(Sev)\textsuperscript{6}} \\ \hline
    \textbf{App 1} &mySugr - Diabetes Tracker Log\textsuperscript{1} & 4.6/5M+& 3k & 208& 7& 12 & 3.1& 1.4\\ 
    \textbf{App 2} &Gluroo: Diabetes Log Tracker \textsuperscript{2} &4.3/50k+& 0.3k& 44& 16& 31 & 3.8& 2.1\\
    % DiabTrend - Diabetes Diary App&4.3/100k+& 1.69k\\
   \textbf{App 3} & Health2Sync - Diabetes Tracker\textsuperscript{3} &4.6/1M+& 0.4k& 59& 6& 3 & 2.9& 1.3\\
   \textbf{App 4} & LibreLinkUp\textsuperscript{4} &4.6/1M+& 0.4k& 32& 4& 3 & 2.8& 1.2\\\hline
     & && 4.3k\footnotesize(Sum)& 343\footnotesize(Sum)&33\footnotesize(Sum) & 49\footnotesize(Sum) &3.2\footnotesize(Ave) &1.5\footnotesize(Ave)  \\
   \hline
 \multicolumn{9}{p{220mm}}{\textit{1.https://www.mysugr.com/en/diabetes-app. 2.https://gluroo.com/. 3. https://www.health2sync.com/. 4.https://www.librelinkup.com/. 5. number of the issues identified by the specific guideline. 6. Average severity rating of issues identified by the specific guideline.}}\\
\hline
    \end{tabular}}

    \label{tab:diabetes_apps}
\end{table}

\begin{figure}[t]
    \centering
    \includegraphics[width=0.9\linewidth]{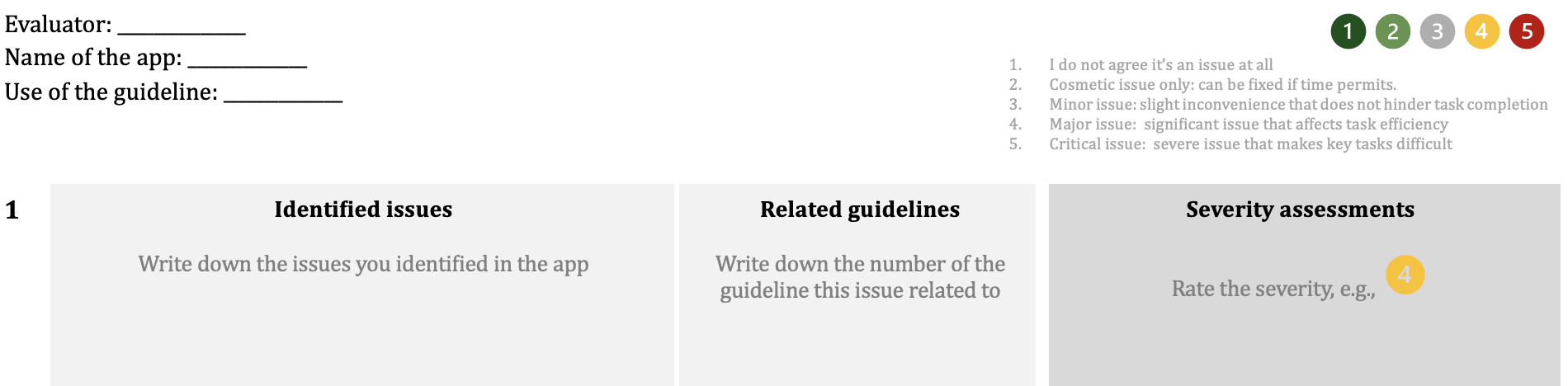}
    \caption{Evaluation note template used by the expert in evaluation session }
    \label{fig:evaluationnote}
\end{figure}

\subsubsection{Application evaluation}
To ensure consistency and clarity in the evaluation process, teams must undergo thorough training and preparation, as emphasized by \citet{nielsen1994usability}. A briefing was conducted the day before the evaluation to review the two sets of guidelines. Each evaluator received two key documents: an \textit{evaluation note} (see Figure \ref{fig:evaluationnote}), and a copy of the \textit{detailed guidelines} to be applied during the evaluation session. The evaluation note includes identified issues, related guidelines, and severity assessments where evaluators assign a severity level of 1-5 (5 being the highest) to reflect the extent to which the issue affects the user's ability to use the application \cite{chen2024case, langevin2021heuristic}. Subsequently, evaluators were requested to outline the issues identified during the evaluation process, taking into account both their severity and frequency of occurrence. As shown in Table \ref{tab:diabetes_apps}, the test guideline identified 33 issues compared to those using control guideline 49 in four applications. While the control guideline uncovered a greater number of issues overall, this is partly because certain test guidelines were \textbf{\textit{not applicable}} in apps lacking adaptation features. For example, App 2 was intentionally selected for its extensive use of adaptation features and is currently open for user feedback and refinement. As a result, it had more issues flagged under the control guideline. However, its misalignment with several test guidelines led to numerous usability issues. This highlights the need to establish a comprehensive guide for adaptations, rather than simply maximizing adaptations, which would inevitably present greater usability challenges to users. The test guideline identified fewer issues in general compared to the control guideline, and the issues it did identify were generally of higher severity. In contrast, the control guideline flagged a larger number of problems, but many of them were rated as low in severity, which explains the difference in perceived impact between the two sets. These \textit{lower-severity ratings} were often associated with \textit{visual design} issues identified by the control guideline, which were considered lower priority. Although acknowledging a wide range of issues is beneficial, prioritizing high-severity problems is essential as they can significantly hinder usability, a critical concern in the mHealth domain \cite{athilingam2018mobile,morey2017managing}.

\subsubsection{User review analysis}
After the human expert evaluation, review comments from the selected mHealth applications were analyzed to determine whether the issues identified during the evaluation process were echoed by end-users (see Figure \ref{fig:comparing}). This process involves categorizing user feedback to identify recurring issues or patterns that align with the issues flagged by the evaluators. By integrating these insights, this step helps bridge the gap between expert evaluations and real-world user experience, ensuring that guidelines address both theoretical challenges and practical usage. Following the process described in Section \ref{sec:reviewanalysis}, the review analysis initially identified 343 user reviews (see Table \ref{tab:diabetes_apps}). After another round of filtering, 131 relevant reviews were retained for analysis across the four selected applications. The evaluation highlighted that the test guidelines successfully identified concerns in key areas that align with the guidelines outlined in Section \ref{sec:guideline}.

\begin{figure}[t]
    \centering
    \includegraphics[width=0.9\linewidth]{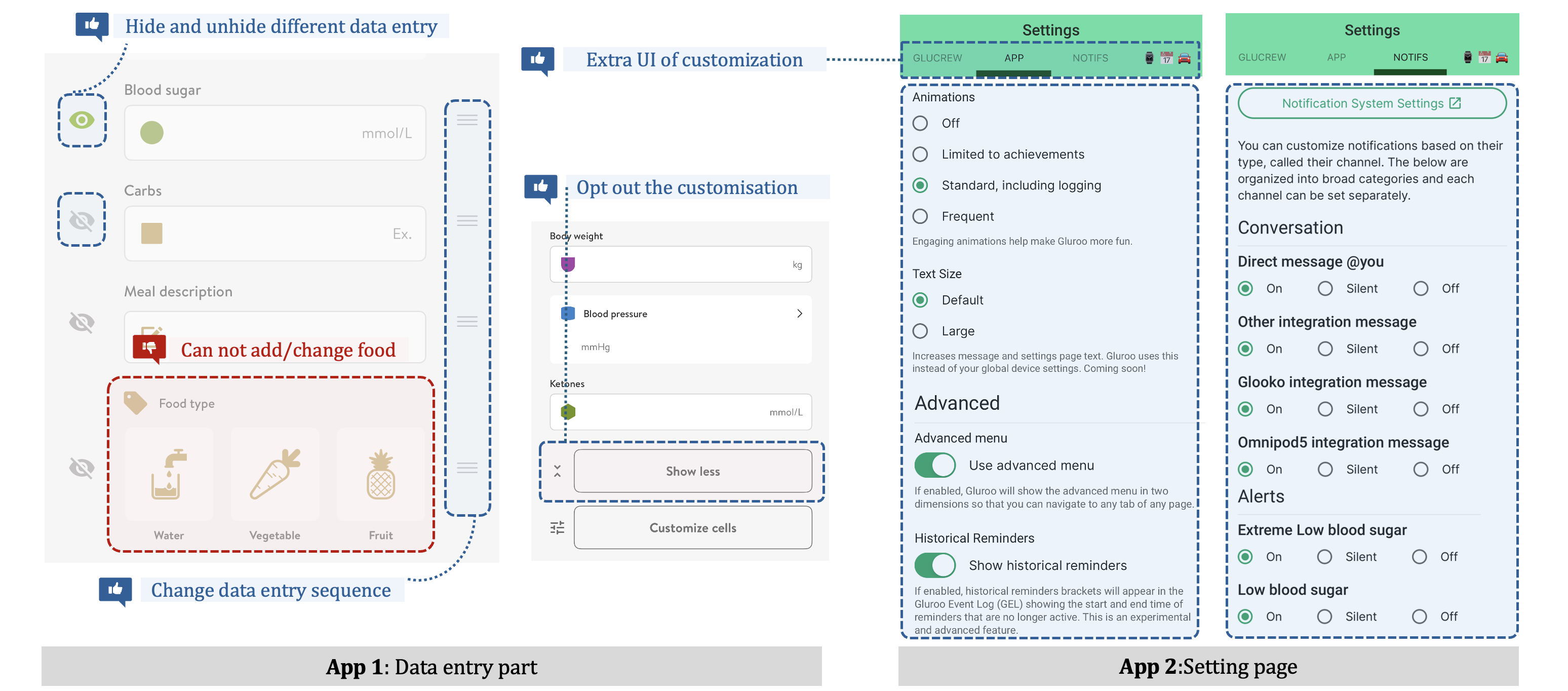}
    \caption{Adaptation design example for empowerment and autonomy}
    \label{fig:app1}
\end{figure}

\textbf{Empowerment and Autonomy.} For this guideline set, it was identified as relevant during expert evaluations of App 1 and App 2, and was reflected in 59 of 131 user review comments. \textbf{App 1} was flagged by evaluators for offering limited controllability over the dashboard and the generation of patient reports for physicians. However, it was positively recognized for its customizable data entry feature (see Figure \ref{fig:app1}). The application allows users to hide, and reorganize data entry fields, with the option to restore the original list via a \textit{“show all fields”} button. Many users described the application as \textit{“customizable”} and praised its \textit{“flexible setup to record details”}. Nonetheless, some users expressed the desire for more food tracking options and the ability to create custom food categories, indicating room for further adaptability. Conversely, \textbf{App 2} offers extensive controllability, allowing users to personalize the UI elements, notifications, navigation methods, and information displays (see Figure \ref{fig:app1}). Despite this flexibility, user reviews revealed mixed feelings. Although many users appreciated the high degree of customization, rated the application highly and described it as \textit{“an excellent resource for individuals managing diabetes with features tailored to their needs”}, others criticized the design for being unintuitive and cluttered, with remarks such as \textit{“the interface is overwhelming and difficult to navigate”.} This divergence in feedback likely reflects the application’s distinctive emphasis on real-time caregiver monitoring, which serves two primary user groups: caregivers and patients. App 2 does not distinguish between caregivers and patients, despite these groups having differing capabilities and preferences that affect their ability to navigate and utilize the app's customizable features. The variation in responses supports the relevance of the test guidelines, \textbf{T5}, \textbf{T6} and \textbf{T7}, which emphasize collaborative adaptation between users and caregivers, as well as the assessment of users’ willingness and capabilities to engage in personalizing the application. The conflicting opinions about the flexibility offered by App 2 can also be linked to guideline \textbf{T8} (\textit{granularity of adaptation}). As shown in Figure \ref{fig:app1}, App 1 provides relatively simple adaptations, such as modifying the number of visible data entry fields, along with a clear opt-out option that helps users anticipate their next steps. In contrast, App 2 implements more complex adaptations, including advanced menu modifications that alter navigation flows and require users to spend time learning the new structure. Notably, App 2 exhibited numerous usability issues identified through the control guideline, and many users still praised its adaptability, highlighting its potential advantages. However, these benefits often come with usability challenges, reinforcing the need to carefully balance the advantages and drawbacks of adaptation. Employing a structured approach such as the one illustrated in Figure \ref{fig:guideflow} can assist developers in identifying appropriate usage contexts and reducing such trade-offs. Without this balance, adaptive features risk introducing further complications rather than improving the user experience \cite{paymans2004usability, jameson2007adaptive, mezhoudi2015toward, deuschel2018influence, hook2000steps}.

\begin{figure}[t]
    \centering
    \includegraphics[width=1\linewidth]{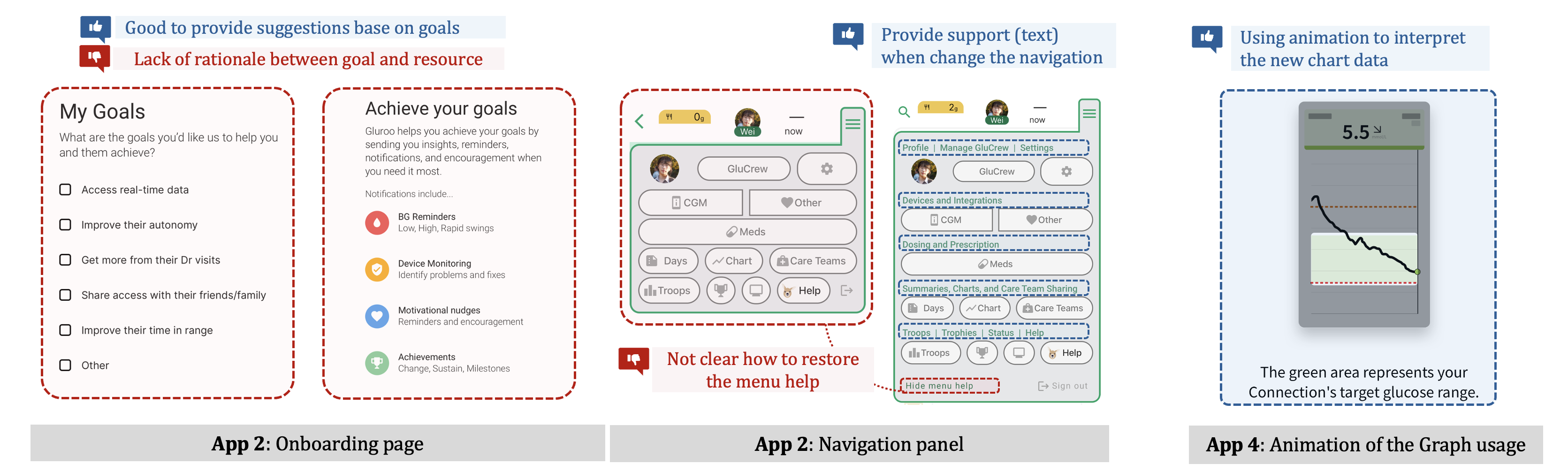}
    \caption{Adaptation design example for user support and interaction}
    \label{fig:app2}
\end{figure}

\textbf{User Support and Interaction.} For this guideline set, it was identified as relevant during expert evaluations of App 2 and App 4, and was reflected in 26 out of 131 user review comments. The App 2 onboarding interface effectively presents goal-oriented prompts such as \textit{“access real-time data”} and \textit{“enhance autonomy” }(see Figure \ref{fig:app2}), aligning with \textbf{T2} (\textit{user support}) by offering users motivation and guidance to personalize their experience. Although the application provides partial assistance following navigation changes, additional support, particularly in restoring or explaining menu options, would improve usability. Although user reviews do not explicitly call for more system support, they do reflect difficulties in locating features and navigating the interface. In contrast, App 4 features a simple tutorial for using the chart function, which users described as \textit{“easy to follow”} (see Figure \ref{fig:app2}).  User feedback further highlights the importance of addressing varying user capabilities, with accessibility concerns taking precedence over requests for adaptive support. For example, older users requested larger fonts and clearer interface elements to accommodate visual limitations, and the desire for customizable alarm settings points to the need for sensory-specific adaptations, consistent with \textbf{T6} (\textit{user capability}).

\begin{figure}[t]
    \centering
    \includegraphics[width=1\linewidth]{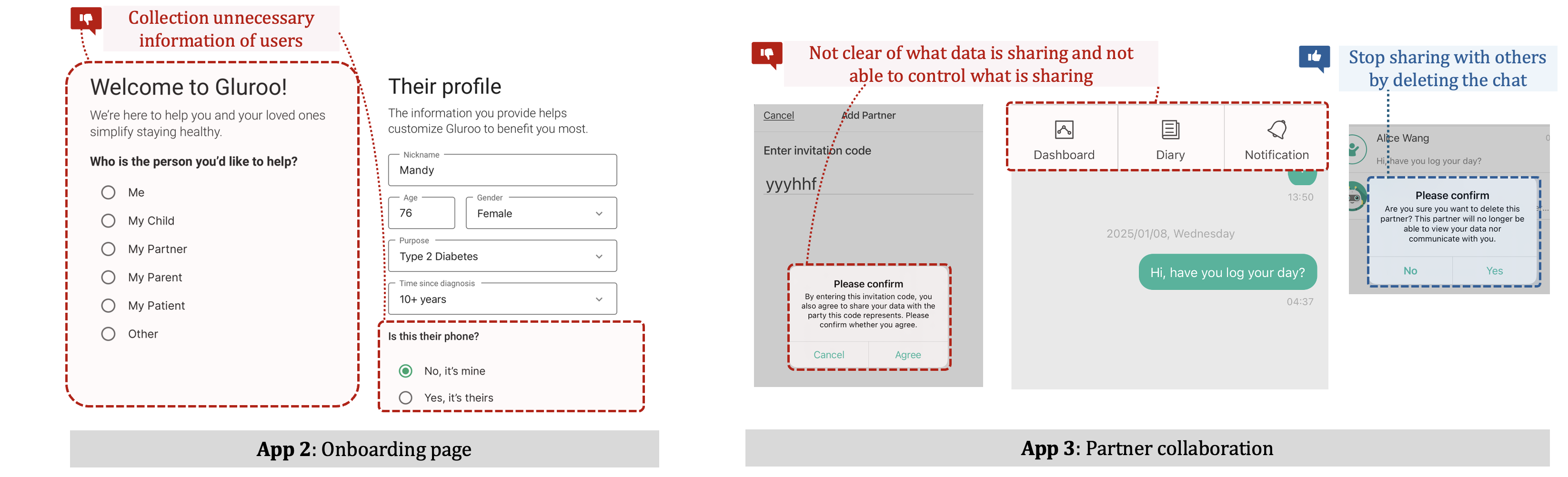}
    \caption{Adaptation design example for caregiver collaboration and adaptation}
    \label{fig:app3}
\end{figure}

 \textbf{Caregiver Collaboration and Adaptation.} For this guideline set, it was identified as relevant during expert evaluations of App 2, App 3 and App 4, and was reflected in 46 out of 131 user review comments. \textbf{App 2} prompts users during the initial login to specify their role (e.g., caregiver or patient) and whether the device is intended for personal use or to monitor another individual (see Figure \ref{fig:app3}). While this information is collected to enable role-based customization, the interface does not visibly adapt based on these distinctions, raising questions about the utility of the data and potential privacy risks. One user review underscores this concern, stating, \textit{“at least give us a privacy policy that we can read before giving personal data up”.} This highlights the importance of transparent data practices and the necessity for clearly differentiated features that reflect user roles, thus fostering trust and improving usability. \textbf{App 3 }similarly lacks differentiated designs tailored to various user roles. It does not require caregivers to provide additional personal information, instead relying on an invitation code provided by the primary user. The application informs users of the sharing of data with partners, but does not specify what data is shared or allow users to control access. However, users retain the ability to remove partners if needed (see Figure \ref{fig:app3}). User reviews reflect concerns about this lack of control, especially about the inability to manage what data caregivers can view. Users also expressed concerns about intrusive notifications, with one stating: 
 
 \boxquoter{Although notifications of the glucose recording are helpful, they are too intrusive, as my partner doesn’t need to know every single detail, especially in work environments.}
 
 These concerns highlight the importance of both \textbf{T9} (\textit{ensuring privacy-conscious mechanisms}) and \textbf{T1} (\textit{promoting user control}), underscoring the necessity of offering more granular controls over data sharing and notification settings to maintain a balance between functionality and privacy. A recurring issue across all three applications is the lack of role-based customization (\textbf{T5}), leading to confusion between personal data and that of a caregiver or partner. For instance, some commented: 

\boxquoter{It’s confusing when partners share accounts—sometimes I don’t know if the logs are mine or others. The application also assumes I have diabetes and asks me a lot of questions as if I’m the one being monitored.}

\noindent These remarks emphasize the need for clear differentiation and data separation of roles so that caregivers can effectively support patients without assuming their identity or navigating irrelevant features. The user reviews corroborated the issues identified by the test guidelines, reinforcing their practical relevance in real-world settings. In addition, several reviews revealed the consequences of overlooking certain guidelines, thereby emphasizing their interrelated nature. This finding further supports the value of our prior guidance (see Figure \ref{fig:guideflow}) in helping software practitioners determine when and how to apply specific guidelines across varying design contexts.

\begin{Summary}{Summary of Stage Two}{guideline}
The guideline development process followed a structured, multi-phase approach, beginning with a comprehensive review of existing literature, followed by evaluation involving both end-users and software practitioners. Feedback from 20 end-users and 43 software practitioners informed several rounds of refinements, resulting in the removal, addition, or modification of specific guidelines, adjustments to the specification format, and the inclusion of contextual usage recommendations. The finalized output consists of nine guidelines, organized into four groups, with accompanying guidance on how each set can be prioritized in different design contexts. These guidelines were then applied to a case study involving real-world mHealth applications, demonstrating their practical relevance and effectiveness in guiding adaptation design and addressing user-specific challenges.
\end{Summary}

\section{Threats to Validity}\label{sec:lim}

\subsection{External Validity} 
\textbf{Stage One.} For participant recruitment in the interview and focus group study, the Socio-Technical Grounded Theory (STGT) theoretical sampling technique was employed. This approach is iterative and supports interleaved data collection and analysis \cite{hargittai2019internet, hoda2021socio}. As a result, the qualitative findings of this study are not intended to be generalizable to larger populations. Instead, they provide in-depth insight into how individuals with chronic diseases perceive and interact with Adaptive User Interfaces (AUIs) in mHealth applications. Future research efforts should seek to validate our findings with more diverse populations, particularly those with lower levels of digital literacy \cite{hargittai2019internet}. Second, the use of a prototype in stage one may have inadvertently introduced \textit{discrepancies between the prototyped adaptations and the practical applications}. This prototype may not fully capture the intricacies of real-world interactions, such as the dynamic interaction between adaptations and other system functionalities or fail to account for users' nuanced allocation of time to specific adaptations \cite{bunt2004role,gajos2006exploring,gajos2008predictability}. These complexities are crucial to consider, as they could influence users' perceptions and behaviors in ways that the prototype might not accurately reflect. Third, we observed \textit{demographic differences in our user survey in stage one study with respect to nationality, age, and the clinic population}. Although we shared the survey on social media to obtain worldwide participation, we were unable to achieve it and found that the majority of the participants (49\%) were from Australia (Table \ref{tab:surveydemographics}). Hence, similar to the observations in \cite{robbins2017health}, the findings of this study may have limited generalizability beyond the specific group of participants. Our user survey sample exhibited a skew towards younger demographics. This demographic disparity is significant as usage and preference patterns are likely to vary between different age groups, and older populations can potentially exhibit different behaviors and preferences. Furthermore, it is important to acknowledge that our study surveyed a general population rather than individuals in hospital or clinic settings. These individuals, who regularly engage with healthcare services, may exhibit different behaviors and usage patterns of mHealth applications compared to those of the general population with milder symptoms.

\textbf{Stage Two.} The guideline evaluation study raises concerns about generalizability due to diverse practitioner backgrounds, organizational practices, and target user groups. There are several limitations identified in this case study. First, the limited number of evaluators, only \textit{ two evaluators}, constrained the study, despite recommendations to use at least \textit{four evaluators} for identifying most issues \cite{nielsen1995conduct}. Second, while both evaluators had experience in UX/UI design and front-end development, they may not represent a broader sample of nonexpert users, limiting the generalizability of the evaluation. Third, evaluating only four \textit{diabetes-related} mHealth applications restricts the extent of insights gained, as a comprehensive evaluation encompassing a wider spectrum of available applications could more effectively evaluate the efficacy of the guidelines. In addition, the diverse nature of chronic diseases and user demographics complicates the creation of universal design guidelines. The guidelines aim to be flexible and cover common needs across chronic diseases, but might lack specificity when applied to particular diseases or specific user groups. We recommend that future studies address these limitations by including a larger and more diverse group of evaluators and testing the guidelines in a broader range of usage contexts and devices to improve their applicability and robustness. Although it recognizes the importance of understanding the underlying \textit{purpose of guidelines} for individuals with chronic diseases, a detailed investigation of this topic lies beyond the scope of this work. Future research should explore how different user-related factors, especially in individuals with chronic diseases, impact the design of adaptations.

\subsection{Internal Validity} 
\textbf{Stage One.} Within the AUI prototype, some participants experienced confusion when interacting with specific adaptive elements. To address this, an \textit{explanatory video} was included to guide users and clarify the purpose and functionality of the different adaptations. Furthermore, we supplemented this with \textit{detailed instructions for reference}, ensuring that participants had readily available guidance should they encounter any challenges while using the system. Similarly, we anticipated potential complexities during the survey phase and thus provided \textit{two concrete examples of AUIs} to clarify AUI concepts for participants. We used the STGT approach for qualitative data analysis and facilitated extensive team discussions to review and refine our analyses, findings, and presentation, thus mitigating potential biases. Although offering compensation to participants for their participation in interviews and focus group studies can raise concerns about the potential of participants to provide false information to qualify for the study \cite{head2009ethics}, it should be noted that four participants declined compensation. Instead, they expressed a preference for the funds to be allocated towards further research endeavors. This underscores the voluntary nature of our study. The survey responses regarding opinions on adaptations for mHealth applications may change based on their evolving needs and experiences as users interact with mHealth applications over time, and they may also stop using them, as long-term use of the mHealth application has always been an issue \cite{ woldaregay2018motivational}.

\textbf{Stage Two.} Software practitioners with different levels of experience in accessibility or adaptive design can provide inconsistent feedback, and personal preferences or biases toward specific design approaches could influence their evaluations. Furthermore, the qualitative nature of the study introduces subjectivity, as interpretations of the guidelines may vary among software practitioners. To mitigate this limitation, the guideline evaluation survey included two \textit{comprehension check} questions designed to assess participants’ understanding of the guidelines. Practitioners who answered incorrectly were given the opportunity to review the question and the guidelines before proceeding, helping to ensure more informed and consistent responses.

\subsection{Conclusion Validity}  
We acknowledge the limitation of using a single prototype to address a wide range of chronic diseases, as different conditions often entail varied user needs, interaction requirements, and adaptation preferences. This choice may constrain the generalizability of the findings. However, our approach is consistent with the common practice of including individuals with multiple chronic diseases in research \cite{elzen2007evaluation}. In fact, some participants in our study presented multiple chronic diseases, mirroring real-life scenarios in which people often contend with comorbidity alongside their primary disease \cite{di2019chronic,islam2014multimorbidity}. Involving the same participants in multiple stages of the study, interviews, focus groups, and end-user guideline evaluations can introduce bias due to their previous exposure to the study context. This repeated participation could influence responses and evaluations, potentially reducing confidence in the generalization and robustness of the results.

\section{Conclusion}\label{sec:con}
This study addressed the pressing need for more user-centered mHealth applications to support individuals managing chronic diseases. Conducted in two stages, the research began with a user study, comprising interviews, focus groups, and surveys, to examine how users perceive and interact with Adaptive User Interfaces (AUIs) in mHealth contexts. Insights from this stage informed the development of a new set of guidelines, which were further refined through feedback from end-users and software practitioners. These guidelines were subsequently validated through case studies involving four real-world mHealth applications, with additional analysis of user reviews to determine alignment between expert-identified issues and end-user experiences.  The results suggest that the proposed guidelines are more effective in uncovering critical usability and adaptation challenges than the existing mHealth usability guidelines. The nine guidelines, organized into four groups, offer software practitioners a structured framework to design adaptive features that accommodate user variability while supporting long-term use. Several avenues remain for further exploration. One potential avenue is the development of automated tools, design templates, or UI component libraries aligned with the guidelines could support their practical integration into development workflows. Another is conducting longitudinal field studies or randomized controlled trials that could evaluate the long-term impact of guideline-based AUI implementations on user engagement, health outcomes, and system usability. 

% This stage identified key challenges users face when interacting with AUIs and proposed a set of recommendations to enhance adaptation design. These recommendations, derived from qualitative insights, revealed complex trade-offs influenced by factors such as user involvement, prior experience with mHealth applications, and individual health conditions. Furthermore, the quantitative survey enriched the analysis by uncovering emerging adaptation categories and capturing participant preferences regarding adaptation types, data sources, collection methods, and desired levels of engagement. 
\begin{acks}
%To Robert, for the bagels and explaining CMYK and color spaces.
Wang, Grundy and Madugalla are supported by the ARC Laureate Fellowship FL190100035. We would like to acknowledge Daniel Gaspar Figueiredo, Elton Lobo, Michael Wheeler, and Paul Jansons for helping us recruit participants for the user study. We extend our sincere thanks to all the end-user participants for sharing their valuable experiences and perspectives, and to the software practitioners who contributed insightful feedback that helped strengthen the proposed guidelines. Special thanks to Rashina Hoda for her guidance in applying Socio-Technical Grounded Theory (STGT) for data analysis and to Md Shamsujjoha for helping refine the presentation of the guideline section in this paper. The following colleagues provided valuable assistance in the form of comments on earlier drafts, data analysis, prototype evaluation, and/or graphics in the study: Dulaji Hidellaarachchi, Elizabeth Manias, Kashumi Madampe, Suyu Ma and Weimin Wang. 

\end{acks}

%%
%% The next two lines define the bibliography style to be used, and
%% the bibliography file.
\bibliographystyle{ACM-Reference-Format}
\bibliography{bibliography}

\appendix
\section{Stage one: User survey questions}
\label{app:A}
\begin{footnotesize}
\textbf{Introduction}: AUIs are software applications where some aspect(s) of the interface is modified to cater for different user needs and/or preferences, e.g. font or button size, color, layout, complexity, interaction style, and so on. We are researching how AUIs can be leveraged to better cater for users with chronic diseases. As part of this survey, you will be asked to answer questions about your perspective toward AUI. On average it will take 10-15 minutes for a participant to complete the survey. You can enter a draw for a AU\$20 voucher if you complete an online survey. Your email address will not be associated with your response since the contact information is collected in another survey. At the end of the survey, you will also be asked to indicate whether you are interested in participating in a focus group study. You may register through this link if you are interested in the focus group study. If you have any questions, please email wei.wang5@monash.edu. Ethical approval has been provided by Monash University. Further details can be provided upon request.
\begin{itemize} [left=0pt]
    \item \textbf{Section 1:Demographic questions.} This section will collect your demographic information. We will not share any identifying information that you submit. The demographic information collected is used only to assess the representatives of the survey participants.
    \begin{itemize}
        \item In what age group are you?
        \item To which gender identity do you most identify as?
        \item In which country do you currently reside?
        \item What is your highest educational qualification?
    \end{itemize}
    \item \textbf{Section 2: Your health status.} This section will collect your health information. 
    \begin{itemize}
        \item Has a health care provider ever told you have a chronic disease?
        \begin{itemize}
            \item I have chronic disease 
            \item Other\_
        \end{itemize}
        \item What chronic disease do you have?\_
    \end{itemize}
    \item \textbf{Section 3: mHealth applications.} This section will collect information for your usage for mHealth applications. 
    \begin{itemize}
        \item Have you used mHealth applications before to manage your chronic disease? \textit{(The most common application of mHealth is the use of mobile devices to educate consumers about preventive healthcare services. However, mHealth is also used for disease surveillance, treatment support, epidemic outbreak tracking and chronic disease management.)}
        \begin{itemize}
            \item	Yes, I have used/currently use the mHealth application 
            \item	No, I never used before 
            \item	Other \_
        \end{itemize}
        \item What kind of mHealth application(s) have you used before?
        \item Why do you mainly use these application(s) for?
        \item How frequently do you use health applications?
        \item How long each time you use the app?
    \end{itemize}
    \item \textbf{Section 4:Adaptive user interface.} This section will collect your perspective on the AUI. An AUI is a user interface which adapts its layout and elements to the needs of the user or context. Here are two examples of AUIs: \label{survey:adaptive}
    \begin{figure}[h]
        \centering
        \begin{subfigure}[b]{0.5\textwidth}
            \centering
            \includegraphics[width=\textwidth]{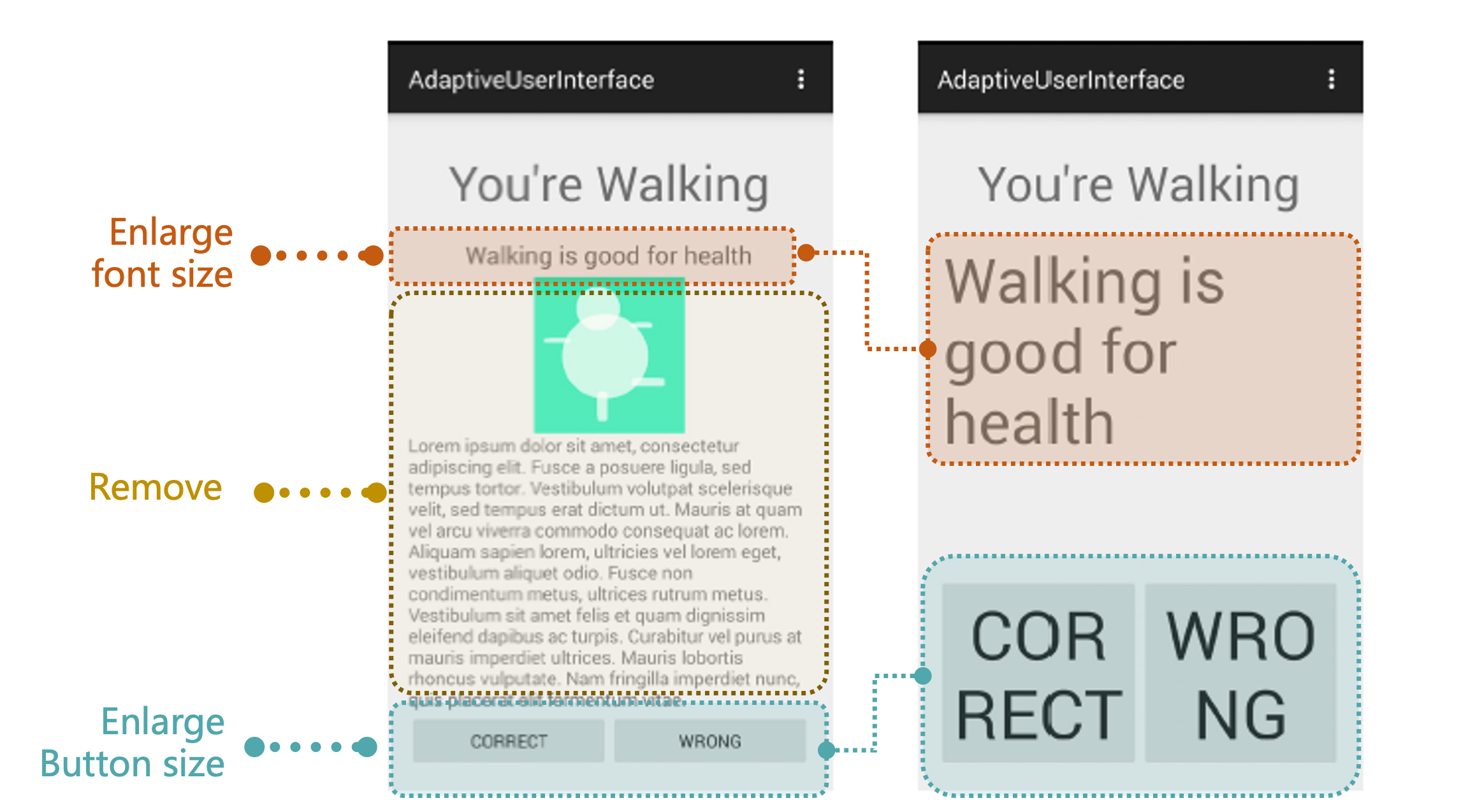}
            \caption{Exercise application with AUI}
            \label{fig:auia}
        \end{subfigure}
        \hfill
        \begin{subfigure}[b]{0.43\textwidth}
            \centering
            \includegraphics[width=\textwidth]{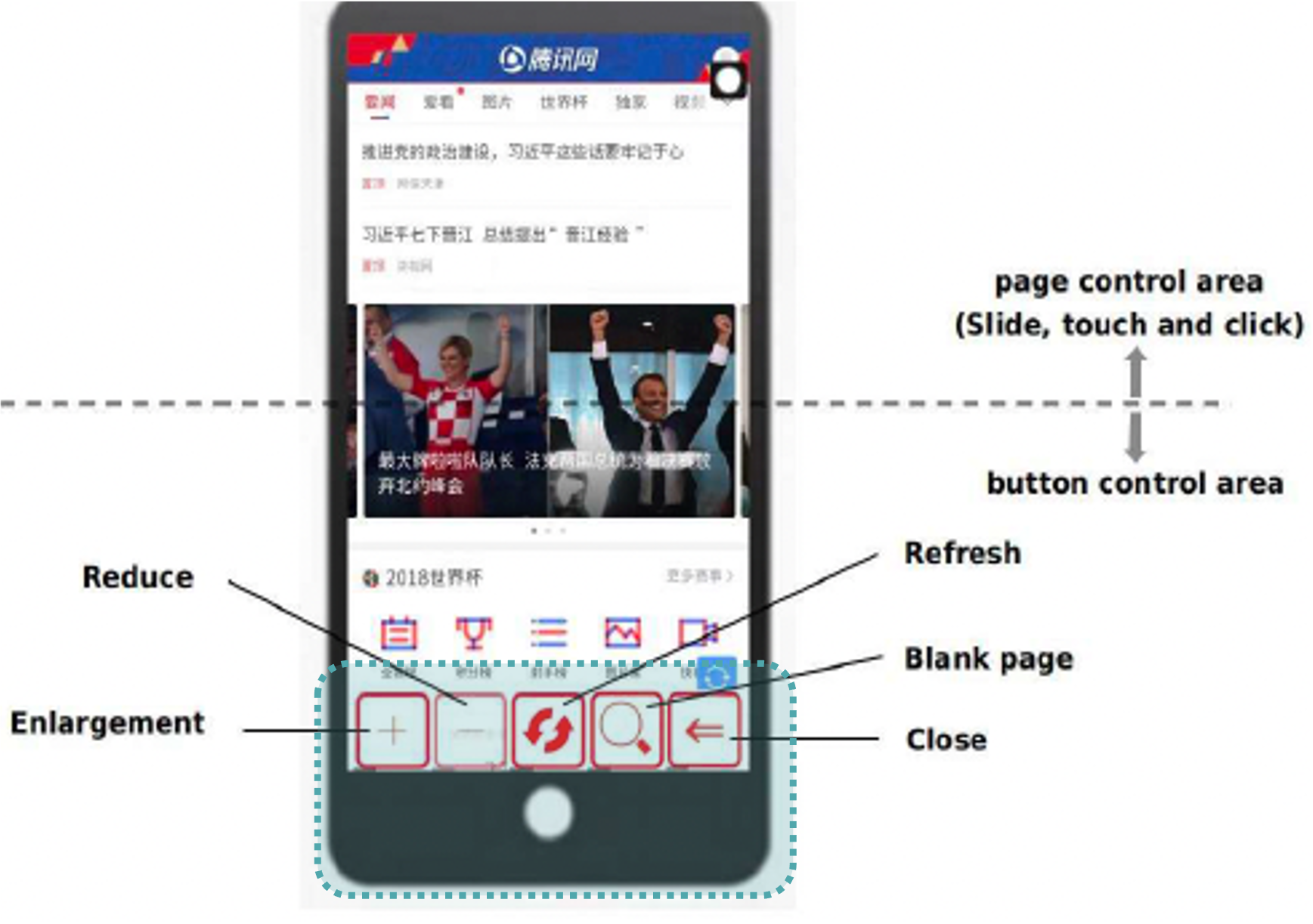}
            \caption{PD-Helper for Parkinson's Disease Patients}
            \label{fig:auib}
        \end{subfigure}
        \caption{Two examples of applications with Adaptive User Interfaces (AUI).}
        \label{fig:aui_examples}
    \end{figure}
    \\\textbf{Exercise application} (Figure \ref{fig:auia}): The left side is a standard version of the interface, to maintain usability and readability of the interface when the user is running, the interface automatically hide images and changes small-font and enlarge the subtitle and two buttons \cite{park2015model}.  
    \\
    \\\textbf{PD-Helper for Parkinson's Disease Patients} (Figure \ref{fig:auib}): Users are allowed to change the font size and other settings in accordance with their own preferences and capabilities \cite{jabeen2019improving}.
    \begin{itemize}
        \item Which kind of adaptation do you prefer for interface adaptation? \textit{(mHealth applications can be adapted based on your needs in different ways. Select all that apply)}
        \begin{itemize}
            \item	Graphic design (layout, font size, colours and themes) 
            \item	Navigation adaptation (only specified functions can be used to provide help in special cases) 
            \item	Different persuasive strategy (to better motivate the desired behaviour change, according to different user types) 
            \item	Content complexity (making content easy to understand and process) 
            \item	Information architecture (giving the user more freedom when navigating through large amounts of textual information) 
            \item	Multimodal interaction (voice/text input, switching of voice/text output in different environments) 
            \item	Interface elements rearrangement (removing, adding or reordering elements on the page) 
            \item	Difficulty level (change difficulty level of game/exercise based on user’s motivation or performance) 
            \item	Add on functions (to better assist the user in using the application, e.g. zoom function for older users) 
            \item	Sound effect (adjust the volume depending on the distance of the person from the device) 
            \item	Other (Please specify) 
        \end{itemize}
        \item Which kind of data do you prefer to use for your interface adaptation? \textit{(Imagine your mHealth application can be adapted based on your characteristics and preference. Please see different types of data that can be used in adapting mHealth applications. Select all that apply)}
        \begin{itemize}
            \item	Your role (you will perform different jobs and be in different situations 
            \item	Your feedback (your likes, dislikes and preferences for the interface) 
            \item	Your motivation (what motivates you to use the application) 
            \item	Your interaction with the interface (e.g. number of clicks, links visited, time spent browsing, etc.) 
            \item	Your game performance (scores, successes and wins in the game) 
            \item	Your physical characteristics (ability to perform different activities in daily life) 
            \item	Your psychological characteristics (thoughts, personality, feelings and other cognitive characteristics) 
            \item	Your demographic information (quantifiable insights of users into the population such as age, gender and education level) 
            \item	Your preferences (preferred layout, input/output, theme and interface design) 
            \item	Your emotions (based on emotions when using the app) 
            \item	Your social activity (the extent to which you interact with others around you) 
            \item	Your physiological characteristics (e.g. stress levels, heart rate, blood pressure and blood oxygen levels) 
            \item	Your goals (the end state you want to achieve by using the app) 
            \item	Other (Please specify) 
        \end{itemize}
        \item How do you wish your data to be collected? (Select all that apply)
        \begin{itemize}
            \item	Analysis of user behaviour through the application (checking the history of phone usage, game performance) 
            \item	Analysis of activities with keyboard 
            \item	Smartphone sensor (phone camera, accelerometer, GPS, microphone) 
            \item	User input through the application (user manually input their interest, gender, preference) 
            \item	External sensor (Kinect sensor, computer/TV camera) 
            \item	Wearable sensor (pedometer, smartwatch/bracelet, medical sensors (blood pressure monitor...) 
            \item	Other (Please specify) 
        \end{itemize}
        \item What level of initiative do you want to take during the adaptation you are preferred? 
        \begin{itemize}
            \item	Manual system (allows you to manually modify certain settings of the user interface) 
            \item	Automatic system (the system adjusts the user interface automatically) 
            \item	Semi-automatic system (you and the system collaborate to achieve adaptation) 
            \item	Other (Please specify) 
        \end{itemize}
    \end{itemize}
    \item How did you hear about us?  
    \item Is there anything else you want to tell us about this survey or our research study?\_
    \item Would you like us to email you the survey result? \textit{(if yes, please leave your email below)}
\end{itemize}
\end{footnotesize}

\section{Stage two: Guideline evaluation survey}
\label{app:B}
\begin{footnotesize}
Thank you for taking the time to participate in our survey. This study aims to validate guidelines for designing adaptive user interfaces in mHealth applications by gathering feedback from software practitioners with experience in developing health-related applications. The survey will assess the applicability, clarity, and practicality of the guidelines, helping us determine their potential for integration into real-world development workflows.
\textbf{Important:} For the best experience, we recommend completing the survey on a desktop device. If you have any questions, please email wei.wang5@monash.edu. Ethical approval has been provided by Monash University. Further details can be provided upon request.
\begin{itemize}
    \item \textbf{Section 1: This section is intended to gather basic information about you.}
    \begin{itemize}
        \item In what age group are you?
        \item To which gender identity do you most identify as?
        \item In which country do you currently reside?
        \item How many employees work in your organisation?
        \item What is your role in the team?(tick all that apply)
        \begin{itemize}
        \item Project manager, Business consultant/Marketing manager/Sales personnel, Requirements analyst, Software architect, Programmer, User interface or Graphical User Interface designer/developer/engineer, App animator or operations developer/engineer, QA engineer, Tester, Other (Please specify)
        \end{itemize}
        \item How many years of experience do you have in development, design, or related fields?
        \item Do you have experience developing health-related applications, particularly those focused on chronic diseases?
        \begin{itemize}
        \item Yes (Please share details about your experience below.)
        \item Other (Please share details about your experience below.)
        \end{itemize}
    \end{itemize}
    
    \item \textbf{Section 2: This section is intended to give you some background information of Adaptive User Interfaces. Details in the User survey \ref{survey:adaptive}}
    \item \textbf{Section 3: Feedback on proposed guidelines.} In this section, we aim to gather your thoughts on the guidelines we’ve developed to improve the design of Adaptive User Interfaces in mHealth applications targeting chronic disease. Here is a link to our proposed guidelines. The subsequent questions will ask for your feedback, understanding, and opinions regarding the proposed guidelines. Please review the guidelines thoroughly before answering the following questions and ensure the guidelines are open in another window for reference. We have included brief comprehension check questions in the survey to ensure that all participants are familiar with the guidelines provided.
    \begin{itemize}
        \item Based on the guidelines you have reviewed, could you identify which guideline aligns with the following example design? \textit{"Customizations are applied as users log into the application for the first time, modifying the application entirely to suit their specific needs."}
        If you’re unsure of the answer, please take a moment to reread the guidelines before responding.
        \item Based on the guidelines you have reviewed, could you identify which guideline aligns with the following purpose? \textit{"Enhance the impact of implemented adaptations and engage users seamlessly without interrupting their normal application usage."}
        If you’re unsure of the answer, please take a moment to reread the guidelines before responding.
        \item Are you able to clearly understand and distinguish between the different guidelines?
        \item How do you perceive the usefulness of each individual guideline?
        \item Would you prefer using our proposed guidelines over existing, standalone guidelines in mHealth applications development process?
        \item What are our proposed guidelines' primary advantages (if any) to support design adaptations in mHealth applications targeting chronic disease (Feel free to choose specific guideline numbers below if applicable)?
        \item Are there any limitations or threats to the proposed guidelines (if they exist) for supporting design adaptations in mHealth applications targeting chronic disease (Feel free to choose specific guideline numbers below if applicable)?
        \item What do you think could be done to accommodate these limitations/threats in the next version of the guidelines (Feel free to choose specific guideline numbers below if applicable)?
        \item Please provide any other suggestions that you may have, e.g., about this research project, the developed guidelines etc. (Feel free to reference specific guideline numbers below if applicable.)
    \end{itemize}
\end{itemize}
\end{footnotesize}
\end{document}